\documentclass[preprint]{elsarticle}
\usepackage{hyperref}
\usepackage{natbib}
\usepackage{geometry}
\usepackage[utf8]{inputenc}
\usepackage[T1]{fontenc}
\usepackage{amsmath,amssymb,amsfonts}
\usepackage{graphicx}
\usepackage{mathalpha}
\usepackage[normalem]{ulem}
\usepackage{cancel}
\usepackage{braket}
\usepackage{booktabs}
\usepackage{bbold}
\usepackage{nicefrac, xfrac}

\numberwithin{equation}{section}

\newcommand{\trev}{t_{\rm rev}}
\newcommand{\tosc}{t_{\rm osc}}
\newcommand{\tdec}{t_{\rm decay}}
\newcommand{\trevo}{t^0_{\rm rev}}
\newcommand{\tosco}{t^0_{\rm osc}}
\newcommand{\ttosco}{\tilde t^0_{\rm osc}}
\newcommand{\tdeco}{t^0_{\rm decay}}
\newcommand{\ttrev}{\tilde t_{\rm rev}}
\newcommand{\ttrevo}{\tilde t^0_{\rm rev}}
\renewcommand{\tt}{{\tilde t}}
\newcommand{\trevauto}{t_{\rm rev}^{\rm auto}}

\newcommand{\ttheta}{\tilde \theta}
\newcommand{\tttheta}{\tilde {\tilde \theta}}

\newcommand{\mean}[1]{\langle \hat #1 \rangle }
\newcommand{\TWA}[1]{\langle \hat #1 \rangle_{\rm TWA} }
\newcommand{\Boot}[1]{\langle \hat #1 \rangle_{\rm emb} }
\newcommand{\round}[1]{\lfloor #1 \rceil}
\newcommand{\floor}[1]{\lfloor #1 \rfloor}
\newcommand{\tf}{\tilde f}

\newcommand{\tH}{\widetilde{H}}
\newcommand{\Nav}{\mathcal{N}_\alpha}  
\newcommand{\ntot}{n_{\rm tot}} 
\newcommand{\barn}{\bar n_2}    
\newcommand{\ntoth}{\hat n_{\rm tot}}
\newcommand{\la}{\mathfrak{k}_\alpha}
\newcommand{\na}{\la}

\newcommand{\coh}{c_2}  
\newcommand{\cohun}{c_1}  
\newcommand{\autocor}{\mathcal{P}} 
\newcommand{\autocormax}{\mathcal{P}^{(1)}_{\rm max}} 

\newcommand{\MMinv}{\mathbf{M}}
\newcommand{\MM}{\mathbf{M}}  
\newcommand{\NN}{\mathbf{N}}
\newcommand{\KKb}{\Delta \mathbf{R}_{\rm em}}
\newcommand{\KK}{\Delta R_{\rm em}}
\newcommand{\AAA}{\mathbf{A}}
\newcommand{\BB}{\mathbf{B}}
\newcommand{\CC}{\mathbf{C}}
\newcommand{\AAt}{\mathbf{A}_t}
\newcommand{\BBt}{\mathbf{B}_t}
\newcommand{\CCt}{\mathbf{C}_t}
\newcommand{\AAb}{\mathbf{A}_{\rm em}}
\newcommand{\BBb}{\mathbf{B}_{\rm em}}
\newcommand{\CCb}{\mathbf{C}_{\rm em}}

\newcommand{\GGS}{\Delta\varphi_S^c}
\newcommand{\GGm}{\Delta\varphi_m^c}
\newcommand{\GGSe}{\Delta\varphi_S^{c,\rm em}}
\newcommand{\GGme}{\Delta\varphi_m^{c,\rm em}}

\newcommand{\GG}{\mathbf{\Delta \Phi}^c}
\newcommand{\DGr}{\GG_{\rm em}}
\newcommand{\DG}{\Delta \Phi^c_{\rm em}}

\newcommand{\DPSe}{\Delta\varphi_S^{\rm em}}
\newcommand{\DPme}{\Delta\varphi_m^{\rm em}}

\newcommand{\cmod}[1]{ \left(#1\right) \mbox{cmod}[2\pi]}
\newcommand{\cmodun}[1]{ \left(#1\right) \mbox{cmod}[1]}

\newcommand{\nW}{[n]^W}
\newcommand{\ntwoW}{[n_2]^W}

\newcommand{\aW}{[a]^W}
\newcommand{\AW}{[A]^W}
\newcommand{\BW}{[B]^W}
\newcommand{\HW}{[H]^W}

\newcommand{\bI}{\mathbf{I}}
\newcommand{\bIa}{\bI_\alpha}
\newcommand{\bPhi}{\mathbf{\Phi}}
\newcommand{\bOmega}{\mathbf{\Omega}_\alpha}
\newcommand{\bk}{\mathbf{k}}

\newcommand{\bp}{\mathbf{p}}
\newcommand{\bq}{\mathbf{q}}

\newcommand{\rep}{r}  

\newcommand{\zee}{z}   
\newcommand{\gs}{\gamma_{\mathfrak{s}} }

\newcommand{\Cstar}{\mathcal{C}_*}
\newcommand{\Ca}{\mathcal{C}_\alpha}
\newcommand{\heff}{ \hbar_{\rm eff} }
\newcommand{\Ia}{\mathcal{I}_\alpha}               
\newcommand{\Ea}{\mathcal{E}_\alpha}               
\newcommand{\Sa}{\mathcal{S}_\alpha}               
\newcommand{\So}{S_0}  
\newcommand{\ma}{m_\alpha}
\newcommand{\hma}{{\hat m}_\alpha}
\newcommand{\Da}{\Delta_\alpha}               
\newcommand{\bDa}{\mathbf{\Delta}_\alpha}     
\newcommand{\heffS}{{\scriptstyle \frac{\heff}{\Sa}}}
\newcommand{\dna}{\delta \mathfrak{l}_\alpha}  

\newcommand{\GTWA}{\mathcal{G}^{\rm TWA}}  
\newcommand{\GBoot}{\mathbf{\mathcal{G}}^{\rm emb}}  

\newcommand{\Wr}{W^{\rm (r)}}

\newcommand{\smeq}{ \! = \! }

\begin{document}

\begin{center}{\Large\textbf{
Multidimensional revival dynamics in a bosonic many-body system
}}
\end{center}

\begin{center}
Denis Ullmo$^{1,*}$, Steven Tomsovic$^{1,2}$, Gabriel M.~Lando$^{1}$ and  Peter Schlagheck$^3$
\end{center}

\begin{center}
$^1$Université Paris-Saclay, CNRS, LPTMS, 91405, Orsay, France \\
$^2$Department of Physics and Astronomy, Washington State University, Pullman, WA 99164-2814 USA \\
$^3$CESAM research unit, University of Liege, 4000 Liège, Belgium \\
\smallskip
* denis.ullmo@universite-paris-saclay.fr
\end{center}

\begin{center}
\today
\end{center}



\section*{Abstract}
The collapse and revival of localized states in many-body systems is a quintessential example of a quantum interference phenomenon that cannot be explained purely within mean-field theories.  Historically, the majority of previous work on quantum revivals has been in a Schrödinger mechanics context, typically for effectively one-degree-of-freedom systems with a focus on the autocorrelation function as the key detection observable. The general revival subject acquires a far greater complexity in the bosonic many-body context that offers a larger variety of observables and intrinsically implies high dimensionality. Here, the focus is on the simplest nontrivial (i.e., not one-dimensional) bosonic many-body system, namely a symmetric two-site Bose-Hubbard model (generally dubbed Bose-Hubbard dimer).  The non-equilibrium situation of a coherent initial state, physically corresponding to a perfect Bose-Einstein condensate, prepared on one of the two sites of this integrable two-degrees-of-freedom system in the strongly self-trapping regime is considered. The revival behaviors emerging from the resulting dynamics within expectation values (e.g.~occupancies), coherences, and the autocorrelation function are theoretically derived, and the entanglement behavior is illustrated.  Several new features, unknown so far in the Schrödinger context and identified here in terms of five ``puzzles'', are encountered and discussed, among them, as pointed out in [Phys. Rev. A {\bf 106}, L051302 (2022)], a synchronized resurgence behavior, differing revivals times depending on the specific observable, and a manifestation of a precession that appears in the underlying mean field dynamics.  These new phenomena can be entirely understood through a comprehensive semiclassical theory, building upon the underlying classical (mean-field) phase-space structure,  thus greatly reducing the complexity as compared to the full quantum many-body dynamics.  A quantitative comparison of the latter with semiclassical predictions and quantum perturbative calculations yields an overall excellent agreement and confirms that a full semiclassical theory is pivotal for the understanding of these genuinely quantum many-body interference phenomena.

\newpage

\vspace{10pt}
\noindent\rule{\textwidth}{1pt}
\setcounter{tocdepth}{2}
\tableofcontents
\noindent\rule{\textwidth}{2pt}
\vspace{10pt}

\newpage

\section{Introduction}
\label{sec:intro}

Revivals of wave packets figure among the most striking and spectacular manifestations of interference in quantum dynamical processes \cite{Robinett04,Frahm66,Narozhny81,Nauenberg90,Bluhm96}.
Originally studied in the framework of atomic and molecular physics \cite{Parker86,Yeazell90,Yeazell91,Meacher91,Wals94,Ergler06,Rudenko06,Seltzer07}, they were, more recently, identified and discussed in a variety of other physical contexts as well, including ultracold gases \cite{Greiner02b,Fischer08,Fischer11}, graphene \cite{Krueckl09,Romera09}, superconducting qubits \cite{Kirchmair13}, spin chains \cite{Genest16,Michailidis20,Manju25}, nanorotors \cite{Stickler18}, as well as cosmology \cite{daSilva15}, to mention a few examples.
Revivals were repeatedly suggested as key tools to be exploited for the purpose of quantum metrology (e.g., for the measurement of magnetic fields \cite{Seltzer07}) and quantum sensing \cite{Stickler18}.
Indeed, the fact that a minimum-uncertainty wave packet can regain again its initial shape at some point in its time evolution, long after having spread out due to the dispersion of the underlying energy eigenspectrum, is a clear signature of quantum coherence being preserved throughout the entire evolution of the packet, and can, in principle, be used to detect the presence (or to confirm the absence) of any coherence-breaking mechanism affecting the system under study.

The phenomenon of revivals is not restricted to the realm of single-particle physics but occurs in many-body systems as well.
However, in most of the few- and many-body revival studies that have been carried out (e.g., \cite{Greiner02b,Ergler06,Rudenko06}), the wave-packet dynamics under consideration can effectively be described in terms of one single degree of freedom, which eliminates complications arising from the high dimensionality of the underlying phase space.
Those complications, when becoming relevant, i.e., in settings that deliberately account for effects beyond the single-mode description (through including, within lattice systems, the presence of hopping \cite{Fischer08,Fischer11,Tomsovic18} or disorder \cite{Manju25}, for instance), are generally found to amount to a degradation of the revival feature \cite{Fischer08,Fischer11,Tomsovic18,Manju25}, thus reinforcing common conviction that the occurrence of revivals is incompatible with the presence of chaos and complexity (see, however, \cite{Tomsovic97} for an example of revivals occurring in a chaotic system).
From that point of view, whereas revivals can generally be considered as a sensitive indicator for long-time coherence and interference in a quantum system, they seem to be unsuited to specifically probe the various Fock-space interference effects that allow one to reveal particle indistinguishability \cite{Tichy12,Giordani18,Dittel18,Dufour20,Brunner23,Dittel24} and that lie at the heart of a number of localization phenomena in many-body dynamical processes, such as coherent backscattering in Fock space \cite{Engl14b,Engl18}, symmetry enhancements of return probabilities \cite{Schlagheck19}, or the formation of genuine many-body scars \cite{Hummel23}, being thus of great relevance for our general understanding of the limitations and breakdowns of the quantum-classical correspondence in many-body systems.

The present article aims at proving the contrary, or, more modestly, at opening a door towards the exploration of complex revival dynamics within genuinely multidimensional many-body systems, beyond simple degradation effects of the revival.
More precisely, we convey the message that revivals can exhibit an astonishing complexity, reflecting interference effects on various levels and within various observables, in the framework of quantum many-body coherent state propagation processes that genuinely involve two degrees of freedom, provided the dynamics within this (possibly reduced) two-degrees-of-freedom system remain integrable.
Specifically, following up on a previous study of us \cite{Schlagheck22}, which was strongly inspired by the revival experiment of Greiner \textit{et al.} using a Bose-Einstein condensate in an optical lattice \cite{Greiner02b}, we consider coherent-state propagations in a bosonic two-mode system, generally dubbed ``Bose-Hubbard dimer'' \cite{Campbell20},
which serves as a model for bosonic Josephson junctions~\cite{Milburn97, Smerzi97}.  
It remains an integrable system as energy and particle number are both conserved.  Thus, chaos and thermalization do not occur in this system, and revival features can therefore be detected at very long evolution times~\cite{Milburn97, Mahmud05} as in the case of the single-site dynamics studied in Ref.~\cite{Greiner02b}, even though they are now modulated by the superposition of two independent dynamical modes, one related to the on-site phase oscillation and another one to the site population exchange.  
This leads to a much richer collection of revival phenomena~\cite{Schlagheck22}.

Going further and attempting to perform these collapse and revival studies in extended lattices with finite hopping does not appear worth the effort at first glance, in view of the huge increase of dimensionality that is thereby introduced, giving rise to a breaking of integrability and thus, most generally, to a destruction of the revival features~\cite{Fischer08, Fischer11}.  However, in specific contexts revivals occur nevertheless, despite the presence of inter-site hopping (which, as demonstrated in \cite{Schlagheck22}, can represent a precious additional ingredient that enables a sensitive detection of revival features through single-particle on-site observables).  
One example is the case in which the dynamics within a symmetry plane in the lattice can be mapped onto a Bose-Hubbard dimer~\cite{Steinhuber20}.
This is possible either through the implementation of a superlattice structure~\cite{Folling07} or through the preparation of a quantum state in the vicinity of a density wave of the form $(\ldots,\psi_1,\psi_2,\psi_1,\psi_2,\ldots)$ under the assumption that deviations from such a density wave structure do not appreciably manifest themselves for the evolution times under consideration (e.g.,calculations in \cite{Tomsovic18} showed revivals within 4- and 6-site Bose-Hubbard rings).  

It is demonstrated in Ref.~\cite{Schlagheck22} that the interplay of  the two dimer modes can, in a parameter regime of macroscopic quantum self-trapping~\cite{Smerzi97}, lead to a series of very intriguing features associated with revivals.  One example is the occurrence of a resurgence phenomenon: while the presence of hopping between the sites generally gives rise to a reduction of the revival amplitude, quantitatively manifested here through the return probability to the initial state, local enhancements of that return probability nevertheless occur at specific values of the hopping strength, which turn out to coincide with a synchronization of that revival with the phase of population exchange dynamics~\cite{Schlagheck22} (further examples will be given in Sec.~\ref{sec:RevInBHdimer}).
Moreover, the involvement of a second site can give rise to a valuable diagnostic tool of the revival dynamics in terms of experimentally accessible single-particle observables, namely the on-site occupancies, which become particularly effective if a maximally imbalanced state (i.e., with one site being totally empty) is initially prepared in the dimer~\cite{Schlagheck22}.

The key objective of the present paper is to provide solid analytical foundations for these resurgent phenomena using quantum perturbative and semiclassical approaches. The latter methods are primarily based on purely classical properties of the system, but nevertheless fully account for non-classical effects such as quantum interference if carried out fully~\cite{Tomsovic18}, and can be very accurate in the deep semiclassical regime when $\hbar \to 0$.  In the context of ultracold bosonic atoms in optical lattices, i.e., the physical system that is most commonly associated with a Bose-Hubbard model, the semiclassical approach can be implemented provided an effective $\hbar$ parameter, given in terms of the inverse of the particle density, is introduced~\cite{Richter22}. 
The semiclassical approach, which can thus be seen as an approximate quantum description valid for small values of the inverse filling factor is benchmarked against a quantum perturbative approach valid for weak inter-site hopping, yielding good quantitative agreement with each other and with the exact quantum dynamics of the system in the parameter regime of interest.  Furthermore, in contrast to approximation-free numerical simulations of the quantum dynamics, it allows one to gain comprehensive quantitative insight into key classical and non-classical mechanisms that are at work in the Bose-Hubbard dimer, in particular, quantum interference and the interplay between on-site coherence oscillations and inter-site hopping dynamics.

The paper is organized as follows: Section~\ref{sec:rev-1d} provides, as an introduction, an extensive discussion of the traditional one-dimensional revival phenomenon, but with a greater emphasis on and review of a complete semiclassical theory and extensions to the bosonic many-body realm.
Section~\ref{sec:RevInBHdimer} combines background necessary for this current work, including aspects of the Bose-Hubbard dimer, and curious revival phenomena recently noted for the dimer in Ref.~\cite{Schlagheck22}.  The section following that one contains the physical picture emerging from semiclassical and quantum perturbative methods that explains all the observed phenomena quantitatively and qualitatively, but without the myriad of details relegated to the later sections and appendices.  The semiclassical theories are contained in Sec.~\ref{sec:classical-dimer} and \ref{sec:semirev}, and the details of the quantum perturbation theory combined with saddle point approximations in Appendices~\ref{app:qpt} and \ref{app:qcf}.  This is followed by concluding remarks in the final section, including a brief discussion of how such revivals could be used as a ``quantum resource''. Discussions of a few fine points in some of the derivations are gathered in Appendices~\ref{app:wigner-function} to \ref{app:DeltaPhiS}.

\section{General theory of one-dimensional wave packet revivals}
\label{sec:rev-1d}

\subsection{Collapse and revivals of wave packets in one-degree-of-freedom systems: Schrödinger}
\label{sec:rev-1d-quant}

For pedagogical purposes, and to properly set the stage of our discussions, let us first review the elementary collapse and revival dynamics that are generically encountered within one-degree-of-freedom ($1D$) systems in Schrödinger mechanics~\cite{Nauenberg90, Bluhm96, Robinett04}.  Those systems are generally characterized by classical integrability (provided they allow for a meaningful semiclassical description, which would not be the case, e.g., for a spin 1/2 degree of freedom) and thus exhibit a good quantum number which we shall assume to be non-negative and which is denoted by $n$ in the following. 
Thus, 
\begin{equation} 
\label{eq:quant-ham-1d}
\hat H \ket{n} = E_n \ket{n} \ , \quad n \in \mathbb{N} \, ,
\end{equation}
with $\ket{n}$ and  $E_n$ being the eigenstates and eigenenergies of  the quantum Hamiltonian $\hat H$, respectively.

Consider a localized wave packet initial state, $\ket{\alpha}$, in such a system prepared so that its decomposition within the system's eigenbasis, as given by the energy eigenstates, $\ket{n}$, is characterized by expansion coefficients $c_n = \braket{n|\alpha}$ whose absolute values as a function of $n$ are strongly peaked about a given (not necessarily integer) reference value $\na  \gg 1$.
The time-dependent quantum state of the system is thus given by 
\begin{equation}
\ket{\alpha_t} = \sum_{n=0}^\infty c_n e^{-i E_n t / \hbar}\ket{n} 
\label{eq:alpha}
\end{equation}
where $E_n$ denotes the energy associated with the eigenstate $\ket{n}$\footnote{An infinite dimensional Hilbert space is implicitly assumed, corresponding to the situation that is most frequently encountered in physics. Quantum tops that rotate with a fixed angular momentum would represent an exception to that rule, since the inclination of their angular momentum vector with respect to a given spatial axis can only acquire a finite number of quantized values. Note that the formalism developed here can be straightforwardly adapted to incorporate the description of such finite systems as well.}.

As $n$ is a good quantum number (and the only one), the eigenenergies $E_n$ vary smoothly with $n$.  Consequently, it is possible to locally approximate the eigenspectrum in terms of a parabolic scaling in $n$, and thus expand them in the vicinity of this reference value $\na$ according to
\begin{equation}
\label{eq:expans-sec2}
  E_n = \Ea + \frac{2\pi \hbar}{\tosc} (n - \na) + \frac{\pi\hbar}{\trev}(n - \na)^2 + O\left[(n - \na)^3\right] \ .
\end{equation}
In the case of the hydrogen atom, one would for instance have $E_n \equiv E(n) = - R / n^2$, with $R$ the Rydberg constant. Thus, $\Ea = - R / \na^2$, $2\pi / \tosc = 2 R / \na^3$ and $2\pi/\trev = - 6 R / \na^4$.
Equation~\eqref{eq:expans-sec2} can be fully justified by the semiclassical theory of Einstein-Brillouin-Keller (EBK) quantization~\cite{Einstein17, Brillouin26, Keller58} (the multidimensional extension of the one dimensional Wentzel-Brillouin-Kramers quantization~\cite{Jeffreys24, Wentzel26, Kramers26}), where the two time scales, $\tosc$ and $\trev$, are fixed by classical quantities (see Eq.~\eqref{eq:ts} and Sec.~\ref{sec:semic-1d} ahead).
Alternatively, and perhaps more often considered~\cite{Nauenberg90, Bluhm96, Robinett04}, the two characteristic time scales can approximately be determined, assuming $\na = n_0$ integer, through discrete energy differences $2\pi / \tosc \simeq (E_{n_0+1}-E_{n_0-1})/2\hbar$ and $2\pi/\trev \simeq (E_{n_0+1}+E_{n_0-1}-2E_{n_0})/\hbar$, and which would respectively result from the first and second discrete $n$-differences of the energies evaluated at $n_0$.

In the special case of a strictly linear spectrum, characterized by the exact expression $E_n = \Ea + (2\pi\hbar/\tosc) (n - \na)$ for the eigenenergies (i.e. the quadratic term and higher-order corrections to Eq.~\eqref{eq:expans-sec2} vanish), the wave packet performs harmonic oscillations yielding perfect recurrences without dispersion, i.e., up to a global (and unimportant) phase $\ket{\alpha_t}$ returns to its initial state $\ket{\alpha}$ at integer multiples of the period $\tosc$, irrespective of the precise values of $c_n$.

Such recurrences occur also in more general integrable systems (that do not correspond to a perfectly harmonic oscillator), but they eventually decay due to the presence of nonvanishing quadratic and higher-order terms in Eq.~\eqref{eq:expans-sec2}.
Those anharmonic terms give rise to an increasing dephasing of the wave packet's components and hence to a \textit{collapse} of the wave packet's motion, it becoming dispersed over the entire classical orbit.
Both the recurrences and the collapse of the wave packet are essentially \textit{quasiclassical} phenomena, in the sense that the time-dependent behavior of all relevant observables displaying those phenomena can be quantitatively reproduced by a purely classical calculation, using the initial uncertainty in position and momentum of the wave packet as the only nonclassical ingredient.

However, for times close to integer multiples of $\trev$ the fixed phase relationship between the wave packet's components is ``magically'' restored, and the periodic oscillations seen at short propagation times become visible again.    
Indeed, defining $\delta n = (n - \round{\na}) \in \mathbb{Z}$ with $\round{\na}$ the nearest integer to $\na$,  $\delta \na = (\na - \round{\na}) \in \mathbb{R}$, and
$K = \Ea \trev/\hbar + 2\pi\delta \na (\delta \na/2 -  \trev/\tosc)$,  the phases at $t=\trev$ in Eq.~\eqref{eq:alpha} making use of the second-order expansion of the eigenenergies, Eq.~\eqref{eq:expans-sec2}, can be exactly rewritten as
\begin{equation} 
\label{eq:Etrev}
  \frac{E_n \trev}{\hbar}   = 
   2\pi\delta n \left(\frac{\trev}{\tosc} + 1/2  - \delta \na \right) + K 
  \quad [{\rm modulo} \; 2\pi]
\end{equation}
up to a correction $\pi \delta n(\delta n + 1)$ that is an integer multiple of $2\pi$ and can thus be neglected upon insertion of the above expression into Eq.~\eqref{eq:alpha}.
Locally near $\trev$ the spectrum is effectively linear, and the evolution behaves as if the wavepacket is following a harmonic-like oscillation except with an extra phase
\begin{equation} 
\label{eq:DeltaTh}
\Da = \pi (1- 2\delta \na)  \; ,
\end{equation}
relative to the linear part of the expansion in Eq.~\eqref{eq:expans-sec2}.

For a short period of time near $\trev$, which is roughly equivalent to the collapse time scale, the wave packet \textit{revives}.  Note that both the collapse and the revival of the wave packet are determined by the quadratic expansion term of the system's eigenenergies.
In contrast to the collapse, however, the phenomenon of the wave packet's revival is intrinsically \textit{nonclassical} as it requires quantum interference in order to emerge (and thus also the absence of any decoherence mechanism that would result from the coupling of the system to an environment), which is impossible to reproduce with classical or quasiclassical methods.

\subsubsection{Return probability}
\label{sec:retprob}

In Schrödinger mechanics, the collapse and revival dynamics of a wave packet is typically revealed by the return probability of the wave packet (autocorrelation function) to its initial state, $\autocor(t) = |\mathcal{A}(t)|^2$, which is given in terms of the absolute square of the temporal amplitude
\begin{equation}
\label{eq:autocs}
    \mathcal{A}(t) = \braket{\alpha|\alpha_t} = \sum_{n=0}^\infty |c_n|^2 e^{-i E_n t / \hbar} \,.
\end{equation}
Considering the strictly quadratic spectrum given by Eq.~\eqref{eq:expans-sec2}, with cubic and higher terms being neglected, following Ref.~\cite{Nauenberg90}
\begin{align}
   & \mathcal{A}(t) =  e^{-i \Ea t / \hbar} \sum_{n=0}^\infty |c_n|^2 
    \exp \left[-2i\pi \left( (n - \na) \frac{t}{\tosc} 
    + \frac{(n - \na)^2}{2} \frac{t}{\trev} \right) \right]
    \label{eq:autocorrelation} \\ 
    & = e^{-i \Ea t / \hbar} \sum_{l=-\infty}^\infty 
    \int_{-\infty}^\infty dn {e^{2\pi i n l}} |c_{n}|^2 
    \exp \left[ -2i\pi \left( (n - \na) \frac{t}{\tosc} 
    + \frac{(n - \na)^2}{2} \frac{t}{\trev} \right) \right] \,, 
    \label{eq:apoisson}
\end{align}
with Poisson summation being employed to yield Eq.~\eqref{eq:apoisson} and the assumption that $\na$ is large allows for the extension of the lower integration limit.

Just to consider a specific example for the expansion coefficients, let
\begin{equation}
c_n = (2\pi\Delta n^2)^{-1/4} \exp\left[-\frac{1}{4\Delta n^2} (n - \na)^2 + i n \theta \right] 
\label{eq:cgauss0}
\end{equation}
where although $\Delta n^2$ could vary over a substantial range $1\ll \Delta n^2 \ll n^2 $, let it equal $\na$.  Under a wide variety of circumstances, this is approximately what emerges asymptotically for minimum uncertainty wave packets, though the discussion in this section is more general and does not require them.  The integral over $n$ in Eq.~\eqref{eq:apoisson} can be performed yielding
\begin{equation}
    \mathcal{A}(t) \simeq \frac{e^{-i \Ea t / \hbar}}{\sqrt{1 + 2i\pi \Delta n^2 \,  t/\trev}}
    \sum_{l = -\infty}^\infty \exp\left[ - \frac{2\pi^2 \Delta n^2}{1 + 2i\pi\Delta n^2 \, t/\trev}
    \left( \frac{t}{\tosc} - l \right)^2  + 2\pi i l \na\right] \,,
    \label{eq:autocorrgauss}
\end{equation}
in agreement with Ref.~\cite{Nauenberg90}.

For short times $0 < t \lesssim \Delta n^{-1} \trev$ the individual Gaussian terms in the sum over $l$ do not appreciably overlap with each other, and $\autocor(t)$ is approximately
\begin{equation}
    \autocor(t) \simeq \frac{1}{\sqrt{1 + (2\pi\Delta n^2 \, t /\trev)^2}} \sum_{l = -\infty}^\infty\exp\left[ -\frac{4 \pi^2 \Delta n^2 }{1 + (2\pi \Delta n^2 \, t/\trev)^2}\left(\frac{t}{\tosc} -  l\right)^2\right] \,. 
    \label{eq:Pcollapse}
\end{equation}
This expression describes a train of recurrence peaks which pop up at integer multiples of $\tosc = 2\pi/\omega_{\rm osc}$ and which decrease in sharpness and amplitude with increasing time.  It gives a highly detailed monitoring of the collapse of the wave packet oscillations.

For significantly longer times $t \gg \Delta n^{-1} \trev$ the evaluation of the absolute square of the expression, Eq.~\eqref{eq:autocorrgauss}, becomes more involved since overlaps between adjacent terms in the sum can then no longer be neglected.  To calculate the return probability for a time $t = \rep \trev + \delta t$, with $\rep\in\mathbb{Z}$,  note that in Eq.~\eqref{eq:autocorrelation}, the resultant term $-\pi i n^2 r \trev/\trev$ is a phase of an integer multiple of $ \pi$, being $(-1)^{nr}$.  In proceeding to Eq.~\eqref{eq:apoisson}, it just shifts $l \to l-r/2$.  Also, Eq.~\eqref{eq:Etrev} leads to
\begin{equation}
\frac{E_n t}{\hbar} = \frac{E_n \delta t}{\hbar} +
   \rep \left[2\pi\delta n \left(\frac{\trev}{\tosc} +\frac{1}{2} -\delta\na  \right) + K \right]
  \quad [{\rm modulo} \; 2\pi]  \; .
\end{equation} 
Thus, for small $\delta t$ Eq.~\eqref{eq:Pcollapse} applies with the substitutions $l \to l-r/2$, $t \to \delta t$ in the normalization and variance term, and $t/\tosc \to [t/\tosc - \rep \na]$ giving 
\begin{align}
    \autocor(t & \smeq 
   \rep \trev + \delta t) \simeq \nonumber \\
   & \frac{1}{\sqrt{1 + (2\pi\Delta n^2 \, \delta t /\trev)^2}} \sum_{l = -\infty}^\infty\exp\left[ -\frac{4 \pi^2 \Delta n^2}{1 + (2\pi \Delta n^2 \, \delta t/\trev)^2} \left(\frac{t}{\tosc} - l -\rep(\na-1/2)\right)^2\right] \; . 
    \label{eq:Previval} 
\end{align}
As in the initial collapse regime, the return probability exhibits a train of recurrence peaks near the integer multiples of the revival time, whose heights fall off and widths increase with $\delta t$ in an identical manner.
A perfect recovery of the initial state (up to a global phase), characterized by $\autocor = 1$, is obtained if 
\begin{equation}
\rep\left(\frac{\trev}{\tosc} - \na+\frac{1}{2}\right) = l\in\mathbb{Z}
\label{eq:perfectrev}
\end{equation}
is an integer, whereas otherwise the recurrence peaks are phase-shifted with respect to the revival peak times.

\subsubsection{Fractional revivals}
\label{sec:fracrev}

Fractional revivals, occurring at non-integer but nevertheless rational multiples of the revival time, manifest themselves through very similar trains of recurrence peaks~\cite{Yeazell91, Robinett04}.  As an example, consider the case of half-revivals, occurring at half-integer multiples of the revival time.  Defining $\delta t = t - (\rep + 1/2) \trev$ for some integer $\rep\in\mathbb{Z}$,  a similar transformation of Eqs.~\eqref{eq:autocorrelation} and \eqref{eq:apoisson} takes place as for the full revival, but with a few key differences.  First, there is a term which is unity for even $n$ terms and $-i(-1)^r$ for odd terms.  Essentially in Eq.~\eqref{eq:autocorrelation} the sum has the structure
\begin{equation}
\label{eq:autodecomp}
\sum_{n=0}^\infty f(n) = \sum_{m=0}^\infty f(2m) - i(-1)^r \sum_{m=0}^\infty f(2m+1) \ .
\end{equation}
and the even and odd terms appear as
\begin{equation}
    \mathcal{A}(t) =\mathcal{A}_{\rm even}(t) + \mathcal{A}_{\rm odd}(t)
\end{equation}
with
\begin{align}
    \mathcal{A}_{\rm even}(t) = & e^{-i \Ea t / \hbar + \pi i \na^2(r+1/2)} \sum_{m=0}^\infty |c_{2m}|^2 \nonumber \\ & \times
    \exp \left[-2i\pi \left( \left[2m - \na\right]\left[ \frac{t}{\tosc} -\na(r+1/2)\right]
    + \frac{(2m - \na)^2}{2} \frac{\delta t}{\trev} \right) \right]
   \label{eq:autocorrelatione} \,, \\ 
    \mathcal{A}_{\rm odd}(t)= & -i(-1)^r e^{-i \Ea t / \hbar + \pi i \na^2(r+1/2)}  \sum_{m=0}^\infty |c_{2m+1}|^2 \nonumber \\
  & \times \exp \left[-2i\pi \left( \left[2m + 1 - \na\right]\left[ \frac{t}{\tosc} -\na(r+1/2)\right]
    + \frac{\left[2m +1 - \na\right]^2}{2} \frac{\delta t}{\trev} \right) \right] \, . 
    \label{eq:apoissono}
\end{align}
Introducing the Poisson summation separately for the even and odd terms generates two nearly equivalent Gaussian integrals except fora small centroid shift, which introduces an additional $l$-dependent phase, but otherwise generates the same result.   That gives
\begin{align}
    \mathcal{A}(t) \simeq &\frac{e^{-i \Ea t / \hbar + \pi i \na^2(r+1/2)} }{2 \sqrt{1 + 2i \pi \Delta n^2 \, \delta t/\trev }}
    \sum_{l = -\infty}^\infty \left(1 - i(-1)^{\rep+l}\right) e^{i\pi l \na} \notag \\
    & \times \exp\left(  \frac{-2\pi^2 {\Delta n^2}}{1 + 2i\pi\Delta n^2 \, \delta t/\trev} \left[\frac{t}{\tosc} - \frac{l}{2} - \left(\rep + \frac{1}{2}\right) \na \right]^2\right) \,.
\end{align}
For small deviations from the half-revival time, with $|\delta t| \lesssim \Delta n^{-1} \trev /2$, the return probability is
\begin{equation}
    \autocor(t) \simeq \frac{1}{2\sqrt{1 + \left( \frac{2\pi\Delta n^2 \,\delta t}{\trev}\right)^2}}
    \sum_{l = -\infty}^\infty\exp\left( \frac{-4\pi^2 \Delta n^2}{1 + \left( \frac{2\pi \Delta n^2 \, \delta t}{\trev}\right)^2} \left[\frac{t}{\tosc}  - \frac{l}{2} - \left(\rep + \frac{1}{2}\right) \na\right]^2\right) \,. \label{eq:Pfracrev}
\end{equation}
As compared to the full revivals [see Eq.~\eqref{eq:Previval}], Eq.~\eqref{eq:Pfracrev} describes a train of recurrence peaks that occur twice as frequently, namely with the period $t_{\rm osc/2}$, but are reduced in height by a factor $2$.
In more intuitive terms, near the half-revival time the wave packet is split into two separate parts that oscillate with opposite phases along the classical periodic orbit. 
A partial recovery of the initial state arises twice as often as in the full revival case, but each time only with half of the total population.

Both integer and fractional revivals are affected by the presence of cubic and higher terms in the expansion of the system's eigenenergies, Eq.~\eqref{eq:expans-sec2}.   As discussed in Sec.~\ref{sec:semic-1d}, quartic corrections and higher terms in the expansion are parametrically small.  Consequently, they give rise to a reduction of the return probability, as compared to Eqs.~\eqref{eq:Previval} and \eqref{eq:Pfracrev}, which are typically negligible, and similar temporal attenuation occurs for the revivals of the coherence observable, Eq.~\eqref{eq:coherenceGauss}. 
The influence of the cubic term, which may in some circumstances be more important, is discussed in more details in Sec.~\ref{sec:semic-1d}.

\subsection{Collapse and revivals of bosonic coherent states in one-degree-of-freedom systems}
\label{sec:rev-1d-quantb}

The revival discussion of Sec.~\ref{sec:rev-1d-quant} can be adapted in a relatively straightforward way to second quantized bosonic localized states.  Nevertheless, there are some distinctions worth noting and some differences in interpretations.  This is reviewed next.

\subsubsection{Coherent states}
\label{sec:gcs}

For simplicity of the discussion, let us consider normalized Glauber coherent states describing initial states of bosonic many-body systems with a single degree of freedom~\cite{Glauber63}.  They are minimum uncertainty states in quadratures and can be expressed compactly in terms of the particle creation operator $\hat a^\dagger$ as
\begin{equation}
\label{eq:cse}
| \alpha \rangle_z = \exp \left(-\frac{\left| z \right|^2}{2} + z \hat a^\dagger \right)| 0\rangle
\end{equation}
with the properties that
\begin{equation} \label{eq:a}
\hat a | \alpha \rangle_z = z | \alpha \rangle_z , \quad \left[ \hat a, \hat a^\dagger\right] = \mathbb{1}\ , \quad \langle \alpha | \hat a^\dagger \hat a | \alpha \rangle_z = |z|^2 \ .
\end{equation}
The $z$ subscript is assumed and dropped from the notation going forward.  Hence, the mean number of bosons of $\ket{\alpha}$ is $\Nav = \langle \alpha | \hat n | \alpha \rangle = |z|^2$.  In general, $\ket{\alpha}$ does not remain a coherent state under evolution, $|\alpha_t\rangle = U(t) | \alpha \rangle$, and the return probability is given by $\autocor(t) = |\braket{\alpha|\alpha_t}|^2$. 

Quadrature (Hermitian) operators may be introduced as follows
\begin{equation}
\label{eq:quadratures}
\hat p  =  i \sqrt{\frac{\heff}{2}}\left( \hat a^\dagger - \hat a \right) \qquad 
\hat q  =   \sqrt{\frac{\heff}{2}}\left( \hat a^\dagger + \hat a \right) \; ,
\end{equation}
where the constant  $\heff$  is written to look like the usual $\hbar$ to emphasize the analogy,  and  is rather usefully set to be inversely proportional to the number of bosons making the $\Nav \rightarrow \infty$ limit equivalent to the $\heff \rightarrow 0$ limit.   The operator pairs $(\hat q,\hat p)$  behave mathematically exactly like the quantized version of a canonical pair of momentum and position variables, $[\hat q, \hat p] = i \heff$, even though they represent real and imaginary parts of an annihilation operator.
Setting $ \langle \alpha | \hat q | \alpha \rangle =  q_\alpha$ and  $\langle \alpha | \hat p | \alpha \rangle = p_\alpha$,  then $z = \sqrt{\Nav}{\rm e}^{i\theta} = (q_\alpha + i p_\alpha)/\sqrt{2\heff}$.  
Using the quadrature operator $\hat q$'s eigenbasis representation $\hat q | q \rangle = q | q \rangle$, the coherent state takes the form
\begin{align}
\label{eq:cohquad}
\langle q | \alpha \rangle =  \left(\frac{1}{\pi \heff}\right)^{1/4} & \exp\left[ - \frac{\left(\ q - q_\alpha \right)^2}{2\heff}  + \frac{i p_\alpha}{\heff} \left(q - q_\alpha \right) + \frac{i}{2 \heff}p_\alpha q_\alpha \right] \ .
\end{align}
Thus, in quadratures the coherent state results in a Gaussian wave packet form with parameters which can be straightforwardly mapped onto those of a wave packet.  One exposition of the parameter mapping is given in Appendix A of~\cite{Tomsovic18b}.

From the quadrature form, it can be deduced that the expansion coefficients needed for the revival analysis starting from Eq.~\eqref{eq:autocs} approximately (asymptotically) follow the form of Eq.~\eqref{eq:cgauss0}.  All the discussion of Sec.~\ref{sec:rev-1d-quant} then follows the exact same mathematical steps with the same appearance of revivals and fractional revivals.

\subsubsection{The coherence}
\label{sec:tc}

In the  context of ultracold bosonic quantum gases, the return probability may be difficult if not to impossible to measure directly.  However, the dynamics can be probed through a variety of observables that are suitable for experimental implementation.  For bosonic atoms in optical lattices, the \textit{coherence} is one example of an observable that can be straightforwardly probed~\cite{Greiner02b} even in the example of a single site; see Sec.~\ref{sec:RevInBHdimer} for the Bose-Hubbard dimer.  This particular quantity can be defined by $|\langle\hat{a}\rangle|^2$, where $\hat{a}$ is the particle annihilation operator.  Its expectation value with respect to the time-evolving state, Eq.~\eqref{eq:alpha}, is straightforwardly evaluated as
\begin{align}
  \braket{\hat{a}}_t & = \braket{\alpha_t|\hat{a}|\alpha_t}  =  \sum_{n=0}^\infty \sqrt{n+1} c_n^* c_{n+1} e^{-i(E_{n+1} - E_n) t / \hbar}  \notag \\
    & \simeq  \sum_{n=0}^\infty \sqrt{n+1} c_n^* c_{n+1} \exp\left[-2i\pi\left(\frac{t}{\tosc} + (n - \na + 1/2) \frac{t}{\trev} \right)\right] \; , \label{eq:acoh}
\end{align}
where this approximate expression follows using Eq.~\eqref{eq:expans-sec2}, which assumes that the coefficients $c_n$ are strongly peaked with $n$ near $\na$.

For the particular case of a Glauber coherent initial state, 
\begin{equation}
  c_{n} = \frac{z^n}{\sqrt{n!}}{\rm e}^{-|z|^2/2}= \frac{(\sqrt{\Nav}e^{i\theta})^{n}}{\sqrt{n!}} e^{- \Nav/2} 
  \label{eq:Glauber}
\end{equation}
for some arbitrary phase $\theta$, the coefficients $c_n$ are peaked around $\Nav-1/2$ and the expansion Eq.~\eqref{eq:expans-sec2} can be performed with $\na = \Nav-1/2$.  Using the exponential series gives
\begin{equation} 
\label{eq:acoht}
  \braket{\hat{a}}_t \simeq \sqrt{\Nav}e^{i\theta} \exp[-\Nav (1 - e^{-2 i \pi  t /\trev})] \exp\left[-2i\pi\left(\frac{t}{\tosc} - (\Nav -1) \frac{t}{\trev}\right) \right] \ . 
\end{equation}
Note that this approximate relation becomes an exact identity in the case of a strictly quadratic spectrum (i.e., featuring vanishing cubic or higher-order terms in Eq.~\eqref{eq:expans-sec2}).
The coherence thus becomes
\begin{equation}
  |\langle \hat{a} \rangle_t|^2 \simeq \Nav \exp[ - 4 \Nav \sin^2(\pi t / \trev)] \,. \label{eq:coherence}
\end{equation}
While being vanishingly small most of the time, it exhibits a periodic sequence of sharp peaks that pop up at integer multiples of the revival time $\trev $ with the temporal width 
\begin{equation} \label{eq:Delta-t}
    \Delta t \sim \Nav^{-1/2} \trev \; .
\end{equation}

As in Eq.~\eqref{eq:cgauss0}, a symptotically the expansion coefficients of Eq.~\eqref{eq:Glauber} can be very well approximated by a Gaussian distribution in the vicinity of $\Nav$
\begin{equation}
  c_n \simeq (2\pi\Delta n^2)^{-1/4} \exp\left[-\frac{1}{4\Delta n^2} (n - \Nav + 1/2)^2 + i n \theta \right] 
  \label{eq:cgauss}
\end{equation}
for some phase $\theta$ and with a variance $\Delta n^2 = \Nav$ (Eq.~\eqref{eq:cgauss0} was specifically chosen to be consistent with this relation).  This expression could be generalized to consider other states with slightly different widths, still with $\Delta n \gg 1$ and close to the same order as $\sqrt{\Nav}$.
Using Poisson summation, which generates a reformulation of the identity, Eq.~\eqref{eq:acoh}, as
\begin{equation}
  \braket{\hat{a}}_t = \sum_{l=-\infty}^\infty  \int_{-0.5}^\infty dn \sqrt{n\!+\!1} c_n^* c_{n+1} \exp\left[-2i\pi\left(\frac{t}{\tosc} + (n - \Nav + 1) \frac{t}{\trev}\right)\right] e^{2 i \pi n l} \,, \label{eq:Poisson}
\end{equation}
and performing the Gaussian integration over $n$ while keeping only leading-order terms in $\Nav$ and $\Delta n$, 
\begin{equation}
  \braket{\hat{a}}_t \simeq \sqrt{\Nav}e^{i\theta} e^{- 2 i \pi t / \tosc} \sum_{l=-\infty}^{\infty} e^{2 i\pi l \Nav} 
  \exp\left[-2 \pi^2 \Delta n^2 \left(\frac{t}{\trev} - l\right)^2  \right]\,,
\end{equation}
which for $\Delta n^2=\Nav \gg 1$ is in excellent agreement with Eq.~\eqref{eq:acoht}.
Using the fact that the mutual overlap between the individual Gaussian terms in the above sum can be safely neglected for $\Delta n \gg 1$, 
\begin{equation}
  |\langle\hat{a}\rangle_t|^2 \simeq \Nav \sum_{l=-\infty}^{\infty} 
  \exp\left[-4 \pi^2 \Delta n^2 \left(\frac{t}{\trev} - l\right)^2  \right]\,,
  \label{eq:coherenceGauss}
\end{equation}
for the coherence, which is in excellent agreement with Eq.~\eqref{eq:coherence} for $\Delta n^2=\Nav$.

\subsection{The semiclassical perspective}
\label{sec:semic-1d}

The goal of this subsection is to further complete the discussion of the one-dimensional revival phenomenology by offering a semiclassical perspective to this phenomenon.  Beyond providing useful semiclassical expressions that can be generalized later on and applied to the Bose-Hubbard dimer, this makes it possible to address the scaling of the coefficients in the expansion, Eq.~\eqref{eq:expans-sec2}, as well as the role of Maslov indices, and to discuss how higher-order terms, beyond the quadratic expansion of  Eq.~\eqref{eq:expans-sec2}, affect revivals.

There are two main tools: the semiclassical theory of Einstein-Brillouin-Keller (EBK) quantization~\cite{Einstein17, Brillouin26, Keller58} and the quasi-classical TWA approximation in the Weyl-Wigner formalism.  Section~\ref{sec:semic-scaling} starts with a discussion of how EBK quantization governs the different scalings in the physics of revivals.  This is followed in Sec.~\ref{sec:wttwa} by a brief introduction to the Weyl-Wigner formalism and the TWA approximation.  However, as noted there, applying the TWA results in the loss of quantum interference information, which is why we refer to this approximation as ``quasi-classical'' and reserve the term semiclassical for approximation schemes that preserve this information. In particular, since revivals are in essence an interference phenomenon, the TWA cannot account for them .  In Sec.~\ref{sec:EBKboot}, an approach is introduced in which EBK quantization is embedded in the TWA,  making it possible to capture all  observed revival phenomena. This approach is generalized ahead in Sec.~\ref{sec:semirev} for the Bose-Hubbard dimer.  First though, a perturbed Bose-Hubbard monomer is presented in the following Sec.~\ref{sec:monomer} as an example of how to employ this formalism and what to expect from it in terms of accuracy.

\subsubsection{EBK quantization and semiclassical scaling}
\label{sec:semic-scaling}

As above, assume that the Hamiltonian $\hat H$ of Eq.~\eqref{eq:quant-ham-1d} admits a well defined classical limit $H$, which because one-dimensional autonomous systems are integrable can be expressed as $H(I)$, with $(I,\varphi)$ the action-angle coordinates.  Within a semiclassical picture the eigenenergies can be derived from an EBK quantization scheme as
\begin{equation} 
\label{eq:EBK}
E_n = H[\hbar (n+ \nu)] \ ,
\end{equation}
where $0 \leq  \nu < 1$ is typically associated with Maslov indices or boundary conditions.

As in Secs.~\ref{sec:rev-1d-quant} and \ref{sec:rev-1d-quantb}, assume that the initial state $\ket{\alpha}$ under consideration is strongly localized in the energy eigenbasis.   Although as noted previously, the precise form of $\ket{\alpha}$ beyond that constraint is of little importance, it is helpful for exposition to pick a specific form, again Gaussian, keeping in mind that its only important characteristics are its mean energy and the width on which the coefficients $c_n$ are spread.  More specifically, introducing a pair of canonically conjugate variables according to $p = \sqrt{2 I} \sin \varphi$ and $q = \sqrt{2 I} \cos \varphi$ (which in the context of a Bose-Hubbard monomer is identified with quadrature variables), assume the initial state to be a minimum uncertainty wave packet in the corresponding representation whose Wigner function (cf.~the discussion below in Sec.~\ref{sec:wttwa}) is given by
\begin{align}
    W_\alpha(I,\varphi) & = \frac{1}{\pi\hbar} \exp \left[ - \frac{1}{\hbar}\left((p - p_\alpha)^2 +  (q - q_\alpha)^2\right) \right] \label{eq:W0-aaa}\\
    & \simeq \frac{1}{\pi\hbar} \exp \left[ - \frac{1}{\hbar}\left(\frac{(I - \Ia)^2}{2 \Ia} +  2 \Ia (\varphi - \varphi_\alpha)^2\right) \right]  \; ,\label{eq:W0-aa}
\end{align}%
with $(q_\alpha, p_\alpha)$ the wave packet centroids, $\Ia = (p_\alpha^2 + q_\alpha^2)/2$ the corresponding classical action, and $\varphi_\alpha = \arctan(p_\alpha/q_\alpha)$. 

Equation~\eqref{eq:W0-aa} shows that $W_\alpha(I,\varphi)$ is significant only for $\delta I \equiv I - \Ia = O(\sqrt{\hbar \Ia})$ and $\delta \varphi \equiv \varphi - \varphi_\alpha = O(\sqrt{\hbar / \Ia})$, which justifies neglecting the  semiclassically small terms, $\delta I^3 / \hbar$ or $\delta \varphi^3 / \hbar$, or higher orders.  Furthermore, $H(I)$ is a classical function, such that it varies smoothly on the scale $\sqrt{\hbar \Ia}$.  Therefore, it can be assumed that on this scale the classical Hamiltonian is well described by the Taylor expansion 
\begin{equation}
\label{eq:H(I)-expanded}
    H(I) = H(\Ia)  + \omega_\alpha \delta I + \frac12 \omega'_\alpha
    \delta I^2 + \frac16 \omega''_\alpha    \delta I^3 + \cdots \; ,
\end{equation}
which can be conceptualized as an expansion in the parameter $\hbar^{1/2}$.  In Eq.~\eqref{eq:H(I)-expanded}, $\omega(I) = \partial H/ \partial I$ is the angular frequency, $\omega'$, $\omega''$, are the corresponding derivatives with respect to the action and $\omega^{(k)}_\alpha$ ($= \omega^{(k)}(\Ia)$) their value for the action $\Ia$.  
Implementing EBK quantization according to Eq.~\eqref{eq:EBK} leads to  
\begin{align}     \label{eq:Hn-expanded} 
    E_n & = \Ea  + \hbar \omega_\alpha (n-\na) 
    + \frac12 \hbar^2 \omega'_\alpha (n-\na)^2
    + \frac16 \hbar^3 \omega''_\alpha   (n-\na)^3 + \cdots \; , 
\end{align}
with $\Ea = H(\Ia)$ and $\na \in \mathbb{R}$ is defined by
\begin{equation} \label{eq:shift}
  \Ia = \hbar( \na+ \nu)  \; ,
\end{equation}
so that $(\Ia - I_{\round{\la}}) = \hbar (\la - \round{\la})$ is the distance of $\Ia$ to the closest quantizing torus.
 Comparing Eq.~\eqref{eq:Hn-expanded} with Eq.~\eqref{eq:expans-sec2} gives
\begin{equation}
\label{eq:ts}
    \tosc = \frac{2\pi}{\omega_\alpha} \; , \qquad 
    \trev = \frac{1}{\hbar}\frac{2\pi}{\omega'_\alpha}  \; .
\end{equation}
The time scale $\tosc$ is thus a purely classical quantity whereas the revival time $\trev$ is intrinsically quantum, being of $O(\hbar^{-1})$, despite the fact that it is otherwise fixed by a classical shearing rate. Hence, the revival time $\trev$ is  significantly longer than $\tosc$ in the semiclassical regime.

Considering the time evolved state, $|\alpha_t\rangle$, given in Eq.~\eqref{eq:alpha}, along with the eigenenergy expansion, Eq.~\eqref{eq:Hn-expanded}, the phase corrections that are relevant to the physics depend on the $\hbar$ scale of the evolved time.  Multiplying the terms in the series by $t/\hbar$ leads to the terms scaling as $\hbar^{-1}\omega_\alpha^{(k-1)} (\delta I)^k t/ k! = O(\hbar^{k/2 - 1} t)$. 
For example, the $\tosc$ time scale is independent of $\hbar\ [t\sim O(\hbar^0)]$ and only the terms up to quadratic order in $\delta I$ are of $O(\hbar^0)$ or greater and can contribute in the semiclassical regime.  The cubic term cannot appreciably contribute until times scales of $O(\hbar^{-1/2})$.  Hence, the initial recurrences and their decay are essentially determined by
\begin{equation}
\label{eq:H2}
    H_2(I) = \Ea  + \omega_\alpha \delta I + \frac12 \omega'_\alpha
    \delta I^2  \; .
\end{equation}
On the other hand, $\trev$ scales as $O(\hbar^{-1})$ and the cubic term
\begin{equation}
\label{eq:H3}
    \delta H(I) =  \frac16 \omega''_\alpha \delta I^3  \; .
\end{equation}
must be included to describe fully quantitatively the behavior of the revivals, and likely even the quartic term.  These two regimes are considered successively in Sec.~\ref{sec:EBKboot}.  Before this, however, we introduce the Weyl-Wigner formalism.

\subsubsection{Weyl-Wigner formalism and truncated Wigner approximation}
\label{sec:wttwa}

An extremely useful framework, relied on extensively ahead, is the construction of a pseudo-quantum phase space provided by the Weyl-Wigner formalism~\cite{Wigner32, Balazs84}.  In the Schr\"odinger context the Weyl symbol of an operator $\hat B$, possibly time-dependent, can be formulated beginning in a configuration space representation as follows,
\begin{equation}
\label{eq:weyl}
\BW(\mathbf{q},\mathbf{p};t)= \int_{-\infty}^\infty {\rm d}\mathbf{x}\ {\rm e}^{i\mathbf{p}\cdot\mathbf{x}/\hbar}\braket{\mathbf{q}+\frac{\mathbf{x}}{2}|\hat B(t)|\mathbf{q}-\frac{\mathbf{x}}{2}}\ ,
\end{equation}
where $D$ is the number of degrees of freedom and $\mathbf{p}$ plays the role of a conjugate momentum variable to the position $\mathbf{q}$.   The same form applies in the bosonic many-body context as well, with $\hbar$ replaced by $\heff$.  For an arbitrary state $\ket{\psi}$ in the Hilbert space, its  Wigner transform $W_\psi$  is simply defined, up to a global prefactor, as the Weyl symbol of the projector $\hat \rho_\psi \equiv \ket{\psi}\bra{\psi}$
\begin{equation}
    \label{eq:wigner}
            W_\psi(\bq,\bp) = \frac{1}{(2\pi\hbar)^D} [\hat \rho_\psi ]^W (\bq,\bp) \; .
\end{equation}
The $(2\pi\hbar)^D$ difference between Eq.~\eqref{eq:weyl} and Eq.~\eqref{eq:wigner} is introduced because the former is such that $[\hat q_i]^W(\bq,\bp) = q_i$, $[\hat p]^W(\bq,\bp) = p_i$,  and quite generally for an operator $\hat B$ with a well defined classical analog $\BW$ is, possibly up to $\hbar$ corrections, the classical symbol associated with $\hat B$, whereas the prefactor in Eq.~\eqref{eq:wigner} guarantees the normalization
\begin{equation} 
 \label{eq:wigner-normalization}
     \int d\bq \, d\bp \, W_\psi(\bq,\bp) =1  \; .
 \end{equation}
In this way, if $W_\psi$ is non-negative, it can be thought of as a probability density in the classical phase space.  In the special case of a multidimensional Gaussian wave packet  $\alpha \rangle$ of the form Eq.~\eqref{eq:cohquad}, the Wigner transform is given by
\begin{equation}
\label{eq:wtgauss}
W_\alpha(\mathbf{q},\mathbf{p})= \left(\pi \hbar \right)^{-D} \exp \left[ - \frac{1}{\hbar} \left( (\vec p - \vec p_\alpha)^2 +  (\vec q - \vec q_\alpha)^2\right) \right]  \; .
\end{equation}
The case for more general forms of wave packets is discussed in Appendix~\ref{app:wigner-function}.

An important property of the Weyl-Wigner formalism is the exact identity valid for any pair of operators $(\hat A, \hat B)$  
\[
{\rm Tr} \hat A \hat B =  \frac{1}{(2\pi\hbar)^D} \int  d\bq \, d\bp \, \AW(\bq,\bp) \,\BW(\bq,\bp) \; .
\]
With $\hat A = \mathbb{1}$ and $\hat B = \hat \rho_\psi$ this gives in particular the normalization, Eq.~\eqref{eq:wigner-normalization}.  Furthermore, this implies that the mean value of an observable $\hat B$ and the autocorrelation of a wave-packet $\psi_0$ are expressed as
\begin{subequations}
    \begin{align}
        \mean{B}_\psi & =  \int  d\bq \, d\bp \, W_\psi(\bq,\bp) \,\BW(\bq,\bp) \; \label{eq:w-meanB}\\
        \autocor_{\psi_0}(t) & = (2\pi\hbar)^D  \int  d\bq \, d\bp \, W_{\psi_0}(\bq,\bp) \, W_{\psi_t}(\bq,\bp) \; . \label{eq:w-P}
    \end{align}
\end{subequations}
Extensive use of these two equations lies in what follows.

Within this framework, the evolution of any operator $B_W(\mathbf{q},\mathbf{p};t)$ is given by the sine function of the Moyal bracket~\cite{Moyal49}, 
\begin{eqnarray}
\label{eq:wto2}
\frac{{\rm d}}{{\rm d}t} \BW(\mathbf{q},\mathbf{p};t) &=& \frac{2}{\hbar} \BW(\mathbf{q},\mathbf{p};t) \  \sin\left[\frac{\hbar}{2}\left(\stackrel{\leftarrow}{\frac{\partial}{\partial \mathbf{q}}}\cdot\stackrel{\rightarrow}{\frac{\partial}{\partial \mathbf{p}}} - \stackrel{\leftarrow}{\frac{\partial}{\partial \mathbf{p}}}\cdot\stackrel{\rightarrow}{\frac{\partial}{\partial \mathbf{q}}}\right)\right] \HW(\mathbf{q},\mathbf{p};t) \nonumber \\
&& \qquad + \frac{\partial}{\partial t} \BW(\mathbf{q},\mathbf{p};t)\ .
\end{eqnarray}
In the Schrödinger context, all the information about the superposition principle and wave interference is contained in some complicated way in the infinite series of differential operators.  Likewise, in the bosonic context, all the quantum interference properties that are required for the quantization of particle number to emerge are somehow buried in the infinite series.  The lowest order approximation of the dynamics, i.e.~the classical~\cite{Moyal49, Groenewold46} or truncated Wigner approximation (TWA)~\cite{Steel98, Sinatra01, Sinatra02}, results from truncating the sine series to the first term giving
\begin{equation} \label{eq:liouville}
\BW(\bq,\bp;t) \approx  \left\{\BW(\bq,\bp;t), \HW (\bq,\bp;t)\right\} + \frac{\partial}{\partial t} \BW(\bq,\bp;t) \ ,
\end{equation}%
where $\{\cdot,\cdot\}$ is the Poisson bracket.  One recognizes in Eq.~\eqref{eq:liouville} the Liouville equation associated with the Hamiltonian flow of $\HW$, and the evolution of $B_W(\bq,\bp;t)$ is thus obtained by the purely classical transport of the initial density $B_W(\bq,\bp;0)$.  As a consequence, quantum interference information is lost, whether at the first or second quantized level.  In the second quantization case, $(\bq,\bp)$ are now the mean field quadratures.  In some sense, the remaining quantum information is reduced to just the uncertainty principle, since no state can generate a single point in the pseudo-phase space, and the symmetry of identical bosons in the latter case is contained in the resulting mean field.

\subsubsection{Embedding EBK into the truncated Wigner approximation}
\label{sec:EBKboot}

As seen in Sec.~\ref{sec:wttwa}, mean field theory leads to  the TWA~\cite{Steel98, Sinatra02} which basically amounts to evaluating the time evolution of the expectation values of observables classically, through the integration of Hamilton's equations of motion, using as initial phase-space probability distribution the Wigner function of the system's initial quantum state. 
As noted, a straightforward implementation of this approach does not account for quantum interference phenomena occurring between different trajectories, which is the reason why TWA is, in essence, a classical or quasi-classical theory, i.e.~not a complete semiclassical theory such as time-dependent WKB~\cite{Maslov81, Huber88}.  
However, as was shown in \cite{Schlagheck19,Hummel22b}, specific interference effects, related, e.g., to the presence of discrete symmetries, can nevertheless be quantitatively incorporated into the framework of the TWA through suitably tailored amendments of its implementation procedure. This also holds for the interference effects giving rise to revival features.
Indeed, it turns out that the latter can be quantitatively captured through a judicious use of EBK quantization embedded within the TWA.
We describe here how to implement this embedding for the analysis of revivals.  First, consider the simplest possible setting, as in Sec.~\ref{sec:semic-1d}, namely a conservative $1D$ Hamiltonian system. 

Even though the following sections consider a variety of measurable quantities (mean values of observables, autocorrelation function, coherence, entanglement), this introductory subsection is limited to only one of them, namely the mean value of some observable $\hat B$, which is assumed to be {\em smooth} in the sense that its Weyl symbol $\BW(p,q)$ varies smoothly on a classical scale and has no $\hbar$-dependent features. 

\subsubsection*{The \texorpdfstring{$t \simeq 0$}{t \~ 0} regime}

For times near $t=0$, as mentioned above, the second order approximation $H_2(I)$, Eq.~\eqref{eq:H2}, of the Hamiltonian is sufficient, and a straightforward TWA is enough to reproduce the time dependence of $\langle \hat B \rangle (t)$.  Using Eq.~\eqref{eq:w-meanB}, it reads
\begin{equation} \label{eq:TWA-oneD-a}
    \TWA{B} (t) = \int {\rm d}\varphi \, {\rm d} I\  W_\alpha(I,\varphi) \BW(I, \varphi + \omega(I) t) \; .
\end{equation}
The goal is to derive an analytical expression for $\TWA{B}(t)$.  To start with, 
since everything takes place in a neighborhood $O(\hbar^{1/2})$ of $\Ia$, the action dependence of $\BW$ becomes negligible in the semiclassical regime. 
Keeping only the $\varphi$ dependence, $\BW(I,\varphi) \simeq \BW(\Ia, \varphi)$ can be written as the Fourier expansion
\begin{equation} \label{eq:fourier-1d}
    B_\alpha(\varphi) :=  \BW(\Ia, \varphi) = b_0 + \sum_{k=1}^\infty  b_k \cos(k \varphi + \chi_k) \; . 
\end{equation}
This gives
\begin{align} 
    \TWA{B} (t) \simeq & b_0 + \sum_k b_k  \Re \Bigg\{ \int \frac {d(\delta \varphi) d (\delta I)}{\pi \hbar}
    \exp \left[ - \frac{\Ia}{\hbar} \Phi_0 + i \Phi_1^{(k)} \right] \Bigg\} \label{eq:TWA-oneD-b}\\
    \Phi_0(\delta I,\delta \varphi)  =     & 
    \frac{(\delta I)^2}{2 \Ia^2} +  2(\delta \varphi )^2 
    \\
    \Phi_1^{(k)}(\delta I,\delta \varphi)  = & 
    k \left[(\varphi_\alpha+\delta \varphi) + \omega(I) t \right] +  \chi_k 
    \; , \qquad 
\end{align}
If one does a naive counting of $\hbar$ factors, only the term $-\Ia \Phi_0 /\hbar$ is identified as rapidly varying, and since $({\pi \hbar})^{-1} \int  {d(\delta \varphi) d (\delta I)} \exp [ - ({\Ia}/{\hbar}) \Phi_0 ] $ is by construction normalized, this amounts simply to setting $\delta I = 0$ and $\delta \varphi = 0$ in $\Phi_1$. This leads to 
\begin{equation} \label{eq:Bclassical}
  \TWA{B} =  b_0 + \sum_{k=1}^\infty  b_k \cos(k\omega(\Ia)t +  k \varphi_\alpha +  \chi_k)
  = B_\alpha(\Ia, \varphi_\alpha + \omega(\Ia)t ) \; ,
\end{equation}  
which is the classical result corresponding to the value of $\BW$ for the classical trajectory lying at the wavepacket's center.

 As time increases, the term $k \omega(I) t$ in $\Phi_1^{(k)}$ eventually displays a significant variation with $I$.  Physically, it is indeed  because different values of $I$ near $\Ia$ have different angular frequencies that the wave packet is spread along the torus, leading to the decay of the oscillations in $\TWA{B}$ around $b_0$.  This term therefore needs to be included in the ``rapidly varying exponent'' when performing a steepest descent approximation.  Approximating the variation of the angular frequency by $\omega(I) = \omega_\alpha + \omega'_\alpha \delta I$ shifts the stationary phase point $\delta I^*$ in the complex plane as 
 \begin{equation}
     \delta I^* = i  k \hbar \Ia \omega'_\alpha t  \, ,
\end{equation}
which implies 
\begin{equation}
      \frac{\Ia}{\hbar} \Phi_0(\delta I^*,0) -  i \Phi_1^{(k)}(\delta I^*,0) = + \frac{(\delta I^*)^2}{2 \Ia \hbar}  -i [k(\omega_\alpha t + \varphi_\alpha) + \chi_k]\; .
 \end{equation}
Keeping again in mind that the Wigner function $W_\alpha$ is normalized to one, performing the Gaussian integral around the stationary phase point leads to 
\begin{align} 
          \TWA{B} & 
       = b_0 + \sum_{k=1}^\infty  b_k 
      \cos[k (\omega_\alpha t + \varphi_\alpha)  + \chi_k] \exp \left[-\frac{\hbar\Ia}{2} (k \omega'_\alpha t )^2\right] \; . \label{eq:BTWA}
\end{align}
Each Fourier component of the classical expectation value Eq.~\eqref{eq:Bclassical} is thus simply damped by the exponential factor $\exp \left[-({\hbar \Ia}/{2 }) (k \omega'_\alpha t )^2\right]$, and disappears on the time scale
\[
t^{(k)}_{\rm decay} = \sqrt{\frac{1}{\hbar \Ia}} \frac{1}{k \omega'_\alpha}  = O(\hbar^{-1/2})\; , 
\]
which generalizes Eq.~\eqref{eq:Delta-t} to an arbitrary observable.

The initial signal is thus characterized by an oscillation on a classical scale $\tosc = 2\pi/\omega_\alpha$, that disappears on the scale $t^{(k)}_{\rm decay} = O(\hbar^{-1/2})$, which is semiclassically much longer than $\tosc$, but much shorter than $\trev = O(\hbar^{-1})$. 

\subsubsection*{The \texorpdfstring{$t \simeq \trev$}{t \~ trev} regime}
\label{sec:OneDtrev}
 
To discuss the behavior of $\mean{B}(t)$ near $\trev = 2\pi/\hbar \omega'_\alpha$, consider the observation made in Sec.~\ref{sec:rev-1d-quant} that for the second order approximation $H_2$ given by Eq.~\eqref{eq:H2}, Eq.~\eqref{eq:Etrev} applies. Thus, at $t=\trev$,
\begin{equation} 
\label{eq:trick}
    \hat U_2(\trev) \equiv \exp \left[- \frac{i}{\hbar} \hat H_2 \trev \right] = \hat R( \omega_\alpha \trev + \Da) \; ,   
\end{equation}
where $\Da$ is the angular shift, Eq.~\eqref{eq:DeltaTh}, and up to an irrelevant global phase
\begin{equation}
    \hat R(\Delta \varphi) := \exp \left[- \frac{i}{\hbar} \hat I \Delta \varphi \right], 
\end{equation}
is the operator performing a rotation of angle $\Delta \varphi$ on the angular variable $\varphi$ conjugated to $I$ (the quantum version of $I$ can be interpreted as $\hat I := -i\hbar \partial_\varphi$, and is such that $\hat I \ket{n} = (n+1/2) \ket{n}$). In the action-angle  variables $(I,\varphi)$, the action  of $H_2$ during the time $\trev$ is thus a rigid rotation of angle $\Delta \varphi = \omega_\alpha \trev + \Da$, where, in addition to the angle $\omega_\alpha \trev$ associated with the linear part of $H_2(I)$, there is the extra term $\Da =  2\pi(\frac12 - \delta\na)$, Eq.~\eqref{eq:DeltaTh}, 
which depends therefore on the action $\Ia$ of the center of the wave packet and on the constant $\nu$ associated with Maslov indices (or boundary conditions). More specifically, $\Da$ is zero if $\Ia$ falls just in the middle of two quantizing actions.

Let us now describe a time interval around $\trev$, and also include the higher order terms $ \delta H = \frac16 \omega''_\alpha (I-\Ia)^3 $ which, as mentioned above, should be taken into account in the expansion, Eq.~\eqref{eq:H(I)-expanded}, since $\trev=O(\hbar^{-1})$. 
 Since $\hat H_2$ and $\hat H$ can be seen as functions of $\hat I$, they commute both together and with $\hat I$, which means that, defining $\delta t \equiv t-\trev$, 
\begin{align}
  \hat U(t) =  \hat U(\trev) \hat U(\delta t) =& \exp \left[- \frac{i}{\hbar} \hat H_2 \trev \right]
\exp \left[- \frac{i}{\hbar} \delta \hat H \trev \right] \exp \left[- \frac{i}{\hbar} \hat H \delta t \right]
 \notag \\
=&  \hat R( \omega_\alpha \trev + \Da)
\exp \left[- \frac{i}{\hbar} \delta \hat H \trev \right] \exp \left[- \frac{i}{\hbar} \hat H \delta t \right] \; .
\label{eq:2cdform}
\end{align}

Here is the subtle part. Start with the expansion Eq.~\eqref{eq:H(I)-expanded} of $H(I)$.  There are two scales for $\delta I$: The smaller one is $\delta I^{(1)} = O(\hbar)$, which corresponds to the difference of action between two quantizing tori; The longer one $\delta I^{(1/2)} = O(\hbar^{1/2})$, which corresponds to the extension of the initial wavepacket.  Furthermore we have also two time scales $\trev = O(\hbar^{-1})$ and $\delta t = O(\hbar^{-1/2})$.  Starting from the expansion Eq.~\eqref{eq:H(I)-expanded} of $H(I)$, it turns out that 
\begin{equation}
        \frac{\trev}{\hbar} \omega^{(n)}_\alpha (\delta I^{(1)})^n = O(\hbar^{n-2}) \qquad \qquad 
        \frac{\delta t}{\hbar} \omega^{(n)}_\alpha (\delta I^{(1)})^n = O(\hbar^{n-3/2})
\end{equation}
Thus, for a propagation time $\trev$ only the linear $(n{=}1)$ and quadratic $(n{=}2)$ terms are sensitive to the quantization of the torus, and for a propagation time $\delta t$ only the linear term is. Now, for $\exp \left[- ({i}/{\hbar}) \hat H_2 \trev \right]$ Eq.~\eqref{eq:trick} is applicable, which eliminates the quadratic term. Since for linear Hamiltonians the TWA is exact, this means that once the system is written as Eq.~\eqref{eq:2cdform} the TWA can be used to propagate the initial wave packet {\em for a time interval $O(\hbar^{-1/2})$ around $\trev$}. Furthermore
\begin{equation}
        \frac{\trev}{\hbar} \omega^{(n)}_\alpha (\delta I^{(1/2)})^n = 0(\hbar^{n/2-2}) \; , \qquad
        \frac{\delta t}{\hbar} \omega^{(n)}_\alpha (\delta I^{(1/2)})^n = 0(\hbar^{n/2-3/2}) \; .
\end{equation}
Keeping only the terms that grow with $\hbar$, anything of order four or higher in $\trev$, and of order three or higher in $\delta t$, can be dropped.

Thus in the end, 
  \begin{equation} \label{eq:Boot-oneD-a}
    \Boot{B} (t) = \int \frac{d\varphi d I}{\pi \hbar} W_\alpha(I,\varphi) B_\alpha \left(\varphi + \omega_\alpha t + \Da    + \frac12 \omega''_\alpha \cdot (\delta I)^2 \trev + \omega'_\alpha \cdot (\delta I) \delta t\right)
    \; .
\end{equation}
At this point, the rest of the calculation follows exactly as for the $t \simeq 0$ case. Performing the integral over $\delta \varphi$ gives
\begin{equation}
\label{eq:TWAEBK_neartrev}
\Boot{B} (t)  \simeq  b_0 + \sum_k b_k  {\rm Re} [B_k] 
\end{equation}
with
\begin{align} 
    B_k & =  
     \int \frac { d (\delta I)}{\sqrt{2 \pi \hbar \Ia}}
    \exp \left[ - \frac{(\delta I)^2}{2 \Ia \hbar} + i k \left( \frac12 \omega''_\alpha \cdot (\delta I)^2 \trev +  \omega'_\alpha \cdot (\delta I) \delta t \right) \right]  \notag \\
    & \qquad \qquad \times \exp \left[ i k \left(\varphi_\alpha + \omega_\alpha t + \Da \right)+ \chi_k   \right]  \\
    & = \frac{\exp \left[ i k \left(\omega_\alpha t + \varphi_k \right) \right]}
    {\sqrt{\zeta_k }}\exp \left[ - \frac{\hbar \Ia}{2 \zeta_k} \left({k\omega'_\alpha \delta t}\right)^2 \right] \; ,
\end{align}
where
\begin{equation}
    \zeta_k \equiv 1 - 2 i\pi k \Ia \frac{\omega''_\alpha}{\omega'_\alpha} \; , \quad \varphi_k = k (\varphi_\alpha + \Da) + \chi_k \; .
\end{equation}

Note that the main difference between $t \simeq 0$ and $t \simeq \trev$, beyond the additional rigid rotation of the wave packet, is the effect of the third order term $\frac16 \omega^{(2)}_\alpha \cdot (\delta I)^3$  acting on the system  in the interval $[0,\trev]$.  This affects the final results through the factor $\zeta_k$, which is complex and of norm larger than one.  The fact that $\zeta_k$ is complex will cause a phase shift in the oscillation (due to the $\sqrt{\zeta_k}$ in the denominator) and an extra oscillating term associated with the imaginary component in the exponent of the second exponential. The fact that the modulus of $\zeta_k$ is larger than one will simultaneously decrease the overall amplitude of the revival and slow down its decay as $\delta t$ increases.  We insist once again on the fact that $\zeta_k$ is a purely classical quantity that remains constant in the semiclassical limit $\hbar \to 0$.  Whether $\zeta_k$ is close to one or large and purely imaginary depends on the ratio $k \Ia \omega''_\alpha/\omega_\alpha$, which is necessarily large for $k\to \infty$, but may be large or not even for $k=1$ depending on the classical Hamiltonian $H(I)$ under consideration.

Later for the dimer, the above description to two dimensional (integrable) systems must be adapted. There the role of this third order term turns out to be largely irrelevant, and the decay of revivals is essentially controlled by the mismatch of the revival times for the different degrees of freedom of the system.

\subsection{Revivals in a perturbed Bose-Hubbard monomer}
\label{sec:monomer}

Consider now the arguments discussed in the earlier subsections with a 1D example. In order to have a system displaying non-trivial revival dynamics while maintaining a connection to cold-atomic systems, focus on the Bose-Hubbard monomer perturbed by a cubic nonlinearity:
\begin{equation}\label{eq:cubicBHmonomer}
    \widehat{H}_{\rm mono} = U \left[ \frac{\widehat{n}(\widehat{n}-1)}{2} + \frac{\varepsilon}{6} \left( \widehat{n} + \frac12 \right)^3 \, \right] \, .
\end{equation}
Such cubic corrections to the total energy, with negative $\epsilon$, arise most naturally for Bose-Einstein condensates that are confined in trapping potentials, where they represent the onset of the transition from the perturbative regime at low condensate populations to the Thomas-Fermi regime at high populations~\cite{Dalfovo99}.  For the case of a spherically symmetric harmonic trap with the trapping frequency $\omega_{\rm trap}$, a variational ansatz yields $\epsilon = - 9 U / 8 \hbar \omega_{\rm trap}$ in leading order.
The first step is to write the classical (mean-field) Hamiltonian associated with Eq. \eqref{eq:cubicBHmonomer}. This procedure is the same as the one that is later employed for the dimer, and is described here in detail. 

Let $\Nav$ be the (conserved) initial state average number of particles, which is assumed to be large.  The small parameter in the mean-field regime can then be taken as the effective Planck constant 
\begin{equation} 
\label{eq:heff}
    \heff \equiv \So \frac{L}{\Nav} \; ,
\end{equation}
with $L$ the number of sites ($L=1$ here for the monomer but is equal to two for the dimer), and where the main role of $\So$ is to carry the dimension of an action so that $\heff$  has the units of the true $\hbar$.  The numerical value of $\So$, as long as it is of order one, is of little importance, and we take simply $\So =1$.

The quadrature operators introduced in Eq.~\eqref{eq:quadratures} lead to $[\hat q,\hat p] = i \heff$, and it is possible to define the mean-field/classical version of the Hamiltonian Eq.~\eqref{eq:cs.BHd} by replacing these operators by the classical variables $(q,p)$.  It is important though to address a slightly subtle point concerning operator ordering. Consider, for instance, the Hamiltonian describing the monomer, which can be written either as $\hat H_{\rm mono} = (U/2) \hat n (\hat n -1) $ or as $\hat H_{\rm mono} = (U/2) \hat a^\dagger \hat a^\dagger \hat a \hat a$, both forms being strictly equivalent through the substitution $\hat{a}^\dagger \hat{a} = \hat{a} \hat{a}^\dagger - 1$.  However, writing $\hat{a}$ and $ \hat{a}^\dagger$ in both expressions in terms of the quadratures $(\hat q, \hat p)$, and replacing them by the classical variables $(q,p)$ leads to classical Hamiltonians differing by $O(\hbar)$ terms.  Of course they are negligible in the purely classical limit, but do however lead to differences of the order of the mean level spacing when these Hamiltonians are EBK quantized. 

The rule of the game defining the correct ordering of the operators {\em to apply EBK quantization} can be shown~\cite{Maslov81} to be that the classical symbol $H_{\rm cl}(q,p)$ associated with a quantum Hamiltonian $\hat H$ should be such that
\[
   \hat H = \frac12 \left( H_{cl} (\overleftarrow{\hat q},\overrightarrow{\hat p}) 
   + H_{cl} (\overrightarrow{\hat q},\overleftarrow{\hat p})  \right) \; .
\]
In other words, the quantum $\hat H$ should be the half-sum of the quantized $H$ for which all $\hat q$'s have been pushed to the left and all $\hat p$'s have been pushed to the right, and the one for which all $\hat q$'s have been  pushed to the right and all $\hat p$'s have been pushed to the left.  For instance, following this prescription for the number operator $\hat n = \hat a^\dagger \hat a$ leads to
\begin{equation}\label{eq:nop}
    \hat n = \frac{\hat p^2 + \hat q^2}{2\heff} - \frac12 \; , 
\end{equation}
so that the classical symbol associated with the number operator is 
\begin{equation}
\label{eq:nclass}
    n(q,p) = \frac{s(q,p)}{\heff} - \frac12 \; ,
\end{equation}
with
\begin{equation}\label{eq:s-action}
    s(q,p) = \frac{p^2 + q^2}{2}  \; ,
\end{equation}
the action variable associated with the variables $(q,p)$.  Note this implies in particular a Maslov index $\mu =2$ for the corresponding quantizing tori.

Turning now to the monomer Hamiltonian $\hat H_{\rm mono}$, and as detailed in appendix~\ref{app:ordering}, the above prescription leads to a classical Hamiltonian associated with \eqref{eq:cubicBHmonomer} in the form
\begin{equation}
    H_{\rm mono}(s) = \frac{U}{\heff^2} \left\{ \frac{s(s-2\heff)}{2} + \frac{\varepsilon}{\heff} \left[ \frac{s(s^2 - 4 \heff^2)}{6} \right] \, \right\} \, .
\end{equation}
The fact that in this expression the interaction energy is multiplied by a factor $\heff^{-2}$ and the coefficient $\epsilon$ by a factor $\heff^{-1}$ signals that to properly address the mean-field/semiclassical limit $\heff \to 0$, one should scale the energies and time as
\begin{equation} \label{eq:rescaling-Et}
    \tH_{\rm mono} := \left( \frac{\heff^2}{U} \right) H_{\rm mono} \; ,  \qquad \tt := \frac{U}{\hbar \heff} t \; ,
\end{equation}
(so that the rescaled time $\tt$ is such that $Et/\hbar = \widetilde E \widetilde t / \heff$), and the perturbation coefficient as 
\begin{equation} \label{eq:rescaling-1d}
    \widetilde{\varepsilon} := \varepsilon/\heff \quad ,
\end{equation}
leading to
\begin{equation}\label{eq:monomer_classical}
    \tH_{\rm mono}(s) = \frac{s(s-2\heff)}{2} + \widetilde{\varepsilon} \left[ \frac{s(s^2 - 4 \heff^2)}{6} \right] \,  \, .
\end{equation}
Further quantities derived from the rescaled Hamiltonian  $\tH_{\rm mono}$ are also labeled with a tilde.

Consider now an initial state that is assumed to be a coherent state $\ket{\zee_\alpha}$ with $\Nav = |\zee_\alpha|^2$ supposed large.  In that case,  the Wigner transform $W_\alpha(q,p)$ of $\ket{\zee_\alpha}$ has exactly the form of Eq.~\eqref{eq:W0-aaa} with $\zee_\alpha = (q_\alpha + i p_\alpha)/\sqrt{2 \heff}$  (and the substitution $\hbar \to \heff$).  Transforming to action angle variables,  $W_\alpha(s,\theta)$ is  centered in $s_\alpha = (q_\alpha^2 + p_\alpha^2)/2 = \heff \Nav$ (with Eq.~\eqref{eq:heff}, numerically $s_\alpha = \So =1$).
The expansion, Eq.~\eqref{eq:H(I)-expanded}, is thus naturally performed around $s_\alpha$, and it gives for Eq.~\eqref{eq:monomer_classical}
\begin{equation}
    \widetilde{\omega}_\alpha = (s_\alpha-\heff) + \frac{ \widetilde \varepsilon}{6} \left( 3 s_\alpha^2 - 4\heff \right) \,,  
    \quad \widetilde{\omega}'_\alpha = 1 + 2 \widetilde \varepsilon s_\alpha \quad \text{and} 
    \quad \widetilde{\omega}''_\alpha = 2 \widetilde{\varepsilon} \, .
\end{equation}
%
Note that with the convention Eq.~\eqref{eq:shift},  the expansion, Eq.~\eqref{eq:expans-sec2}, is performed around $\na = \Nav \! - \! \frac12$, so that there is in Sec.~\ref{sec:rev-1d-quant}  a $-\frac12$ shift with respect to the usual practice (see text between Eqs.~\eqref{eq:Glauber}-\eqref{eq:acoht}). 
We provide in Appendix~\ref{sec:shift} a rather extensive discussion about the origin of this shift, about the reason why it does not change the end results, and finally why it is nevertheless a good idea to keep it. 
The revival time in the original monomer,  $\ttrevo \smeq 2\pi /\heff \mapsto \trevo \smeq 2\pi \hbar /U$ is seen to be shifted\footnote{Ahead, it is seen that the shift in revival time is dependent on the circumstances and upon the precise measure of interest.  In fact, some measures lead to shifts later in time.} to an earlier time $\ttrev \smeq 2\pi /\heff(1+2\epsilon s_\alpha) \mapsto  \trev = 2\pi \hbar/ U(1+2\epsilon s_\alpha)$. 

Focusing on the mean value of the quadrature $q$-operator, which in action-angle coordinates is associated with the classical function $q = \sqrt{2 s} \cos \theta$, and for which  the Fourier decomposition in Eq.~\eqref{eq:fourier-1d} therefore reduces to a single term, Eq.~\eqref{eq:BTWA} straightforwardly results in 
\begin{equation}\label{eq:TWA_monomer_near0}
    \langle \widehat{q} (t) \rangle_\mathrm{TWA} = \sqrt{2 s_\alpha} \exp \left[- \frac{\heff s_\alpha \big( \widetilde{\omega}'_\alpha  \tt \big)^2}{2} \right] \cos \widetilde{\omega}_\alpha \tt \, .
\end{equation}
Equation~\eqref{eq:TWAEBK_neartrev} is also very simple: 
\begin{align}\label{eq:TWA_monomer_neartrev}
    \Boot{q(t)} &= \sqrt{2 s_\alpha} 
    \exp \left[ -\frac{\heff s_\alpha}{2 \zeta} \left(\widetilde{\omega}'_\alpha (\tt- \ttrev)\right)^2\right] 
    {\rm Re} \left( \frac{e^{ i \widetilde{\omega}_\alpha \tt  }}{\zeta} \right)  
    \, ,  \\
     \zeta & = 1 - 2\pi i s_\alpha \left( \frac{\widetilde{\omega}_\alpha''}{\widetilde{\omega}'_\alpha} \right) \, . \notag
\end{align}

\begin{figure}
    \centering
    \includegraphics[width=0.9\linewidth]{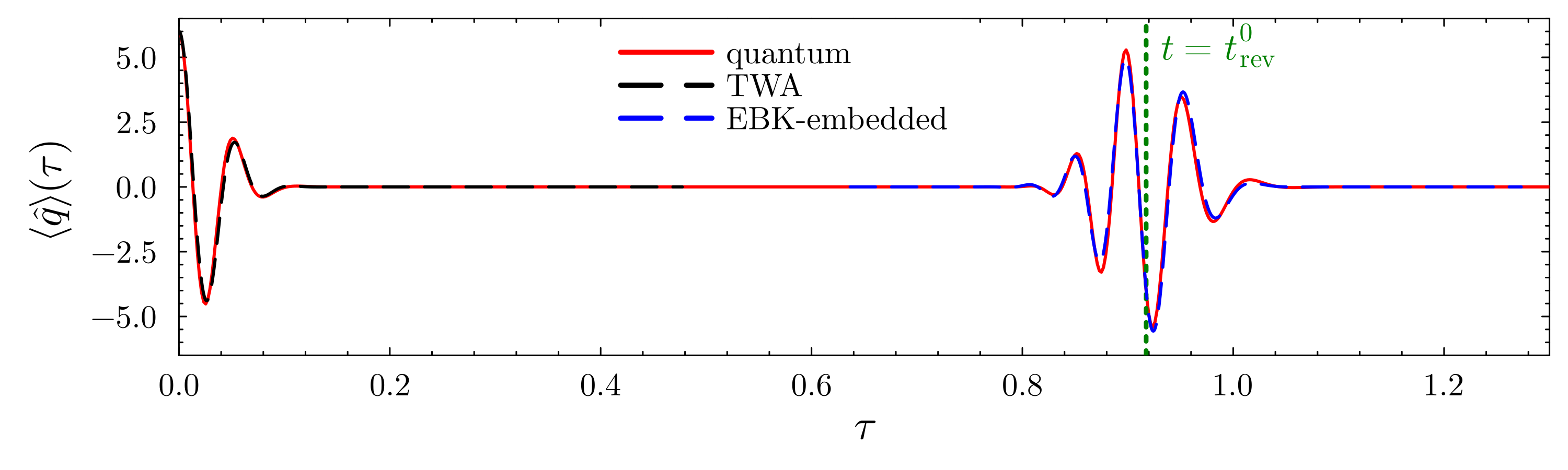}
    \caption{
    Mean value of position for the single-site Bose-Hubbard model perturbed by a cubic nonlinearity, Eq.~\eqref{eq:cubicBHmonomer}, as a function of the scaled time $\tau = t/\trevo$, with $\trevo = 2\pi\hbar/U$.  The initial state is the coherent state $\ket{\zee_\alpha}$ with $\zee_\alpha = 6$, and the Hamiltonian parameters are $\varepsilon=0.005$, $U=1$. 
Solid red curve : exact quantum calculation;  Dashed black curve (near the origin): TWA approximation Eq.~\eqref{eq:TWA_monomer_near0}; Dashed blue curve (near the revival time):  EBK-embedded TWA approximation Eq.~\eqref{eq:TWA_monomer_neartrev}.  
    Notice how the revival time in the original system, at $\tau=1$, is pushed to earlier due to the nonlinear term (dotted green line).  When the revival takes place, there is also a visible decrease in the mean-value's amplitude, which in the unperturbed case would rise up to its initial value.
    }   \label{fig:oneD}
\end{figure}

Figure~\ref{fig:oneD} shows, for $\zee_\alpha = 6 + i 0$, i.e.\ $\Nav = 36$, the time evolution of the mean value of the ``position'' (i.e.~quadrature centroid) of the evolved coherent state of the form of Eq.~\eqref{eq:W0-aa} in the perturbed monomer, Eq.~\eqref{eq:cubicBHmonomer}. What is seen corresponds to the generic scenario: near $t=0$, $\langle \widehat{q}(t) \rangle$ oscillates and converges to a constant value which is maintained for a very long time, but around $\ttrev \simeq 2\pi /(\heff \widetilde \omega'_\alpha )$ a new burst of oscillations takes place. Formulae \eqref{eq:TWA_monomer_near0} and \eqref{eq:TWA_monomer_neartrev} are clearly able to perfectly capture the mean-value's behavior both near the origin and the revival time, respectively.

\section{Revivals in the Bose-Hubbard dimer}
\label{sec:RevInBHdimer}

As detailed in Ref.~\cite{Schlagheck22} and pointed out in the introduction, revival phenomena have a much richer character in the Bose-Hubbard dimer than for a single site, and the overarching goal of this paper is to elucidate the origins of the observed behaviors.  Toward this end, first the various behavioral manifestations of the revivals due to the more complicated dynamics of the dimer are recalled.

\subsection{The configuration under study}

Consider a gas of ultracold bosonic atoms in a one-dimensional (1D) optical lattice.  The initial state of this gas is conceptualized as a perfect Bose-Einstein condensate which is prepared about a condensate wavefunction of the form $(\ldots,\psi_1,\psi_2,\psi_1,\psi_2,\ldots)$, with $\psi_1,\psi_2\in\mathbb{C}$ the odd- and even-site amplitudes.
Under some dynamical circumstances, the above density-wave structure of the wavefunction is sufficiently preserved that the system's dynamics can be mapped~\cite{Steinhuber20} onto the two-site Bose-Hubbard Hamiltonian
\begin{equation} 
\label{eq:cs.BHd}
\hat{H} = - J \left( \hat{a}_1^\dagger \hat{a}_2 + \hat{a}_2^\dagger \hat{a}_1 \right) + \frac{U}{2} \sum_{j=1}^2 \hat n_j \left(\hat n_j - 1 \right) 
\end{equation}
with $\hat{a}_j^\dagger,\hat{a}_j$ the particle creation and annihilation operators on site $j=1,2$ and $\hat{n}_j = \hat{a}_j^\dagger\hat{a}_j$ the occupancy operator on that site.
Here, $U$ is the on-site interaction strength and $J$ is twice the nearest-neighbor hopping strength of the 1D lattice under study (a similar relation can be derived for higher dimensional lattices with the only difference being that the factor is not two~\cite{Steinhuber20}).  Assume in the following that owing to some specific dynamical constraints, such as the imposition of a superlattice structure~\cite{Folling07} (in which case $J$ would be the hopping strength within each two-site sublattice), the Bose-Hubbard dimer, Eq.~\eqref{eq:cs.BHd}, maintains a valid and complete description of the system's time evolution.

The initial condensate state $|\boldsymbol{\zee}\rangle$ within this two-site system can be rather accurately modeled by means of the coherent state
\begin{equation}
|\boldsymbol{\zee}\rangle = |{\zee_1, \zee_2}\rangle = \exp \left(-\frac{1}{2}\left(\left|\zee_1\right|^2 + \left|\zee_2\right|^2\right) + \zee_1  \hat a^\dagger_1 + \zee_2 \hat a^\dagger_2 \right)|{0, 0}\rangle \,, \label{eq:cs.BEC}
\end{equation}
where $|{0, 0}\rangle$ denotes the vacuum state in the dimer. We suppose an integer mean population of this system, i.e., $(\zee_1,\zee_2) \in  \mathbb{C}^2$  with 
$|\zee_1|^2 + |\zee_2|^2 = \langle \boldsymbol{\zee}| \ntoth |\boldsymbol{\zee}\rangle = \Nav \in \mathbb{N}$  where $\Nav$ is the (conserved) average number of particles in the dimer.  A particularly interesting case is the one where the second site is initially empty, $\zee_2 = 0$, and, owing to self-trapping, 
becomes only marginally populated during time evolution.   In the following, we therefore always consider initial states of the form
\begin{equation} \label{eq:PSI0}
    |\alpha \rangle =  |{\sqrt{\Nav}, 0}\rangle \; .
\end{equation}
This altogether gives rise to a scenario that is rather similar to the one that is studied in Ref.~\cite{Greiner02b}. 
The only difference is that each initially occupied site featuring coherence oscillations transfers a small part of its population to a secondary partner site instead of retaining its full occupancy throughout the entire time evolution.  To render this marginal second-site occupancy detectable, mean dimer populations $\Nav$ that are significantly larger than the site occupancies realized in Ref.~\cite{Greiner02b} are considered, namely of the order of $\Nav \sim 10^2$.

As is argued in Ref.~\cite{Richter22} and worked out below in more detail, the dimensionless parameter $\Nav$ plays the role of an inverse reduced Planck constant.  In fact, it turns out to be convenient to set $\heff = 2/\Nav $, the inverse mean filling factor.  It is complementary to another dimensionless parameter that characterizes the dynamics of the system, namely the ratio
\begin{equation}
\label{eq:gamma}
\gamma = \heff \frac{J}{U} = \frac{2 J}{\Nav  U} \ ,
\end{equation}
which is the product of the inverse filling factor ($\heff$) times the ratio of the hopping strength $J$ to interaction strength $U$. The dynamical range of greatest interest involves small values of $\gamma$, say $\lesssim 0.3$, which in combination with a large value of $\Nav $ (small $\heff$) allows for the application of both quantum perturbation theory and semiclassical theory in order to understand the revival dynamics in the Bose-Hubbard dimer. 

\subsection{Collapse and revivals in the coherence and the return probability}
\label{sec:coherence-return}

Let us first consider the special case of an optical lattice with infinitely high barriers and hence isolated lattice sites.
The system's eigenenergies associated with on-site phase oscillation dynamics read 
\begin{equation}
E_n = \frac{1}{2} U n (n - 1) \label{eq:En}
\end{equation}
where the quantum number $n$ can be identified with the occupancy of the site.
In this case, the expansion of the energies to quadratic order, according to Eq.~\eqref{eq:expans-sec2}, is exact, with the characteristic time scales 
\begin{align} 
    \tosco &= \frac{2\pi \hbar }{U \Nav } \label{eq:tosco} \; ,\\
    \trevo &= \frac{2\pi \hbar }{U} \label{eq:trevo} \; .
\end{align}

A coherent state of the type in Eq.~\eqref{eq:cs.BEC}, but restricted to a single site with the mean occupancy $|\zee_1|^2 = \Nav $, features a root mean square width $\Delta n = \sqrt{\Nav }$.
It thus gives rise, according to Eq.~\eqref{eq:Pcollapse}, to  oscillations of period $\tosco$ which decay on the scale
\begin{equation}
        \tdeco = \frac{2\pi \hbar }{ \sqrt{\Nav} U} \label{eq:tdeco} \; .\\
\end{equation} 
Revivals take place at integer multiples of $\trevo$.
Each time they lead to a perfect recovery of the initial coherent state as they are not subject to temporal attenuation due to the absence of cubic or higher order terms in Eq.~\eqref{eq:En}.  Furthermore, the revivals and recurrence remain perfectly synchronized
due to the relation of Eq.~\eqref{eq:perfectrev} being always fulfilled\footnote{Note that for integer $\Nav$, this is seen immediately as the consequence of the fact that $\trevo/\tosco = \Nav  \in \mathbb{N}$ and $\la = \frac12$ (since for 1D system the Maslov index is $\frac12)$.  However, for the isolated lattice, choosing a non-integer $\Nav$ would of course not destroy the perfect revival, as the shift of $\Nav$ from an integer would be exactly compensated by a shift of $\la$ away from $\frac12$.}.

As demonstrated in Ref.~\cite{Greiner02b}, the presence or absence of coherence in the on-site phase oscillation dynamics of an optical lattice can be revealed by performing time-of-flight absorption imaging, which essentially amounts to monitoring the atomic momentum distribution after having switched off the lattice potential and allowing the atoms freely move for a suitable time.
This momentum distribution can be written as
\begin{equation}
\rho(\mathbf{k}) = |\widetilde{\chi}(\mathbf{k})|^2\left(\sum\nolimits_{\mathbf{l}} (n_{\mathbf{l}} - |a_{\mathbf{l}}|^2) + |\widetilde{a}(\mathbf{k})|^2 \right)
\end{equation}
in a $D$-dimensional lattice,  with $\mathbf{l} \equiv (l_1,\ldots,l_D)\in\mathbf{Z}^D$ the lattice site indices and $\mathbf{k} \equiv (k_1,\ldots,k_D)$ the wave vector corresponding to momentum.  
Here, $\widetilde{\chi}$ denotes the (rather broad) Fourier transform of the (spatially rather narrow) Wannier orbital describing the single-particle wave function on a lattice site, $n_{\mathbf{l}} = \langle \hat{a}_{\mathbf{l}}^\dagger \hat{a}_{\mathbf{l}} \rangle$ is the mean population on the site $\mathbf{l}$, $a_{\mathbf{l}} = \langle\hat{a}_{\mathbf{l}}\rangle$ is the expectation value of the bosonic annihilation operator on that site, and $\widetilde{a}(\mathbf{k}) = \sum_{\mathbf{l}} a_{\mathbf{l}}e^{-i\Delta\mathbf{l}\cdot\mathbf{k}}$ with the lattice spacing $\Delta$ is the Fourier series over those expectation values.
This latter Fourier series gives rise to a sequence of narrow peaks at integer multiples of $2\pi/\Delta$ in reciprocal space (i.e., at $\mathbf{k} = 2\pi\boldsymbol{\nu}/\Delta$ with $\boldsymbol{\nu} \in \mathbb{Z}^D$), whose width can be related to the presence of a shallow overall confinement potential inducing an inhomogeneity of the lattice.
The presence of perfectly coherent states on the lattice sites implies $n_{\mathbf{l}} = |a_{\mathbf{l}}|^2$, thus allowing for a maximal resolution of those momentum distribution peaks, whereas in the total absence of coherence, $a_{\mathbf{l}} = 0$, a rather flat and structureless momentum distribution $\rho(\mathbf{k}) = |\widetilde{\chi}(\mathbf{k})|^2\sum\nolimits_{\mathbf{l}} n_{\mathbf{l}}$ would be observed.

\begin{figure}[t]
\begin{center}
\includegraphics[width=\linewidth]{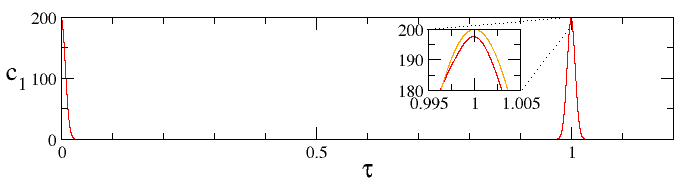}
\caption{\label{fig:coh} Coherence on the initially populated site as a function of the scaled time $\tau = t / \trevo$  ($\trevo = 2\pi\hbar/U$), calculated for $\Nav  = 200$ and $\gamma = 0.14$ (red line).
As highlighted in the inset, the restoration of coherence at the revival time is nearly 99 percent.
A perfect revival of the coherence would be obtained for an isolated-site time evolution, as shown by the dashed orange line in the inset. }
\end{center}
\end{figure}

A similar diagnostic, based on superlattice configurations combined with Brillouin zone mapping techniques \cite{Folling07}, can be implemented to experimentally probe the presence or absence of coherence within each one of the two sites of the Bose-Hubbard dimer.
As in the case of isolated lattice sites, this diagnostic would yield as key observable the absolute square of the coherence amplitude $a_l = \langle \hat{a}_l \rangle$, for $l=1,2$, which can be seen as a measure for the closeness of the system's wave function on site $l$ to a coherent state.
Figure \ref{fig:coh} displays the time evolution of $\cohun = |a_1|^2$, which, for short, is denoted as \textit{site 1 coherence}, for the case of a dimer that has a dimensionless hopping strength $\gamma = 0.14$ and is initially loaded with a Bose-Einstein condensate [described via Eq.~\eqref{eq:cs.BEC}] featuring $\Nav =200$ atoms on site $1$ and no atoms on site $2$.
The collapse and revival features are very clearly visible in this observable and display no significant difference to the case of an isolated site, given by Eq.~\eqref{eq:coherence}, with the only exception being that the revival peak does not attain the maximally possible height $\cohun = \Nav $ for $\gamma > 0$.

\begin{figure}[t]
\begin{center}
\includegraphics[width=\linewidth]{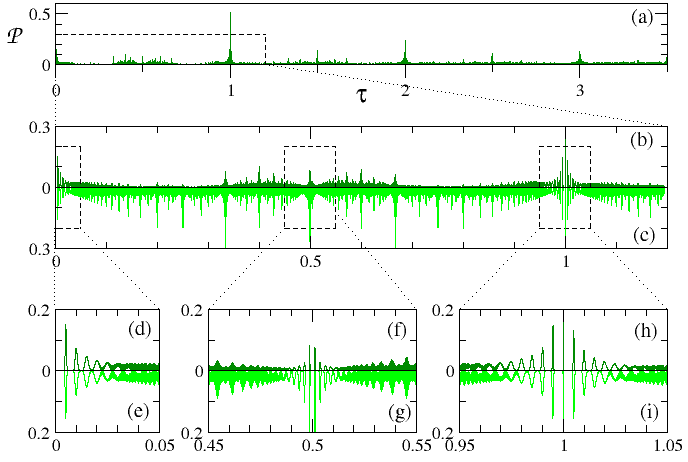}
\caption{\label{fig:probs} Return probability to the initial state as a function of the scaled time $\tau = t /\trevo$, calculated for $\Nav  = 200$ and $\gamma = 0.14$ (a,b,d,f,h) and compared with the corresponding isolated-site return probability (calculated for $\gamma = 0$) (c,e,g,i).
The zoomed panels highlight the first collapse of the recurrence oscillations (d,e), the first half-revival near $\tau=0.5$ (f,g), as well as the first full revival near $\tau = 1$ (h,i).}
\end{center}
\end{figure}

More detailed information on the time evolution of the system is yielded by the return probability $\autocor(t) = |\langle \boldsymbol{\zee}|\boldsymbol{\zee}(t)\rangle|^2$ of the time-evolving wave function $|\boldsymbol{\zee}(t)\rangle$ to the initial state, Eq.~\eqref{eq:cs.BEC}, even though this particular observable is much harder to access in an experiment.
Figure \ref{fig:probs} displays the time evolution of the return probability for an isolated site (lower panels) as well as for the dimer in the presence of a weak dimensionless hopping strength $\gamma = 0.14$ (upper panels), considering an initial condensate state given by $|\boldsymbol{\zee}\rangle = |\sqrt{200},0\rangle$.
Besides the revival peak at $\trevo$, a number of additional peaks structures are visible, corresponding to fractional revivals that occur at $r \trevo/s$ where $r$ and $s$ are coprime integers (see~\cite{Yeazell91} for an experimental realization and \cite{Robinett04} for a more in-depth discussion of fractional revivals).
As pointed out in Ref.~\cite{Tomsovic18}, all the peak structures are signatures of \textit{quantum interference} that can, in principle, be reproduced with properly implemented semiclassical approaches, whereas quasiclassical (e.g., Truncated Wigner) methods are not able to capture them \cite{Lando19}.

\subsection{Interplay with hopping dynamics}

\begin{figure}[t]
\begin{center}
\includegraphics[width=14 cm]{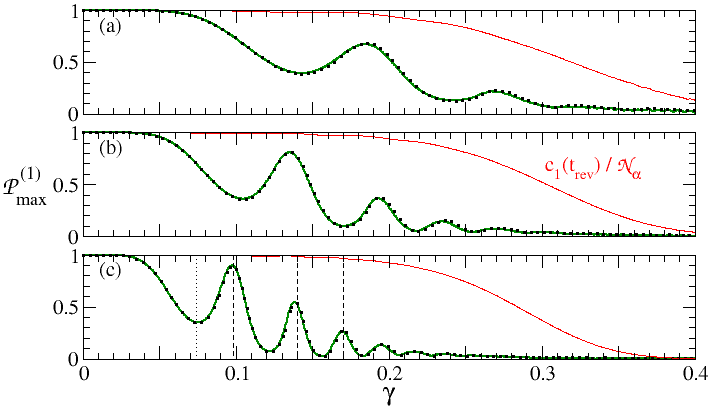}
\end{center}
\caption{Maximal value of the revival peak for the renormalized occupied-site coherence (red line) and the return probability to the initial state (green line) as a function of $\gamma$, for (a) $\Nav  = 50$, (b) $\Nav  = 100$, (c) $\Nav  = 200$. On panel (c), the  dashed lines at $\gamma = 0.098, 0.14$, and   $0.17$ indicate local maxima  of the curve associated with the return probability. 
These maxima are essentially perfectly reproduced by a high order quantum perturbative calculation, indicated by the black dots, which is based on the quantum perturbative framework developed in Appendix \ref{app:qpt}. 
Similar, but far smaller variations of the renormalized occupied-site coherence (red line) exist, but are effectively imperceptible in this figure. 
\label{fig:max}}
\end{figure}

Rather naturally, it would be expected that the introduction of particles hopping to a neighboring site would reduce the maximal return probability $\autocormax= {\rm max}_{t \simeq \trevo} \autocor(t)$ at the revival time as compared to the isolated-site case, even if only a tiny fraction of the population is ever actually transferred to a second site (the return probability is an extremely sensitive measure in a many-body system).  
However, as demonstrated in Ref.~\cite{Schlagheck22}, $\autocormax$ is far from being a monotonously decreasing function of the hopping strength. Figure~\ref{fig:max} illustrates this feature by focusing on the peak return probability $\autocormax(\gamma)$ as a function of the coupling parameter $\gamma$, Eq.~\eqref{eq:gamma}.  One observes a modulation which is quite pronounced, and which, curiously, intensifies as the number of peaks increases, with increasing mean particle number $\Nav $.

\begin{figure}[t]
\begin{center}
\includegraphics[width=\linewidth]{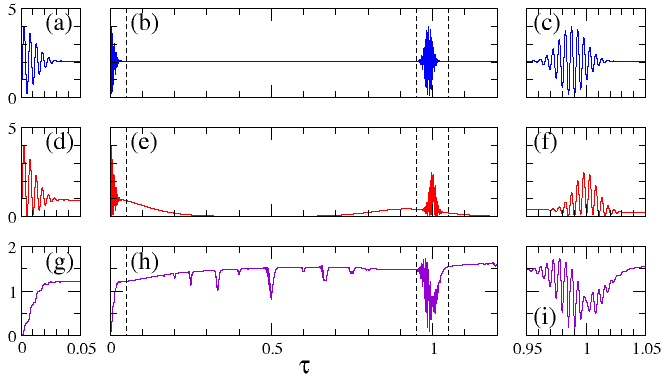}
\caption{\label{fig:hop}
Signatures of the collapse and revival dynamics on the  initially unpopulated site.  Panels(a,b,c) show the mean occupancy, panels (d,e,f) show the mean coherence, and panels (g,h,i) show the entanglement entropy between the two sites, plotted as a function of $\tau = t /\trevo$ for $\Nav  = 200$ and $\gamma = 0.14$.  Panels (a,d,g) and (c,f,i) show zooms on the initial collapse and the revival feature, respectively.
}
\end{center}
\end{figure}

Consider now the dynamics through a variety of observables related to the initially unoccupied site (site-2).  To begin with, Fig.~\ref{fig:hop} panels (a,b,c) plot the time evolution of the mean occupancy $n_2 = \langle \hat{a}_2^\dagger \hat{a}_2\rangle$ for a particular value of the hopping strength, $\gamma=0.14$.
Clearly the occupancy exhibits an oscillatory, yet rapid relaxation to a long time average.  Then just slightly before $\tau \equiv t/\trevo = 1$, this is followed by a revival of short lived oscillatory behavior that seems to just reappear mysteriously.  
Just as for the revivals of the single site dynamics, this reappearance of oscillations is a signature of genuine quantum interference and cannot be reproduced by a quasiclassical Truncated Wigner approximation, which otherwise yields excellent quantitative agreement with the exact quantum behavior of $n_2$ for the initial oscillations, their damping, and the subsequent long term average value~\cite{Schlagheck22}.

As illustrated in Fig.~\ref{fig:hop} panels (d,e,f), the coherence on site $2$, $\coh = |\langle \hat{a}_2\rangle|^2$, displays similar collapse and revival characteristics as $n_2$: 
initial oscillations of $\coh$ fade away, and the oscillations reappear in the vicinity of $\trevo$, again due to quantum interference, though without attaining the the same amplitude as in the initial transient regime.  However,  $n_2$, away from the initial decay of the oscillations and from  their revivals near $\trevo$,  saturates to a constant value.  Remarkably, in the case of $\coh$ this is superimposed on a very slow secular decrease taking place on a time scale much larger than the one of the decay of the oscillations, and leading eventually to  nearly vanishing values. 
Both $n_2$ and $\coh$ correspond to one-body observables that can be accessed in a quantum gas experiment, $n_2$ via quantum gas microscopy \cite{Bakr09,Sherson10} and $\coh$ via time-of-flight absorption imaging, similar to the procedure in~\cite{Greiner02b} except combined with Brillouin zone mapping techniques in order to single out the second site~\cite{Folling07}.

Signatures of the more subtle partial revivals are practically not at all visible in the above one-body observables.  However, they are revealed in the entanglement entropy between the two sites.
For the sake of technical simplicity, consider the \textit{linear entropy} 
\begin{equation}
    S(t) = - \ln \mathrm{Tr}\left[\rho_t^2\right] \,,
\end{equation}
corresponding to the case $\alpha = 2$ of the family of R\'enyi entropies $S_\alpha = (1-\alpha)^{-1} \ln \mathrm{Tr} [\rho^\alpha]$, where 
\begin{equation}
\rho_t = \mathrm{Tr}_2 |\Psi_t\rangle\langle\Psi_t|
\end{equation}
is the reduced density matrix associated with site $1$.
Decomposing the system's state vector in the two-site Fock basis as
\begin{equation}
|\Psi_t\rangle = \sum_{n_1,n_2=0}^\infty c_{n_1 n_2}(t) |n_1,n_2\rangle \,,
\end{equation}
gives
\begin{equation}
\mathrm{Tr}\left[\rho_t^2\right] = \sum_{n_1,n_1',n_2,n_2'=0}^\infty c_{n_1 n_2}^*(t) c_{n'_1 n_2}(t) c_{n'_1 n'_2}^*(t) c_{n_1 n'_2}(t) \,.
\end{equation}
Figure~\ref{fig:hop} panels (g-i) display the behavior of this entanglement entropy as a function of time. 
Full and partial revivals are clearly visible, giving rise to local minima in $S(t)$, indicating the partial disentanglement of the evolving state near times $\tau = r\tau_1/s$.  Again, quantum interference is essential to produce these results.

\begin{figure}
\begin{center}
\includegraphics[width=\linewidth]{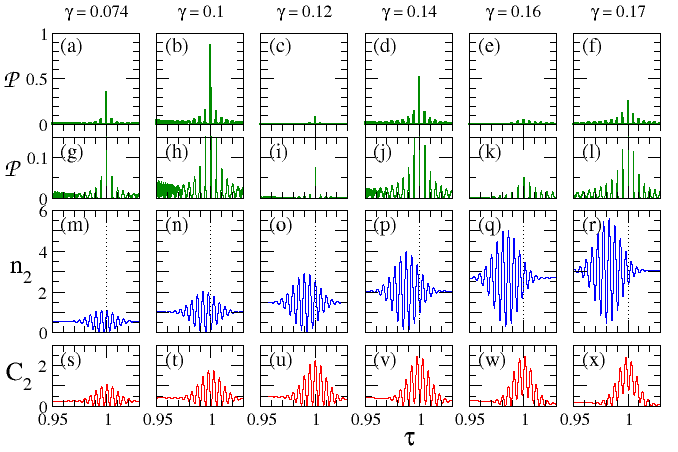}
\end{center}
\caption{\label{fig:sync}
Correlation effects between the revival peak of the phase evolution dynamics on site 1 and the oscillations in the hopping dynamics. Shown are, for $\Nav  = 200$ and panels (a,g,m,s) $\gamma = 0.074$, panels (b,h,n,t) $\gamma = 0.1$, panels (c,i,o,u) $\gamma = 0.12$, panels (d,j,p,v) $\gamma = 0.14$, panels (e,k,q,w) $\gamma = 0.16$, and panels (f,l,r,x) $\gamma = 0.17$, the return probability $\autocor$ panels (a-l), the site 2 occupancy $n_2$ panels (m-r), and the site 2 coherence $\coh$ panels (s-x), as a function of the scaled time $\tau = t/\trevo$.
Consistent with Fig.~\ref{fig:max}, panels (a-f) clearly illustrate that the revival peak height varies non-monotonously with $\gamma$.
Locally maximal peak heights in panels (b,d,f) are encountered if the site 2 occupancy is minimal exactly at the revival time (marked by dotted vertical lines), corresponding to local maxima in the site $1$ occupancy, whereas a locally maximal site $2$ occupancy (and hence a minimal site $1$ occupancy) leads to a local minimum in the revival peak height panels (a,c,e).
}
\end{figure}

On the face of it, it might have been expected that the impact of adding hopping to the dynamics would mainly just degrade the revival features of the single site system, and that the aim of the quantum perturbation and semiclassical theories  developed in the next sections would be to describe quantitatively this phenomenon.  However, it appears that the presence of a second site gives rise to a much richer physics than the one of the one site revivals, and that a number of striking qualitative phenomena call for a physical explanation.  

To start with, it is worthwhile to explore the origin, and the characteristics, of the resurgent behavior observed for the maximum return probability $\autocormax$. In particular, as pointed out in Ref.~\cite{Schlagheck22} and demonstrated  in Fig.~\ref{fig:sync}, the presence of hopping gives rise to intriguing cooperative effects, manifested in the form of enhancements or suppressions of the revival peak height, depending on whether the population on the second site is minimal or maximal at $\trev$.
In other words, the revival peak height exhibits local maxima (as a function of $\gamma$) that correspond to a minimal transfer of particles to the initially unoccupied site $2$.
Incidentally, this effect of synchronization correlates with the presence of integer or half-integer time-averaged populations on the second site, as if some sort of quantization law was at work. 
As is clearly visible in Fig.~\ref{fig:sync}, an integer population on the second site is, on average, encountered whenever, at the phase oscillation revival time, the instantaneous mean occupancy on that site is minimal, and thus the revival peak height in the return probability is maximal, whereas a minimal revival peak height, corresponding to a maximal instantaneous mean occupancy on the second site at the revival time, is linked with a half-integer value for the time-averaged population of that site.  
A second puzzling feature that requires explanation is the difference of behavior between an observable like $n_2$, for which nothing happens except near $t \simeq r \trev$, and $a_2$, which is characterized by a very slow decay of its secular component, implying the existence of a time scale much longer than $\tosco$, Eq.~\eqref{eq:tosco}, and $\tdeco$, Eq.~\eqref{eq:tdeco}.

To avoid being drawn too quickly into the technical aspects of the quantum perturbation and semiclassical theories,  the next section, Sec.~\ref{sec:explanation}, provides these explanations at a descriptive level together with a selected number of analytical expressions given without derivation. The
derivations are detailed in the remainder of the manuscript (Secs.~\ref{sec:classical-dimer} and \ref{sec:semirev} and the appendices).

\section{Physical interpretation}
\label{sec:explanation}

The preceding section revealed the presence of a number of novel revival phenomena that are specific to the system under study, namely the fact that a two-degrees-of-freedom system with integrable dynamics is considered, as well as to the observables that the physical context of ultracold bosonic atoms offers, and that did not occur in conventional collapse-and-revival investigations focusing essentially on the behavior of the autocorrelation function in effective one-degree-of-freedom systems. Specifically, there are at least five remarkable ``puzzles'' that we would like to elucidate:
\begin{itemize}
\item[P1] First of all, there is not one single revival time, but rather several of them are encountered.  For example, Fig.~\ref{fig:sync} shows that $\trevauto$ (the time at which $\autocormax(\gamma)$ occurs) remains essentially pinned at $\trevo$ (it has a tiny shift to later times), and in the same way the location of the maxima of the envelop of the site-2 coherence $\coh$ has  little variation with $\gamma$, whereas on the contrary the maxima of the envelope of the site-2 occupation $n_2(t)$ has a visible shift toward shorter times.
\item[P2] As observed in the middle pannel of Fig.~\ref{fig:hop}, the site-2 coherence features also some sort of a collapse and revival behavior in its smooth background. Again, this phenomenon is totally absent in conventional revival dynamics; it does not arise in the site-1 coherence (see Fig.~\ref{fig:coh}) that was studied in previous experiments \cite{Greiner02b}.
\item[P3] As for the autocorrelation function, a most remarkable feature, visible for instance in Fig.~\ref{fig:max}, is the pronounced oscillation as a function of  the perturbation parameter of the revival peak height $\autocormax(\gamma)$. This phenomenon was already highlighted in Ref.~\cite{Schlagheck22}. 
\item[P4] Associated with this phenomenon, it turns out (cf.~Fig.~\ref{fig:sync}) that the  peaks in $\autocormax(\gamma)$  are encountered if  $n_2$ has a local minima exactly at $\trevauto$, whereas a locally maximal $n_2$  leads to a dip of  $\autocormax(\gamma)$.
\item[P5] Again visible on Fig.~\ref{fig:sync} and rather intriguingly, peaks in $\autocormax(\gamma)$ occur whenever the time-averaged $n_2$ is an integer, whereas half-integer  averaged $n_2$ correlate with local minima of the revival peak. Again, this apparent (pseudo)-quantization rule was already encountered in Ref.~\cite{Schlagheck22}.
\end{itemize}

To shed light on these new revival phenomena, and understand the basic mechanisms
that underlie their existence, the semiclassical framework evoked in Sec.~\ref{sec:rev-1d}
becomes critically important. Indeed, it can, and will, be shown that these phenomena all have classical roots and can be quantitatively understood from a suitable phase-space analysis of this system, using standard semiclassical theory for their proper translation into the quantum realm.  This is done in detail in Secs.~\ref{sec:classical-dimer} and \ref{sec:semirev}.
This particular section, Sec.~\ref{sec:explanation}, is devoted to providing an ``executive summary'' of the detailed analysis, focusing on physical interpretations and leaving out technical details that are postponed to the subsequent sections.
We also compare here semiclassical findings with those obtained from quantum perturbation theory, detailed in Appendix~\ref{app:qpt}, which represents an alternative (semi-)analytical route to the understanding of the system that can provide complementary insight --- albeit of lower intuitive and explanatory power than the semiclassical approach.

\subsection{Classical frequencies, shearing rates, and precession}
\label{sec:freqshear}

It is mentioned in Sec.~\ref{sec:semic-1d} that in a $1D$ system partial initial state recurrences, which appear with a classical frequency, $t_{\rm osc}$ (Eq.~\eqref{eq:ts}), are distinguished from revivals, which appear with a frequency, $t_{\rm rev}$ (Eq.~\eqref{eq:ts}), by the fact that revivals occur after the initial state has completely dispersed.  The revivals occur on a much longer quantum time scale, which nevertheless is determined by the rates of shearing in the classical dynamics.

Not surprisingly then, the first step in interpreting the phenomena of Sec.~\ref{sec:RevInBHdimer} is to identify properly the relevant classical frequencies and shearing rates. For the Bose-Hubbard dimer, there is, in addition, a relevant  dynamical feature, which cannot be present in $1D$ systems, namely precession, which is responsible for part of the wealth of phenomena noted in the previous section, and that is discussed below.

As in Sec.~\ref{sec:monomer}, the (mean field) semiclassical limit considered here corresponds to $\heff \to 0$  where the effective Planck constant $\heff \sim \Nav^{-1}$ is defined by Eq.\eqref{eq:heff}.  
Furthermore, to define properly this semiclassical limit, it is necessary to rescale energies and times as in  Eq.~\eqref{eq:rescaling-Et} and the coupling constant as in Eq.~\eqref{eq:gamma},  giving for the rescaled classical Hamiltonian associated with Eq.~\eqref{eq:cs.BHd}
\begin{equation}
\label{eq:ham2PI}
    \tH (s_1,s_2,\theta_1,\theta_2) = \frac{1}{2} \left[ s_1(s_1-2\heff) + s_2(s_2-2\heff) \right] - 2 \gamma \sqrt{s_1 s_2} \cos(\theta_1 - \theta_2) \quad .
\end{equation}
As in Eq.~\eqref{eq:nop}, $s_{1,2}/\heff  \equiv \nW_{1,2} + \frac12$, where $\nW_j$ is (the Weyl symbol of) the particle number operator $\hat n_j$ on site $j=1,2$, and the  $\theta_j$ are the corresponding angular variables (mean field phases).  These scalings justify that $\gamma$ is the relevant parameter measuring the strength of the coupling between the two sites.  Furthermore, introducing $\Sa \equiv \heff \Nav$ ($\Sa$ thus has the dimension of an action, and with our convention a numerical value $2$) makes apparent that the rescaled times 
\begin{align}
    \ttosco &= \frac{2\pi}{\Sa} \label{eq:ttosco} \; ,\\
    \ttrevo &= \frac{2\pi }{\heff}\label{eq:ttrevo} \; .
\end{align}
(see Eqs.~\eqref{eq:tosco}-\eqref{eq:trevo}) are respectively $O(\heff^0)$ and $O(\heff^{-1})$.   
Below, all ``classical'' quantities (in particular the  frequencies and their derivatives) are always discussed in the context of the {\em scaled} Hamiltonian $\tH$. 
However to simplify the notation, the tilde is dropped except for the Hamiltonian itself and scaled time variables (e.g. to distinguish the rescaled time $\ttrevo$, Eq.~\eqref{eq:ttrevo}, which is a pure number divided by $\heff$, from its (experimentally relevant) unscaled analog Eq.~\eqref{eq:trevo}). 

Because the total number of particles $[\ntot]^W = \nW_1 + \nW_2$ is a conserved quantity, it is often more appropriate to describe the classical motion in term of $(S=s_1+s_2, s= (s_1-s_2)/2)$), which are such that $S/\heff= [\ntot]^W + 1$ and $s/\heff \equiv (\nW_1 - \nW_2)$ is the particle imbalance.  $S$ is therefore also a conserved quantity.   The  corresponding mean field phases are $\Theta = (\theta_1 + \theta_2)/2$ and $\theta = \theta_1 - \theta_2$,  where the latter can be understood as a Josephson phase. 
Furthermore, since the initial state,  Eq.~\eqref{eq:PSI0}, has all the particles in site $1$, this classical imbalance is initially maximal, which means that in the $(s,\theta)$ coordinates the classical motion starts from the top of the allowed region of phase space (see top row of Fig.~\ref{fig:level_curves}), or in the equivalent spherical representation, Eq.~\eqref{eq:spherical}, from the north pole of the sphere (see bottom row of Fig.~\ref{fig:level_curves} ahead in Sec.~\ref{sec:classical-dimer}).   

Owing to the fact that the Bose-Hubbard dimer has two constants of the motion, the energy and the total number of particles,  the Hamiltonian Eq.~\eqref{eq:ham2PI} is integrable, which implies that trajectories are confined on two-dimensional tori that can be characterized by a pair of action $\bI^T := (S,m)$ such that $\tH(\bI)$ is independent of corresponding angular variables $\mathbf{\Phi}^T = (\varphi_S,\varphi_m)$.   The second action variable $m$, defined by Eq.~\eqref{eq:action-m}, is the symplectic area enclosed by the trace  of the torus in the  $(s,\theta)$ plane. 
In the limit of uncoupled sites, $\gamma\to 0$,  it  corresponds to the action variable $s_2$ associated with the number of particles in the second site, and 
$\varphi_m$ is identical to $\theta_2$,  whereas $\varphi_S$ corresponds to the mean angle $(\theta_1 + \theta_2)/2$.

Because the initial coherent state has a maximum imbalance, a central role is played by the torus to which the north pole belongs.  We shall call it the `{\em north-pole torus}, and furthermore refer to the intersection of this torus with the $(s,\theta)$ plane, (or for the equivalent spherical coordinates) as the {\em north-pole trajectory}.
The north-pole torus is characterized by $\bIa^T = (\Sa,\ma)$ where $\Sa/\heff = \Nav +1$ and where $\ma$, given by Eq.~\eqref{eq:ma},  corresponds to the symplectic area enclosed by the north-pole trajectory (red curve in the top row of Fig.~\ref{fig:level_curves}). Furthermore, one should keep in mind in the rest of the discussion that, the north-pole trajectory contracts to a fixed point as $\gamma \to 0$, $\ma (\gamma \smeq 0) = 0$, and more precisely  $\ma(\gamma) =  \gamma^2/\Sa + O(\gamma^4)$  (see Eq.~\eqref{eq:ma}).
In practice, many results below are expressed in terms of the slightly different quantity $\hma$ defined 
by Eq.~\eqref{eq:hma}.  The action $\hma$  is not strictly speaking the north-pole trajectory action $\ma$, but it can be identified with  it in the perturbative regime ($\hma(\gamma) = \ma(\gamma) + O(\gamma^4)$), and remains numerically very close across the entire $\gamma$-domain of interest (see inset of Fig.~\ref{fig:autocorr-max}).

The Wigner transform of the initial coherent state Eq.~\eqref{eq:PSI0} is centered on the ``north-pole torus'' $\bIa^T = (\Sa,\ma)$.  Following the spirit of Secs.~\ref{sec:semic-1d}-\ref{sec:EBKboot}, the main inputs within the semiclassical description are, beyond the action $\Sa$ and $\ma$ themselves,  the frequencies $(\omega_S := \partial_S \tH,  \omega_m := \partial_m \tH)$ and their derivatives, {\em computed on the torus $\bIa$}, and given by Eq.~\eqref{eq:omegas-5}.
Below we use the more compact notation 
\begin{equation} 
\label{eq:Omegas}
 \bOmega =\begin{pmatrix}
        \omega_S \\ \omega_m
    \end{pmatrix} \; \qquad
    \partial \bOmega =\begin{pmatrix}
        \partial_S \omega_S & \partial_m \omega_S \\
        \partial_S \omega_m & \partial_m \omega_m 
    \end{pmatrix} \; .
\end{equation}
(Note that to lighten the notations we have dropped the index $^\alpha$ on the frequency $\omega_{S,m}$ and their derivatives, but in this section all theses quantities are assumed to be computed on the north-pole torus $\bIa$.)
In the perturbative regime $\gamma \ll 1$ we can Taylor expand these quantities near $\gamma \smeq 0$ (unperturbed system) as
\begin{align}
\label{eq:omegaexpan}
\bOmega(\gamma) &= \bOmega^0 + \gamma^2 \bOmega^1 + O(\gamma^4)\nonumber \\
\partial \bOmega(\gamma) &= \partial \bOmega^0 + \gamma^2 \partial \bOmega^1 + O(\gamma^4) \ ,
\end{align}
with the unperturbed frequencies and their derivatives
\begin{equation} 
\label{eq:Omegas0}
 \bOmega^0 =\begin{pmatrix}
        \Sa -\heff \\ -\Sa 
    \end{pmatrix} \; \qquad
    \partial \bOmega^0 = \begin{pmatrix}
        1 & -1 \\
        -1 & 2 
    \end{pmatrix} \; ,
\end{equation}
(note that the rescaled frequency derivatives are pure numbers) and their first variation with the coupling given by
\begin{equation} 
\label{eq:Omegas1}
 \bOmega^1 = \frac{1}{\Sa}\begin{pmatrix}
        -1  \\ \;\; 4 
    \end{pmatrix} \; \qquad
    \partial \bOmega^1 = \frac{1}{\Sa^2}\begin{pmatrix}
        0  & -2  \\
        -2  & 12 
    \end{pmatrix} \; .
\end{equation}
Note that Eqs.~\eqref{eq:Omegas0} and \eqref{eq:Omegas1} can be exactly deduced from quantum mechanical discrete first and second energy differences using the quantum perturbative results; see Eq.~\eqref{eq:dOmegas00} and the last paragraph of Appendix~\ref{app:qcf}.  For all those quantities that depend on the frequencies and shearing rates, there is guaranteed consistency with semiclassical theory to the order calculated.

\begin{figure}[tb]
\centering
     \includegraphics[width=\linewidth]{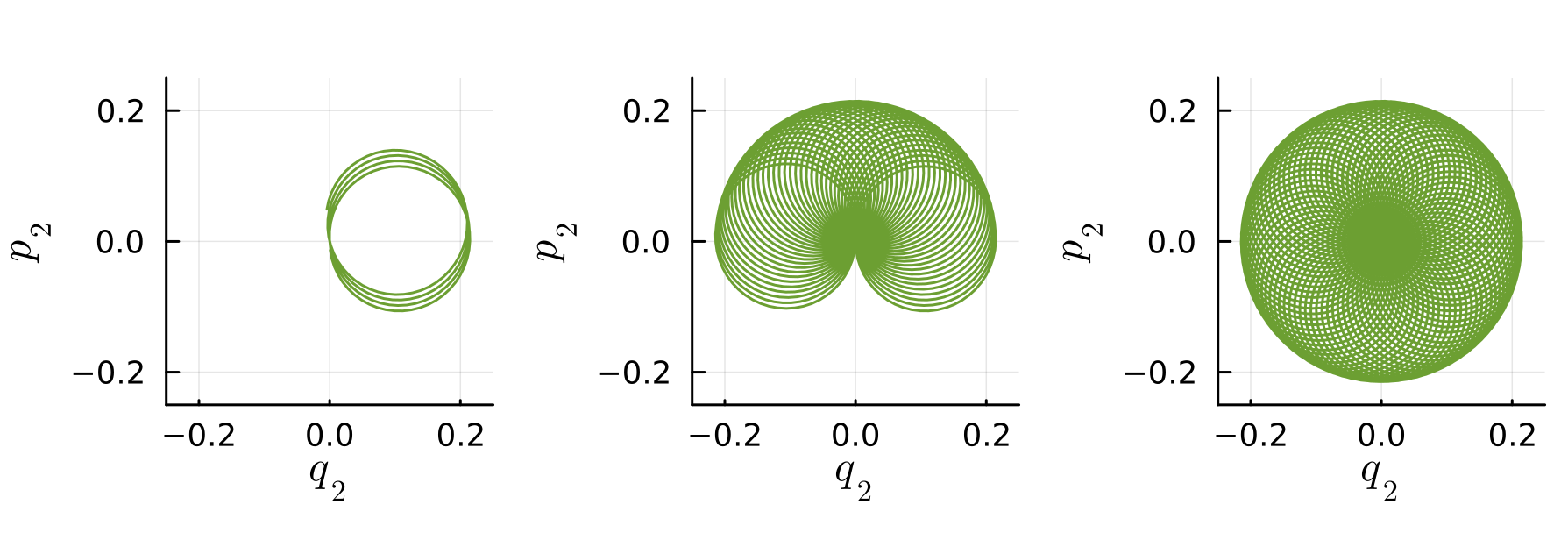} 
\caption{Projection on the $(q_2,p_2)$ plane of a trajectory initiated at $(q_1,p_1,q_2,p_2) = (\sqrt{2 \Sa}, 0, 0, 0)$, and evolved during a time $2 \, \tosco$ (left), $20 \, \tosco$ (middle), and $50 \, \tosco$ (right), for $\gamma = 0.3$.  The rather fast oscillation corresponding to the $\theta$ dynamics is superposed on a very slow precession of the ``guiding center''
.  As the operator $\hat a_2$ is semiclassically associated with $\frac{1}{\sqrt{2 \heff}}(q_2+i p_2)$ (cf Eq.~\eqref{eq:quadratures}), the very slow decay of $\mean{a_2}$ observed in panels (d), (e), and (f) of Fig.~\ref{fig:hop} is associated with the time necessary for this guiding center to fully circle around the origin.  The associated time scale, although of purely classical origin, can be comparable to the quantum time scale $\trevo$ for the values of $\Nav$ and $\gamma$ that are considered in this paper.}
\label{fig:precession}
\end{figure}

For the unperturbed system $\omega_S + \omega_m$ nearly vanishes, and more precisely, neglecting the $\heff$ corrections  $\omega_S + \omega_m \approx 3\gamma^2/\Sa + O(\gamma^4)$ (Eqs.~\eqref{eq:omegaexpan} -- \eqref{eq:Omegas1}). This has an important consequence readily visible in Fig.~\ref{fig:precession} in which the dynamics for the classical variables $(q_2,p_2)$ (the classical symbols of the quadratures $(\hat q_2, \hat p_2)$ associated with the coherence $\hat a_2$) is displayed for the trajectory initiated at the center of an initial coherent state specified in the caption.  Indeed, the quasi-elliptical motion shown in the left panel can be seen as taking place with the relatively fast frequency $\omega_m$.  However, the guiding center of the quasi-ellipse can be seen to precess around the point $(0,0)$ with the very small frequency $(\omega_S + \omega_m)$, leading for much later time to what is seen on the middle and right panels of Fig.~\ref{fig:precession}.  Because of the deep relation between the coherence $\hat a_2$ and the classical variable $(q_2,p_2)$, this has a significant effect on the time dependence of $\mean{a_2}$; see Sec.~\ref{sec:IV-a2} ahead.

\subsection{Number oscillations}
\label{sec:IV-n2}

\begin{figure}
    \centering
     \includegraphics[width=\linewidth]{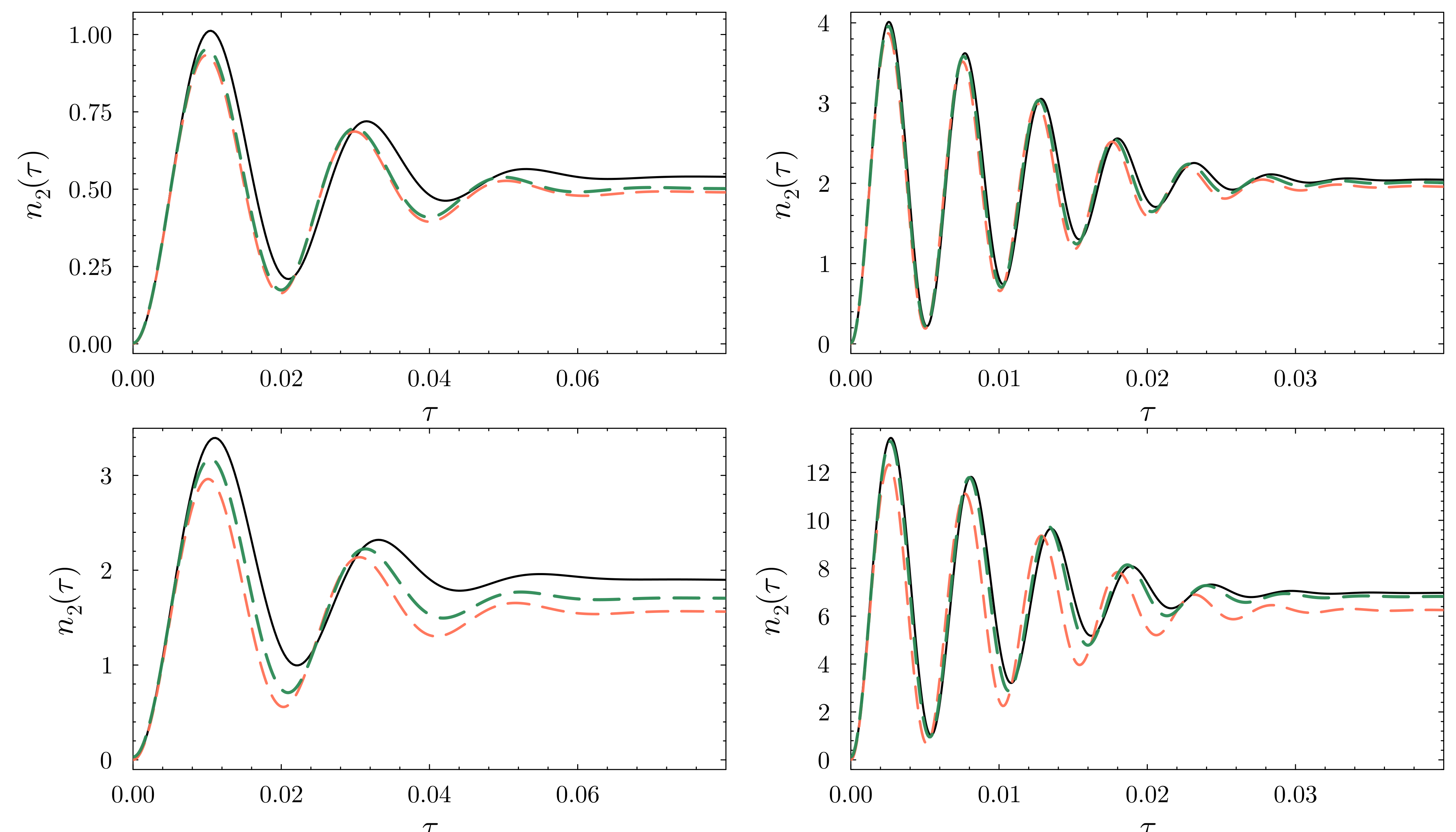} 
     \caption{Number of particles $n_2$ in the second site as a function of the rescales time $\tau = t/\trevo$, near $t=0$.  Solid black : exact quantum ; Dashed  green : semiclassical (TWA) Eq.~\eqref{eq:n2t0-f0f1}; Dashed red : quantum perturbation theory.  Left column : $\Nav= 200$, right column $\Nav=50$, upper row :$\gamma = 0.14$, lower row :$\gamma = 0.25$}
    \label{fig:number-t0}
\end{figure}

Consider first the time evolution of the average number of particles on site 2.  In that case, the angular variation of the Weyl symbol  of the operator $\hat n_2$ can be shown to be expressed on the torus $\bIa$ as $\nW(\bIa,\bPhi) \equiv  n_2^\alpha(\varphi_m)$ given by
\begin{equation} \label{eq:n2Wigner}
    n_2^\alpha(\varphi_m)   = \left(\frac{2\gamma^2}{\Sa \heff}\right)  \tf(\varphi_m) \;  ,
    \qquad
    \tf(\varphi_m) := \left(\frac{\Sa}{2\gamma} \right)^2 \left(1-\sqrt{1 - \left(\frac{4 \gamma}{\Sa}\right)^2 \cos^2 \widetilde \theta \left(  \varphi_m \right) } \right) \quad , 
\end{equation}
with the function $\ttheta(\varphi_m)$ given by Eq.~\eqref{eq:tilde-theta}, and which remarkably depend only on the second component of $\bPhi^T = (\varphi_S,\varphi_m)$.     In the small  $\gamma$ regime, $f(\varphi_m)$ simplifies to 
\begin{equation} \label{eq:n2WignerPert}
\tf(\varphi_m) =   (1 - \cos \varphi_m)  + O(\gamma^4) \; ,
\end{equation}
which implies that for most values of the coupling within the self-trapping regime $\tf(\varphi_m) = \sum_k \tf_k  \exp(ik \varphi_m)$ is dominated by its $0$ and $\pm 1$ Fourier components.  Keeping only these Fourier components near $t=0$ gives 
\begin{equation} \label{eq:n2t0-f0f1}
   \TWA{n_2}(\tt) = \frac{2 \gamma^2}{\Sa \heff} \left(\tf_0
    +  
    2  \tf_1 \cos \left[  \omega_m  \tt \right]
    \exp \left[ -\frac{\heff}{2} 
    \left(\Sa (\partial_S \omega_m)^2 
    + \hma (\partial_m \omega_m )^2 \right) \, \tt^2 \right]  \right) \; , 
   \end{equation}
with $\lim_{\gamma \to 0} \tf_0 = 1$, $\lim_{\gamma \to 0} \tf_1 = -1/2$, and where the parameter $\hma \simeq \ma = O(\gamma^2)$  is defined in Eq.~\eqref{eq:hma-sec6}.  As  seen in Fig.~\ref{fig:number-t0}, this expression and the quantum perturbative one, Eq.~\eqref{eq:pt.n2pert}, give excellent agreement with the quantum data.   The behavior of $\mean{n_2}(\tt)$ near $t=0$ is characterized by oscillations of frequency $\omega_m$ and amplitude $\simeq  \gamma^2 / 2\heff$, that, neglecting $\hma = 0(\gamma^2)$ in front of $\Sa$, decay on a time scale $\tdec \simeq 2/ (\partial_S \omega_m \sqrt{\Sa \heff})$, and saturate to a constant value $ \simeq  \gamma^2 / 2\heff$.

\begin{figure}
    \centering
     \includegraphics[width=\linewidth]{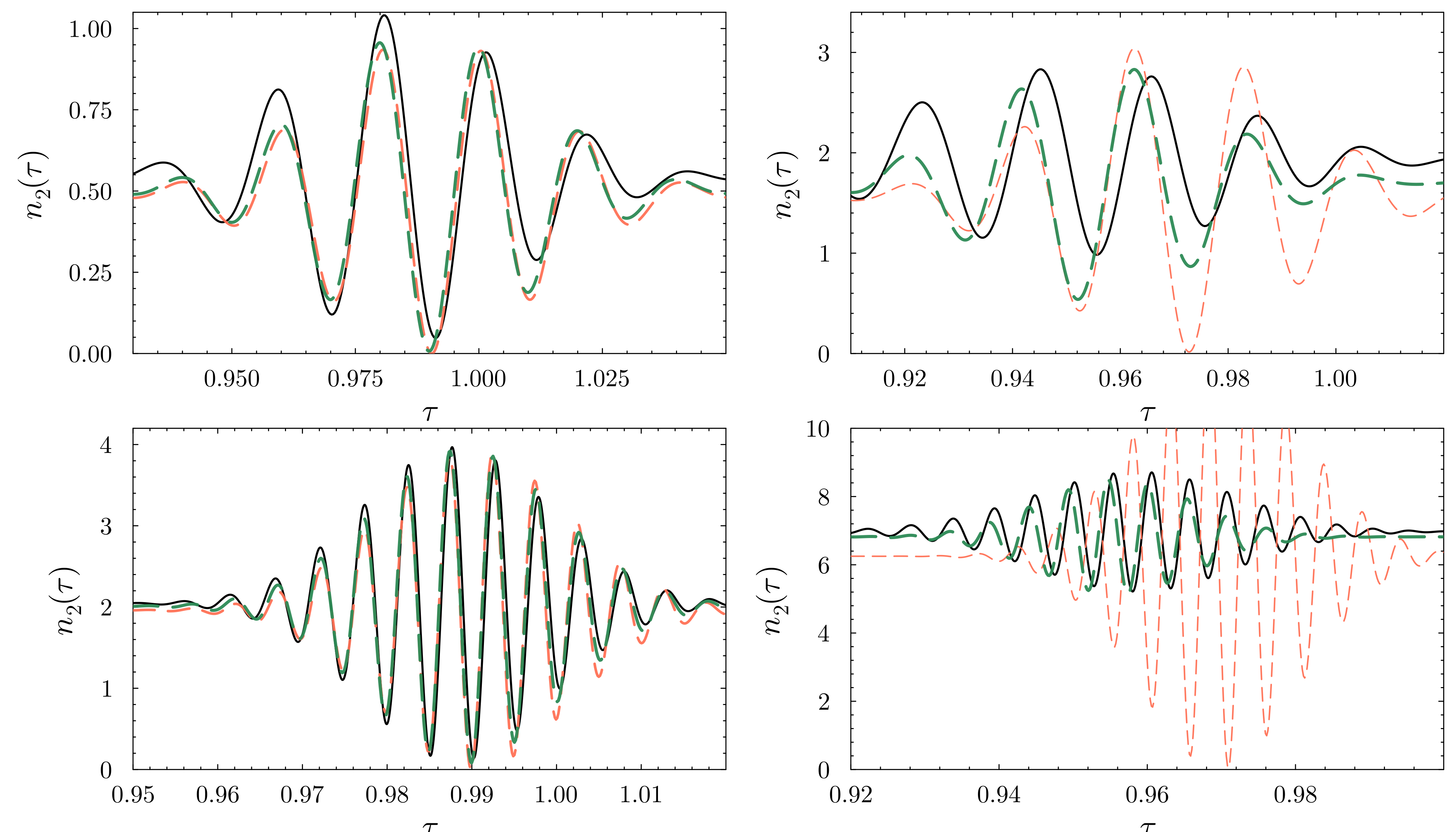} 
     \caption{Number of particles $n_2$ in the second site as a function of the rescales time $\tau = t/\trevo$, near $t=\trevo$.  Solid black : exact quantum ; Dashed  green : semiclassical (EBK-embedded-TWA) Eq.~\eqref{eq:n2trev-f0f1}; Dashed red : quantum perturbation theory.  Left column : $\Nav= 200$, right column $\Nav=50$, upper row :$\gamma = 0.14$, lower row :$\gamma = 0.25$}
    \label{fig:number-trev}
\end{figure}

Near $\trevo$ we get using the EBK-embedded TWA approximation
\begin{align} 
   \Boot{n_2}(\tt) &= \frac{2\gamma^2}{\Sa \heff}
   \left(\tf_0     +  2 \tf_1 \cos \left[   (\omega_m  \tt + \Delta_m) \right]
       \exp \left[-\frac{\heff}{2} \GBoot_{22}(\tt) \right] \right)\; ,
       \label{eq:n2trev-f0f1} \\  
\mbox{with} \quad  \GBoot_{22}(\tt) :& = 
\Sa  (\partial_m \omega_S)^2 (\tt -  \ttrev^{mS})^2 
+ \hma (\partial_m \omega_m)^2 (\tt - \ttrev^{mm})^2
 \; , \label{eq:G22}
 \end{align}
where $\Delta_m = - 4 \pi \la$,  with\footnote{Using the general definition: $x\, \mathrm{cmod}[y] := [(x+y/2)\mod{y}]-y/2$.} $\la := \cmodun{\ma/\heff - 1/2}$, and 
\begin{align*}
    \ttrev^{mS} &= \frac{(\partial_m \omega_S)_0 }{(\partial_m \omega_S)} \ttrevo \; , \\
    \ttrev^{mm} &= \frac{(\partial_m \omega_m)_0 }{(\partial_m \omega_m)} \ttrevo   \; .
\end{align*}
As seen in Fig.~\ref{fig:number-trev}, this also gives excellent agreement with the quantum data.  The quantum perturbative expression, Eq.~\eqref{eq:pt.n2pert}, is also very accurate for $\gamma = 0.14$, but shows significant deviations for $\gamma=0.25$.

Equation~\eqref{eq:n2trev-f0f1} indicates that the evolution of $\mean{n_2}(\tt)$ near the revival time is very similar to that near $t=0$, and is characterized in particular by  oscillations on a classical time scale of $2\pi/\omega_m$ under  an envelope of width $\sim \heff^{-1/2}$. However, there are a few differences.  One of them  is the existence of the phase shift $\Delta_m$ in the classical oscillations.  Another difference is  that the maximum of the envelope no longer needs to correspond to a minimum of these oscillations.  Furthermore, the maximum of the envelope  shifts slightly in position because  the difference between   $(\partial_m \omega_S,\partial_m \omega_m)$ and their unperturbed values causes  $\ttrev^{mS}$ and $\ttrev^{mm}$ to differ from their unperturbed value $\ttrevo$. Additionally, as $\ttrev^{mS}$ and $\ttrev^{mm}$ have different variations with the parameter $\gamma$, the two terms of  $\GBoot_{22}(\tt)$ cannot be zero at the same time.  Each of them would tend to shift the maximum of the envelope  to a different place, leading to an incomplete revival.  However, since $\hma = O(\gamma^2)$, the second term of $\GBoot_{22}(\tt)$ remains relatively small in the perturbative regime, and the revival in this regime is essentially  complete.

\subsection{Coherence}
\label{sec:IV-a2}

\newcommand{\tg}{\tilde g}

\begin{figure}
    \centering
     \includegraphics[width=\linewidth]{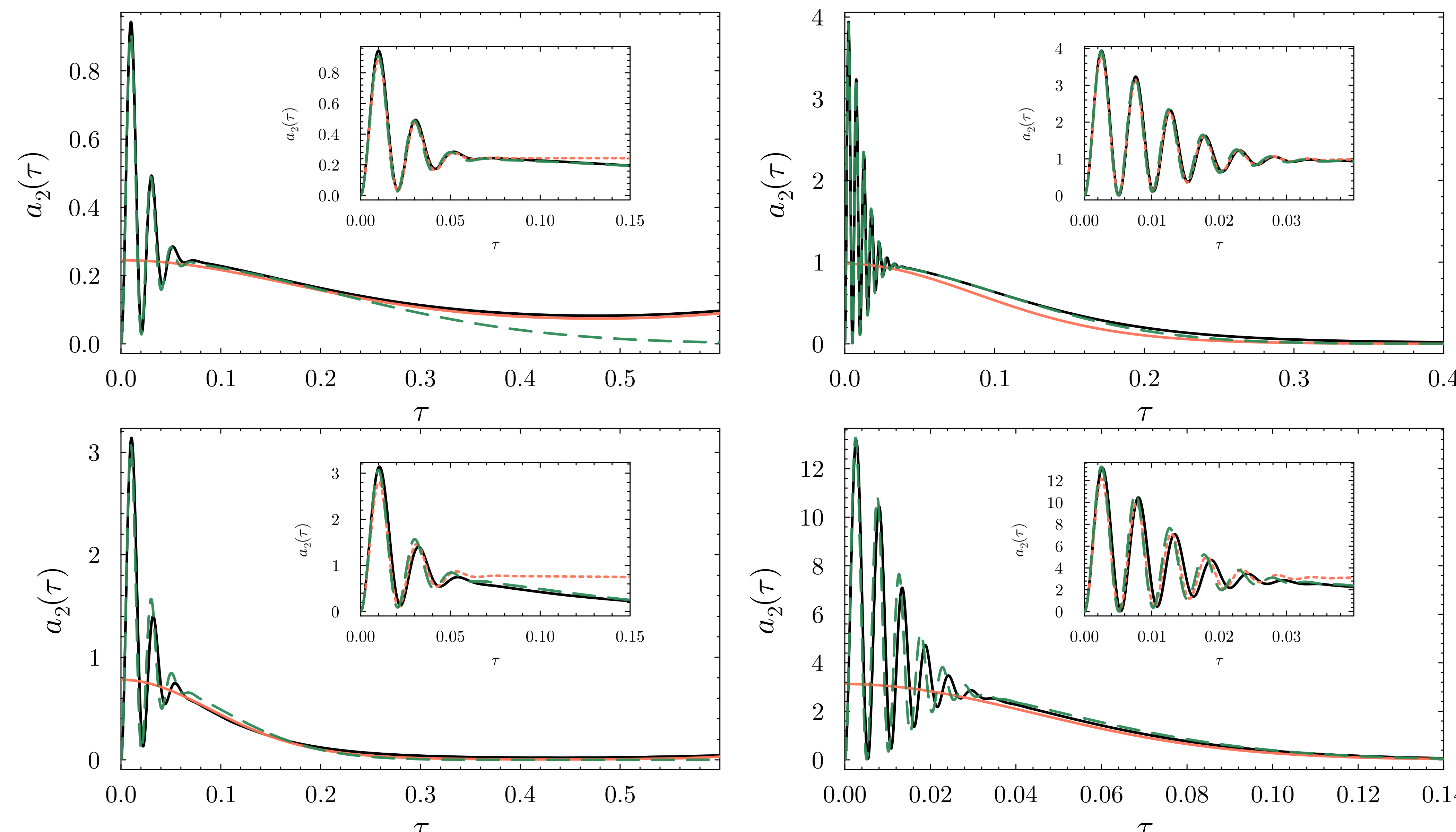} 
     \caption{Site two coherence $a_2$  as a function of the rescales time $\tau = t/\trevo$, near $t=0$.  Solid black: exact quantum ; Dashed  green : semiclassical (TWA) Eq.~\eqref{eq:n2t0-f0f1};  Red: quantum perturbation theory (solid [main panel] Eq.~\eqref{eq:A55}, dotted [inset] Eq.~\eqref{eq:cohamp}).  Left column: $\Nav= 50$, right column $\Nav=200$, upper row: $\gamma = 0.14$, lower row :$\gamma = 0.25$.  Inset : blow up of the for short times.}
    \label{fig:coherence-t0}
\end{figure}

The calculation of the coherence, $\mean{a_2}$, follows essentially the same route as that of $\mean{n_2}$, except that the angular dependence of the Weyl symbol of $\hat a_2$ on the north-pole torus reads $\aW(\Ia,\bPhi) \equiv a_2^\alpha(\varphi_S,\varphi_m)$ with
\begin{equation} 
\label{eq:a2-phis}
    a^\alpha_2(\varphi_S,\varphi_m) = i \sqrt{\frac{\frac12 \Sa - s^\alpha(\ttheta(\varphi_m))}{\heff}}   
    (-1)^{\floor{-\varphi_m/2\pi} }
    \exp \left[ -\frac{i}{4} \left(\varphi_m - 2 \tttheta(\varphi_m) \right) \right] 
    e^{-i \varphi_S} 
    \; ,
\end{equation}
(cf.~Eq.~\eqref{eq:a2-explicit}), with the functions $\ttheta (\varphi_m)$ and $\tttheta (\varphi_m)$  given by Eqs.~\eqref{eq:tilde-theta}-\eqref{eq:tilde-tilde-theta}, and $s^\alpha(\theta)$ by Eq.~\eqref{eq:stheta}.  Keeping only the two dominant Fourier components of  $a^\alpha_2(\varphi_S,\varphi_m)$ leads near $t=0$ to  
\begin{equation} 
\label{eq:mean-a2t0-s4}
\begin{aligned}
   & \TWA{a_2} (\tt)  = \sqrt{\frac{2 \gamma^2}{\heff \Sa}}
    \Bigg[ 
    \tg_0 \exp(-i\omega_S  \tt) \exp \left[ - \frac{\heff}{2} 
    \left(\Sa (\partial_S \omega_S)^2  + \hma (\partial_S \omega_m )^2 \right) \,\tt^2
 \right] \\
& +
        \tg_{-1} \exp(-i(\omega_S + \omega_m) \tt) \exp \left[ - \frac{\heff}{2}  \left( \Sa (\partial_S(\omega_S+\omega_m))^2  + \hma (\partial_m(\omega_S+\omega_m))^2 \right) \, \tt \right]
    \Bigg]
    \end{aligned}
    \qquad  ,
\end{equation}
where $\lim_{\gamma \to 0} \tg_0 = -1$, $\lim_{\gamma \to 0} \tg_{-1} = 1 $. 
Two remarks are in order before we consider the behavior of $\mean{a_2}$ near $\ttrevo$.  First, note that because of the factor $\exp(-i\varphi_S)$ in \eqref{eq:a2-phis}, there is no constant term in the Fourier decomposition of $a^\alpha_2$.  We therefore do not expect $\mean{a_2}$ to saturate to a constant when the classical oscillations are damped, it should simply vanish.  On the other hand, although the first term of Eq.~\eqref{eq:mean-a2t0-s4} behaves essentially in the same way as what was observed for $\mean{n_2}$, namely oscillations with the classical angular frequency $\omega_S$ and amplitude damped on the time scale $\tdec \simeq 2/ (\partial_S \omega_S \sqrt{\Sa \heff}) = O(\heff^{-1/2})$, the second term (proportional to $g_1$) behaves in a completely different way.

Indeed, from Eqs.~\eqref{eq:Omegas0}-\eqref{eq:Omegas1}, we see that for the unperturbed system $(\omega_S+\omega_m) = O(\heff)$, and in the perturbative regime
\begin{align}
    (\omega_S+\omega_m)  & = -\heff + 3\frac{\gamma^2}{\Sa} + O(\gamma^4)\\
    \partial_S(\omega_S+\omega_m)  & = - 2 \frac{\gamma^2}{\Sa^2} + O(\gamma^4)\; .
\end{align}
For $\gamma = 0.14$, as in Fig.~\ref{fig:hop}, both the classical frequency $(\omega_S+\omega_m)$ and the decay rate $( \Sa \heff/2)^{1/2}  \, \partial_S(\omega_S+\omega_m)$ are therefore quite small.  Classically, the smallness of $(\omega_S+\omega_m)$ is associated with the precession motion displayed on Fig.~\ref{fig:precession}.

For the parameters values of Fig.~\ref{fig:hop}, the $O(\gamma^2)$ ``perturbative smallness'' of $(\omega_S+\omega_m)$ and $\partial_S(\omega_S+\omega_m)$ is somewhat smaller, but on the same scale, as the ``semiclassical smallness'' of $\heff \sim 1/\Nav$. 
This results in the variations associated with the component $g_{-1} $ taking  place on a scale much longer than the one characteristic of $\mean{n_2}$ or of the component $g_0$ of $\mean{a_2}$, and appearing as a slow decay on a scale comparable to $\trevo$.  This explains the characteristic features observed on the panel (e) of Fig.~\ref{fig:hop}, which, as seen in Fig.~\ref{fig:coherence-t0}, is nicely reproduced by the semiclassical expression Eq.~\eqref{eq:mean-a2t0-s4} and by the quantum perturbative expressions Eqs.~\eqref{eq:cohamp}-\eqref{eq:A55}.

\begin{figure}
    \centering
     \includegraphics[width=\linewidth]{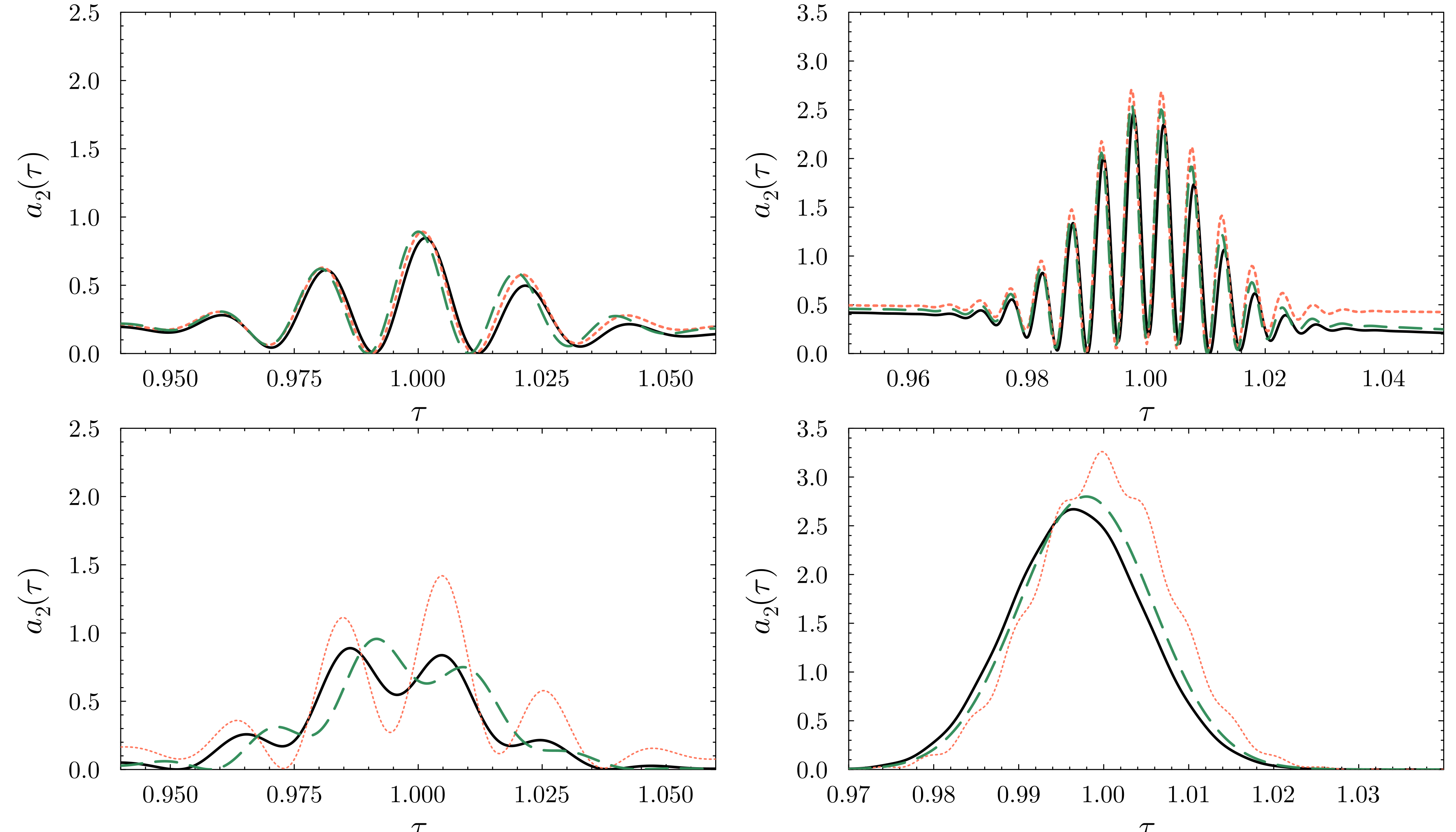} 
     \caption{Site two coherence $a_2$  as a function of the rescales time $\tau = t/\trevo$, near $t=\trevo$.  Solid black : exact quantum ; Dashed  green : semiclassical (EBK-embedded-TWA) Eq.~\eqref{eq:n2t0-f0f1}; Dotted red : quantum perturbation theory.  Left column : $\Nav = 50$, right column $\Nav=200$, upper row :$\gamma = 0.14$, lower row :$\gamma = 0.25$}
    \label{fig:coherence-trev}
\end{figure}

Going now to the neighborhood of $\ttrevo$, the time dependence of the site-2  coherence is given by
\begin{equation}
\begin{aligned}
    \mean{a_2}_{\rm emb}(\tt) & =   
        \sqrt{\frac{2 \gamma^2}{\heff \Sa}}
    \Bigg\{
    \tg_0 \exp[-i\omega_S \tt + i\pi  (1 + 2 \delta \la)] \exp \left[ - \frac{\heff}{2} \, \GBoot_{11} (\tt) \right] \\
& +
        \tg_{-1} \exp[-i(\omega_S + \omega_m) \tt + i\pi  (1 - 2 \delta \la)]  \exp \left[ - \frac{\heff}{2} \, \GBoot_{+} (\tt) \right]
    \Bigg\}
    \end{aligned}
    \qquad  ,
\end{equation}
with
\begin{subequations} \label{eq:G11+} 
 \begin{align} 
    \GBoot_{11}(\tt) :& = 
\Sa  (\partial_S \omega_S)^2 (\tt -  \ttrev^{SS})^2  + \hma (\partial_S \omega_m)^2 (\tt - \ttrev^{Sm})^2 \label{eq:G11} \\
\GBoot_{+}(\tt) :& = 
\Sa  \,(\partial_S (\omega_S + \omega_m))^2 (\tt -  \ttrev^{S+})^2  + \hma \, (\partial_S (\omega_S+\omega_m))^2  \, (\tt - \ttrev^{m+})^2 \label{eq:G+} 
\end{align}
\end{subequations}
and where 
\begin{align*}
    \ttrev^{SS} &= \frac{(\partial_S \omega_S)_0 }{(\partial_S \omega_S)} \ttrevo \; , & 
    \ttrev^{Sm} &= \ttrev^{mS} = \frac{(\partial_S \omega_m)_0 }{(\partial_S \omega_m)} \ttrevo   \; , \\
    \ttrev^{S+} &= \frac{(\partial_S (\omega_S+\omega_m))_0}{(\partial_S (\omega_S+\omega_m))} \ttrevo \; , &
    \ttrev^{m+} &= \frac{(\partial_m (\omega_S+\omega_m))_0 }{(\partial_m (\omega_S+\omega_m))} \ttrevo   \; .
\end{align*}
As long as $\hma = O(\gamma^2)$ can be neglected in front of $\Sa$, what we observe is the superposition of two phenomena: a shift of the revival time to $\ttrev^{SS}$ for the classical oscillation pattern, and the effect of the shift $\pi  (1 - 2 \delta \la)$ on the slow, secular like oscillation.   As seen in Fig.~\ref{fig:coherence-trev}, this expression very nicely reproduces the exact quantum result near $\trevo$.

\subsection{Autocorrelation function}
\label{sec:autofunc}

In Sec.~\ref{sec:autocor}, it is discussed that the semiclassical evaluation of the autocorrelation function is on one hand ``simpler'' than that of expectation values, but on the other hand somewhat more ``fragile''.

Indeed, computing the autocorrelation function does not require expressing the Weyl symbol of some operator in the perturbed action-angle basis, and once the initial Gaussian state is expressed in this basis, one has essentially to perform Gaussian integrals leading to 
\begin{equation} \label{eq:C(t)}
    \autocor(t) = \frac{1}{\sqrt{ \det(\MMinv \cdot  \AAA(t))}}
    \exp \left[
    \frac{1}{\heff }\left(\BB^T(t) \AAA^{-1}(t) \, \BB(t)- \CC(t) \right)
    \right]
\end{equation}
with $\MM = \begin{pmatrix} \Sa & 0 \\ 0 & \hma \end{pmatrix}$ (see Eq.~\eqref{eq:M}) the matrix describing the sheering of the initial coherent state in the action-angle coordinates, and the matrices $\AAA$, $\BB$, $\CC$,  given by Eqs.~\eqref{eq:auto-t0} near $t=0$ and by Eqs.~\eqref{eq:auto-trev} near $t=\trevo$. Assuming the matrix $\NN$, Eq.~\eqref{eq:NN}, is very close to the identity (which is essentially true for all the $\gamma$-range of interest),  $\CC$ takes the form
\begin{equation}
    \CC(t) = \Sa \GGS(t)^2 + \hma \GGm(t)^2 \; ,
\end{equation}
where $\GG(t) = (\GGS(t),\GGm(t))^T$  is the location of the center of the evolved coherent state at time $t$, given simply by $\GG_0(t) \simeq \left(  \cmod{\omega_S \, \tt}  ,    \cmod{\omega_m \, \tt}  \right)^T$ near $t=0$, and where the EBK-embedding near $\trevo$ just induces the phase shift $(\Delta_S,\Delta_m)$, Eq.~\eqref{eq:Deltas}, leading to 
\begin{equation} 
\label{eq:DeltaPhis}
\DGr(t)
= \begin{pmatrix}  \cmod{\Delta_S + \omega_S \, \tt} \\ \cmod{\Delta_m + \omega_m \, \tt}  \end{pmatrix} \; .
\end{equation}
Physically,  the Gaussian $\exp(-\CC(t)/\heff)$ just corresponds to what would be the overlap between the displaced ``frozen'' coherent state. 
The prefactor $1/\sqrt{ \det(\MMinv \cdot  \AAA(t))}$ expresses that this overlap has to be reduced because of the spreading of the evolved coherent state, and $\BB^T(t) \AAA^{-1}(t) \, \BB(t)$ on the other hand takes into account that the spreading may increase this overlap when the centers of the evolved coherent state 
do not match.

\begin{figure}
    \centering
     \includegraphics[width=\linewidth]{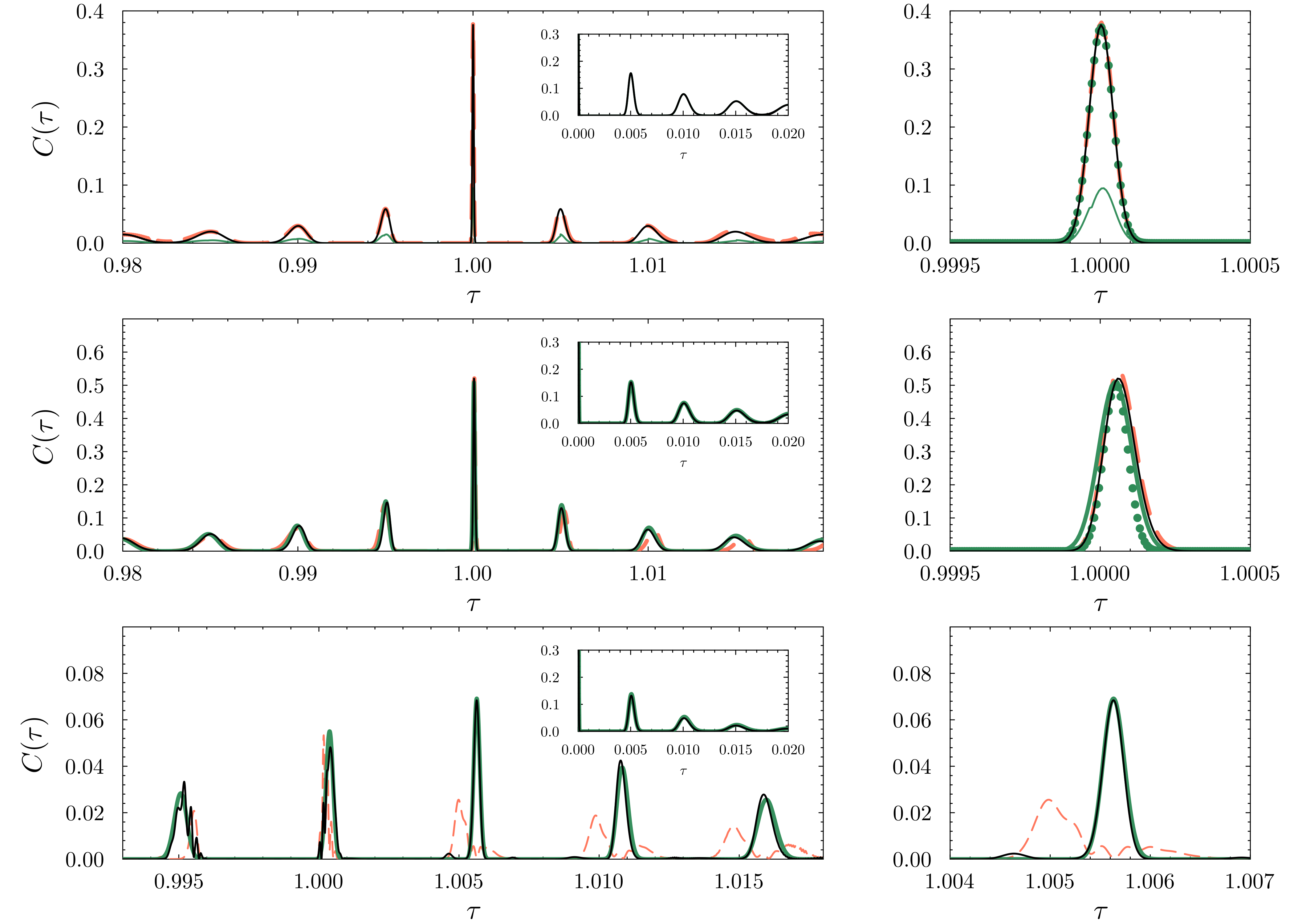} 
    \caption{Comparison between exact quantum (solid black),  semiclassical approximation (solid green) and quantum perturbation theory (dashed red)  for the autocorrelation function for $N=200$ and $\gamma = 0.07$ (top), $\gamma = 0.14$ (middle) and $\gamma = 0.21$ (bottom).  Note that $\gamma = 0.07$ corresponds to a dip and  $\gamma = 0.15$ and $0.21$ correspond to peaks in Fig.~\ref{fig:max}c.  Right column: blow up of the central peak, the green dotted line corresponds to the use of Eq.~\eqref{eq:Ctilda} instead of Eq.~\eqref{eq:C(t)} in the semiclassical expression.  Insets : autocorrelation near $t=0$.  Note the difference of scale between the different panels.}
    \label{fig:autocorr-trev}
\end{figure}

Figure~\ref{fig:autocorr-trev} shows a comparison between these semiclassical expressions, and both the exact and perturbative quantum results for the autocorrelation, for a total number of particles $N=200$ and three values of the coupling parameter, $\gamma = 0.07$ (corresponding to a dip of Fig.~\ref{fig:max}) and $\gamma = 0.14$ and $0.21$ (corresponding to peaks of Fig.~\ref{fig:max}).  It shows that the semiclassical expression always works very well near $t=0$ (see inset), and that it reproduces extremely precisely the exact quantum autocorrelation for the two latter values, but underestimates significantly the amplitude of the autocorrelation for the ``dip'' at $\gamma = 0.07$.    As illustrated in Sec.~\ref{sec:autocor}, this is a general feature that the semiclassical approximation near $\trevo$ is more fragile for the smaller $\gamma$ values corresponding to dips of Fig.~\ref{fig:max}.  The origin of this behavior is discussed in the next subsection.   We note already however that, as visible in the right column of Fig.~\ref{fig:autocorr-trev}, just substituting $\CC(t)$ by 
\begin{equation}
\label{eq:Ctilda}
    \tilde \CC(t) = \Sa \GGS(t)^2 + 2 \,\hma  (1 - \cos(\GGm(t)) \; ,
\end{equation}
which can be justified  for small enough $\gamma$ and sufficiently close to $\trevauto$, is enough to reproduce nicely the central peak of the autocorrelation.  The quantum perturbative result works as well for $\gamma$ values corresponding to dips or to peaks, but quickly become much less accurate as  $\gamma$ increases beyond $0.14$.

\subsection{Synchronization and resurgence}
\label{sec:syncres}

Figures~\ref{fig:number-trev},~\ref{fig:coherence-trev}, and~\ref{fig:autocorr-trev} demonstrate that for the quantities considered, the semiclassical approximations based on an EBK-embedded TWA approach are  extremely accurate once the total number of particles $N$ in the dimer is large enough, with the notable exception of the autocorrelation for small values of $\gamma$ corresponding to dips in Fig.~\ref{fig:max}c.   As these semiclassical results depict the observed quantum phenomenon around the revival time in terms of the properties of the classical Bose-Hubbard dimer, the necessary tools are now at hand to provide a qualitative understanding of the observed behaviors. In particular, it is now possible to explain the five ``puzzles'' [P1-P5] formulated at the beginning of the section, to which we add a sixth: [P6] why is the is the amplitude of the semiclassical approximation for the autocorrelation significantly smaller than the exact amplitude for small values of $\gamma$ corresponding to the dip in Fig.~\ref{fig:max}c,  despite it being so precise at the peaks of that figure?

The first puzzle of multiple revival times, [P1], is actually not too much of a surprise upon reflection.  Indeed, it is known that in the one-dimensional case $\trev$ is governed by the shearing rate $\omega'_\alpha$ (cf.~Eq.~\eqref{eq:ts}).  However, the direct analog of this $1$D shearing rate is the $2 \times 2 $ matrix $\partial \bOmega$ (cf.~Eq.~\eqref{eq:Omegas}), which being symmetric, has three independent entries with rather different behaviors since $\partial_S \omega_S$   has no first order dependence in $\gamma^2$, contrary to $\partial_m \omega_S = \partial_S \omega_m $ and $\partial_m \omega_m $ (cf.~Eq.~\eqref{eq:Omegas1}).  This is directly visible in the difference between expressions for the function $\GBoot_{22}$, Eq.~\eqref{eq:G22}, governing the behavior of $\mean{n_2}$ and the functions $\GBoot_{11}$ and $\GBoot_{+}$ governing $\mean{a_2}$.  More generally, for the expectation value of smooth operators, and as seen in Eq.~\eqref{eq:LambdaBoot}, the nature of the relevant Fourier component of the operator under consideration determines which entries of $\partial \bOmega$ plays the major role.

The second puzzle of an additional background feature in the coherence, [P2], is clearly much less expected.  However, the corresponding explanation has been given already in Sec.~\ref{sec:IV-a2}: for the second Fourier components of $a_2^\alpha(\varphi_S,\varphi_m)$, the relevant frequency is $\omega_+ = \omega_S + \omega_m$, and the relevant shearing rate is $\partial_S \omega_+$, both of which are zero in leading $\gamma^2$ order. The $\gamma^2$  smallness of $\omega_+$ results in the precessional motion displayed in Fig.~\ref{fig:precession}.  For the choice of parameters here, this  is comparable with the smallness associated with $\heff  \sim 1/\Nav$.  From the semiclassical analysis, we can therefore understand the very atypical behavior of $\mean{a_2}$ as originating from the fact that a classical time scale $t^+_{\rm osc} = 2\pi/\omega_+$ is, for the choice of parameters considered, similar to the quantum time scale associated with revivals $\trevo$. 

Understanding the remaining puzzles, [P3-P6], requires disentangling somewhat the semiclassical expression, Eq.~\eqref{eq:C(t)}, which being the result of a Gaussian integral is somewhat less transparent than its physical content.  The main physical mechanism here is indeed contained in Eq.~\eqref{eq:PhiBoot}, which expresses, after the EBK-embedding has been performed, how the initial coherent state has evolved near $\trevo$.  Decomposing Eq.~\eqref{eq:PhiBoot} into the two components, this reads 
\begin{subequations} 
\label{eq:Phi_emb}
    \begin{align} 
   \DPSe (\delta \bI,t) &=  \GGSe + (\partial_S \Omega_S \delta S +  \partial_m \Omega_S \delta m) \tt 
   - (\partial_S \Omega^0_S \delta S + \partial_m \Omega^0_S \delta m )\ttrevo \; , \\
   \DPme (\delta \bI,t) &=  \GGme + (\partial_S \Omega_m \delta S +  \partial_m \Omega_m \delta m) \tt 
   - (\partial_S \Omega^0_m \delta S + \partial_m \Omega^0_m \delta m )\ttrevo \; ,
\end{align}
\end{subequations}
where $(\GGSe,\GGme)^T$ given by Eq.~\eqref{eq:DeltaPhis} correspond to the center of the evolved coherent state and with the phase shifts $\Delta_{S,m}$ given by Eq.~\eqref{eq:Deltas}. What Eq.~\eqref{eq:C(t)} quantitatively expresses is that to have a significant autocorrelation, one needs to have the center of the evolved coherent state $(\GGSe (t) ,\GGme (t))$ near its starting point $(0,0)$ while its spreading due to the extension in action space remains reasonably small.  This is particularly visible in the 1D analog,  
Eq.~\eqref{eq:autocor_trev_1d}, of the autocorrelation function, Eq.~\eqref{eq:C(t)}, where the rapid oscillations (in the $1$D version) associated with $     \Delta \Phi_{\rm rec } (t) =   \cmod{\Delta + \Omega \tt}$ and the slower variation associated with the spreading of the coherent state, $\Delta R_{\rm rec } (t)= [\partial \Omega \tt -\partial \Omega^0 \ttrevo ]$, are particularly transparent. 

\subsubsection{Analysis of the perturbative regime}

To gain a clearer understanding of the physical mechanisms behind the remaining puzzles [P3-P6], this subsection begins by discussing the perturbative regime, in which these mechanisms are rather transparent. We then move on to the next subsection to explore what can be said beyond this regime.

\begin{figure}
    \centering
     \includegraphics[width=0.8\linewidth]{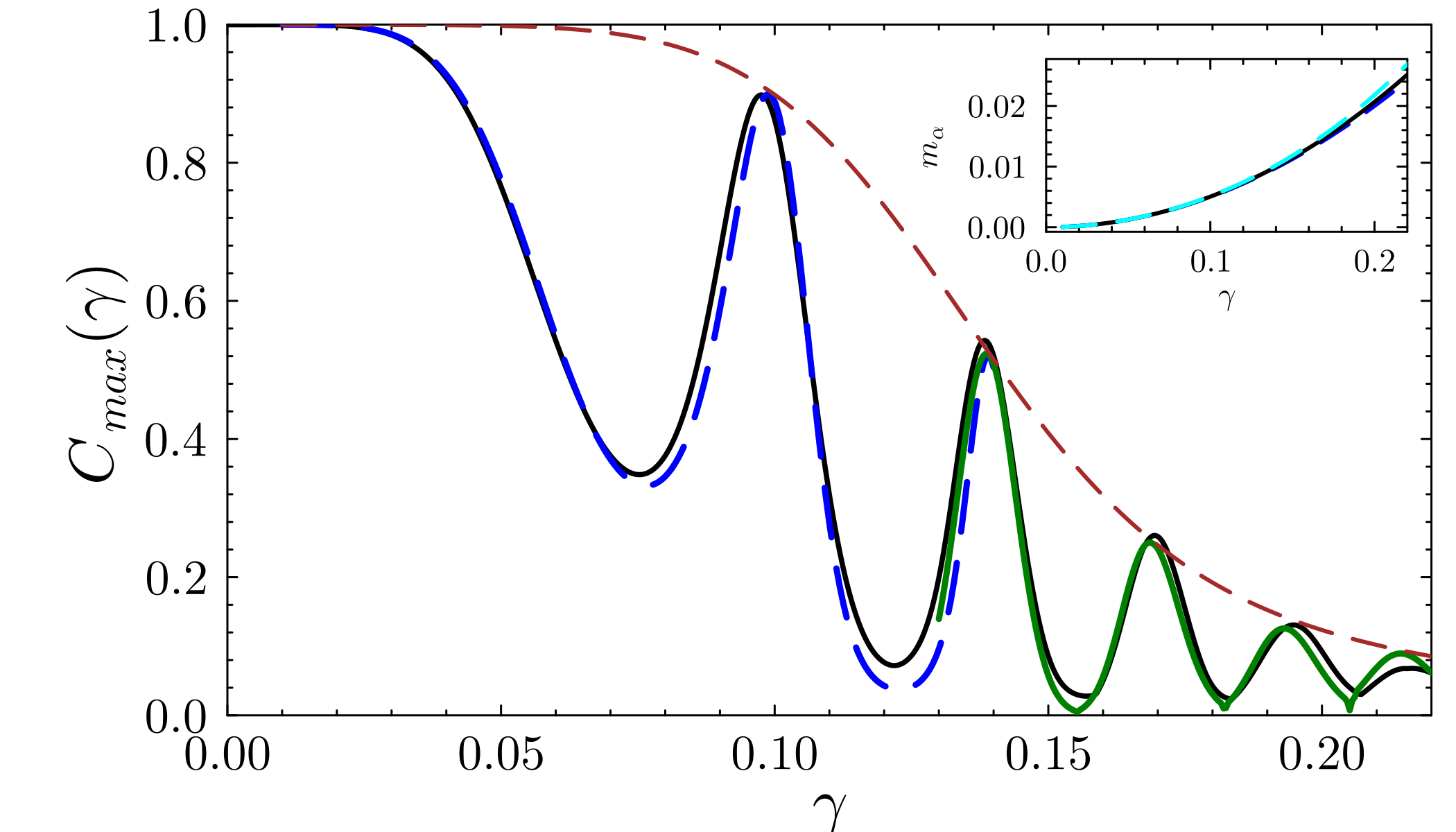} 
    \caption{Maximum of the autocorrelation function $\autocor(t)$ for $\Nav=200$ as a function of $\gamma$:  solid black: exact quantum; solid green:  semiclassical expression Eq.~\eqref{eq:C(t)},  and dash blue: perturbative approximation of this expression  Eq.~\eqref{eq:C(trevo)}.   The dash brown curve is the value of the prefacor ${1/\sqrt{\det{(\MMinv \cdot \AAb(\trevauto))}}}$.
    Inset: three versions of $\ma(\gamma)$.  Solid black: exact action of the north-pole trajectory, Eq.~\eqref{eq:ma}; dash blue: leading-order approximation $\ma^0$, Eq.~\eqref{eq:ma0}; and dash cyan: $\hma$ given by Eq.~\eqref{eq:hma}.  The three curves are  indistinguishable until $\gamma \simeq 1.2$, and remain very close in the entire $\gamma$-domain of interest.}
    \label{fig:autocorr-max}
\end{figure}

In the perturbative regime, and for times not too far from $\trevauto$,  it can be shown that the term $\BBb^T(t) \AAb^{-1}(t) \, \BBb(t)$ in the exponent of Eq.~\eqref{eq:C(t)} is negligible compared with $\CCb(t)$, in which case 
\begin{equation} \label{eq:Cpert}
    \autocor(t) \simeq  \frac{1}{\sqrt{ \det(\MMinv \cdot  \AAb(t))}}
    \exp \left[-   \frac{1}{\heff }\left(\Sa [\GGSe(t) ]^2 + \hma [\GGme(t) ]^2 \right)
    \right] \; .
\end{equation}
Remarkably, in this expression it happens that whereas $\Sa = 2 = O(1)$, the parameter $\hma \to 0$ as $\gamma \to 0$, because in this limit the north-pole trajectory reduces to a fixed point, and $ \hma = O(\gamma^2)$ can be extremely tiny.    As a consequence, $\Sa/\heff = \Nav \gg 1$ is very much in the semiclassical regime, even though $(\hma/\heff )$ can remain smaller or not too much larger than unity throughout the whole perturbative regime, even for $\Nav \simeq 200$.  The semiclassical expressions would be expected to be less well founded there.  There is, nevertheless, a saving feature, which is that, as discussed in Appendix~\ref{app:top-dyn}, the dynamics near the north pole is essentially harmonic-oscillator like for small $\gamma$, and the harmonic oscillator, being a linear system, renders semiclassical dynamics exact, independent of the existence of a large parameter (e.g.~small $\hbar$).  Expectation values of operators such as $\mean{n_2}$ or $\mean{a_2}$ remain therefore very precise even in the deep perturbative regime.  However, as seen in Fig.~\ref{fig:autocorr-trev}, when $\GGme \simeq \pi$  the autocorrelation function becomes sensitive to the smallness of the action.   As is discussed in greater detail in Sec.~\ref{sec:autocor}, what happens then is that the Gaussian integral leading to Eq.~\eqref{eq:C(t)}, which in the deep semiclassical regime is independent of the canonical coordinates used to perform the integral, becomes sensitive to this if $\ma/\heff \lesssim 1$.   Computing these Gaussian integrals in the action angle variables $(m,\varphi_m)$ leads to Eq.~\eqref{eq:Cpert}, but doing it in the corresponding quadrature coordinates leads to Eq.~\eqref{eq:Ctilda}, which as shown in Fig.~\ref{fig:autocorr-trev} perfectly reproduces the central peak of the autocorrelation function.  Clearly, the two expressions are equivalent in the deep semiclassical regime, since if $\heff \to 0$ only the neighborhood of $\GGme = 0$ gives any significant contribution.  However, as illustrated in Fig.~\ref{fig:autocorr-max}, using Eq.~\eqref{eq:Cpert}, with the substitution $(\GGme)^2 \to 2 (1 - \cos(\GGme))$ reproduces very nicely the curve $\autocormax(\gamma)$ up to a range of $\gamma$ where the original expression, Eq.~\eqref{eq:C(t)}, can be used.  This in some sense solves the last puzzle [P6].

As seen in Fig.~\ref{fig:autocorr-max}, the prefactor, $1/\sqrt{ \det(\MMinv \cdot  \AAA(\trevauto))}$, is a smooth monotonous decreasing function of $\gamma$.  The origin of the reentrant behavior of the maximum of the autocorrelation near $\trevo$ displayed in Fig.~\ref{fig:max} is thus rather straightforward: when the angles $\GGSe (t)$ and $\GGme(t)$ are in phase, both of them can be equal to zero at the same time, giving a large autocorrelation value corresponding to the peaks in Fig.~\ref{fig:max}.  Conversely, dips in that figure correspond to configurations where  these two angles are out of phase (which, of course, cannot occur near $t=0$).
As a consequence, as clearly visible from Eq.~\eqref{eq:Cpert} the condition that $\GGSe (t) \simeq 0$ is much more impactful than the condition $\GGme (t) \simeq 0$.  Furthermore,  the term in $\delta m$ can be neglected in Eqs.~\eqref{eq:Phi_emb}, or in other words  the coherent state spreads much more rapidly in the $(S,\varphi_S)$ coordinates than in $(m,\varphi_m)$.

A second idiosyncracy of the Bose-Hubbard dimer is that in the perturbative $\gamma \to 0$ limit, both $\GGSe (\trevo)$ and $\partial_S \omega^\alpha_S$ have only a very weak $O(\gamma^4)$ dependence in the coupling $\gamma$.  This is directly visible in Eq.~\eqref{eq:Omegas1} for $\partial_S \omega^\alpha_S$, and is true for $\GGSe (t)$ for a somewhat more subtle reason discussed in Appendix~\ref{app:DeltaPhiS}, but which eventually boils down to the fact that for all $\gamma$'s and all values of $S$, the fixed point energy is given by $\tilde H(S, m \smeq 0) = S(S-2\heff) + \gamma^2$, so that $\omega_S (S,m \smeq 0) \equiv \partial_S H(S, m  \smeq 0) $ has no dependence on $\gamma$.  As a consequence, since the location  of the maximum of the autocorrelation is mainly fixed by the reformation and recurrence of the coherent state in the $(S,\varphi_S)$ sector, this explains why the maximum of the autocorrelation function remains relatively unchanged at $\trevo$ in the perturbative regime.   With these remarks in mind, it is possible to derive a rather transparent expression for the value of the maximum of the autocorrelation function (the highest value of $\autocor(t)$ in Fig.~\ref{fig:autocorr-trev}).  To leading $\gamma^2$ order at $t = \trevo$, Eq.~\eqref{eq:C(t)} reads 
\begin{equation}  \label{eq:C(trevo)}
\begin{aligned}
     \autocor(\trevo) \simeq & \frac{1}{\sqrt{\mathcal{D}}}     \exp \left[ -\hma (\GGme(\trevo))^2\right] \; , \\
     \mathcal{D} = & \det{(\MMinv \cdot \AAb(\trevo))} \simeq 1 + 2 \Sa \hma (\gamma \,\ttrevo (\partial \bOmega^1)_{S,m} )^2 \; .
\end{aligned}
\end{equation}
Comparing this expression with Eq.~\eqref{eq:n2trev-f0f1} sheds light on  the explanation of the puzzle [P4].  Indeed, we recognize in that equation that the argument of the cosine is simply the phase $\GGme(t)$ that occurs  in  the exponent of Eq.~\eqref{eq:C(trevo)}. Thus, if $\GGme (\trevauto \! \simeq \! \trevo)= 0$, the cosine is maximal, and $\Boot{n_2}(t)$ shows a local minimum (since $\tf_1  < 0$), whereas $\autocormax(\gamma)$ is maximal. Conversely, when $\GGme = \pi$, $\autocormax(\gamma)$ is minimal and $\Boot{n_2}(t)$ shows a local maximum.  That the same phase $\GGme$ governs both the oscillations of $\Boot{n_2}$ and the height of the maxima of $\autocor$ is furthermore expected.  Indeed, $\GGme(t)$  characterizes the motion of the center of the Wigner transform of the evolved coherent state on the north-pole trajectory (shown for instance as the red curve in the top row of Fig.~\ref{fig:level_curves}).  The phase $\varphi_m = 0$ corresponds to the north pole, i.e.\ to the top of the $(s,\theta)$ plane where the imbalance $(n_1 - n_2) = 2s/\heff$ is maximal, and $\varphi_m = \pi$ corresponds to the lowest point on that curve and thus to a minimum of the imbalance (and thus a maximum of $n_2$) on that curve.  As discussed in~\cite{Schlagheck22}, it is thus natural to expect that a peak of $\autocormax(\gamma)$, associated with a north-pole trajectory being back to the north pole, would correspond to a local minimum of $\Boot{n_2}(t)$, and that conversely, a dip of  $\autocormax$ would correspond to a local maximum of $\Boot{n_2}(t)$.

At this point, the only mystery remaining before moving beyond the perturbative regime is the puzzle [P5], which relates to the correlation between $\barn$, the {\em average value} of $\mean{n_2}$ away from a revival, and the presence of a peak or a dip of $\autocormax(\gamma)$.  From the above discussion, we expect this to be related to the connection between the value of $\barn$ and the phase $\GGme(\trevo)$.  
This connection is, on the one hand, based on the generic
relationship between the phase shifts $(\Delta_S,\Delta_m)$ being $0$ or $\pi$ and the action $\ma$ fulfilling the EBK quantization.   More specifically in the case $\Delta_m$,  it turns out  $\Delta_m = -4\pi \la$ (see Eq.~\eqref{eq:Deltas}) with $\la = (\ma/\heff - 1/2)$, which has to be an integer for the torus $\ma$ to quantize.  Dropping the $-1/2$ which becomes irrelevant once multiplied by $-4\pi$ gives 
\[ \Delta_m = \cmod{4\pi\ma/\heff } \; . \]
Therefore, there is a direct link between, for instance, having an integer or half-integer $\la$ and having $\Delta_m = 0$.   

The second ingredient needed to resolve the puzzle [P5] is the link between this latter property and the value of $\barn$.
This particular link, and the fact that the term $\omega^\alpha_m \ttrevo$ of $\GGme$ does not come into play here, can be related to a particular feature of the Bose-Hubbard dimer.
To see this, let us first note that $\ma(\gamma) = \ma^0 + O(\gamma^4)$, with 
\begin{equation} 
\label{eq:ma0}
    \ma^0 \equiv \frac{\gamma^2}{\Sa}
\end{equation} 
the leading order approximation of the north-pole trajectory action.  In the perturbative regime under consideration,  $\ma \simeq \hma \simeq \ma^0$, and it is not necessary to distinguish between these three quantities (see inset of Fig.~\ref{fig:autocorr-max}). Then, neglecting the $\heff$ correction, note that the left equalities of Eqs.~\eqref{eq:Omegas}-\eqref{eq:Omegas1} read:
\begin{equation}
\bOmega(\gamma) =  
    \Sa \begin{pmatrix}  + 1 \\ -1    \end{pmatrix} 
    + \ma^0 \begin{pmatrix} -1 \\ +4 
    \end{pmatrix} \; .
\end{equation}
Remarkably, apart from the action scales, $(\Sa,\ma)$, the matrix elements are integers, which is clearly an idiosyncrasy of the Bose-Hubbard dimer.  As a consequence $\omega^\alpha_m \ttrevo = (-\Sa + 4 \ma^0) (2\pi/\heff) = 2\pi(- (\Nav+1) + 4 (\ma^0/\heff))$.  Dropping the first term, which is a multiple of $2\pi$, gives 
\begin{equation}
\GGme \simeq  \cmod{\Delta_m + \omega^\alpha_m \ttrevo} \simeq \cmod{4\pi \ma} \; .
\end{equation}
Thus,  $\GGme = 0$ (maximum of $\autocormax(\gamma)$ or local minimal of $\mean{n_2}$)  corresponds to $\la$ integer of half-integer, and $\GGme = \pi$ (minimum of $\autocormax(\gamma)$ or local minimum of $\mean{n_2}$)  corresponds to $\la$ quarter-integer.

To connect to $\barn$, note that up to $O(\gamma^4)$ corrections,  Eq.~\eqref{eq:n2WignerPert} reads  $n^\alpha_2(\varphi_m) = 2 \frac{\ma^0}{\heff }(1 - \cos \varphi_m)$,  and thus 
\begin{equation}
    \barn \simeq  2 \frac{\ma^0}{\heff }  = 2\la + 1\; .
\end{equation}
Thus $\GGme = 0$  corresponds to $\barn$ integer , and $\GGme = \pi$  corresponds to $\barn$ half-integer, which resolves the last puzzle.

\subsubsection{Beyond the perturbative regime}

Beyond the perturbative regime, the status of the five puzzles [P1-P5] is, of course, not the same.  The first two, [P1-P2], are extremely generic and rely neither on the perturbative approximation nor on any peculiarity
of the Bose-Hubbard model.  Because a torus in a two-dimensional integrable system is characterized by three independent shearing rates, one expects the exact ``revival time''' to be  observable-dependent; and because a strong autocorrelation requires both angles $\GGSe$ and $\GGme$ to be zero, successive reduction and reentrance of $\autocormax(\gamma)$ is generically expected.  The only quirk of the perturbative regime in this respect is the observation of any significant autocorrelation at all when these two angles are out of phase.

On the other hand, the puzzles [P3-P5] are very much system specific and a consequence of the perturbative regime.  In particular [P4] and [P5] are exact statements in leading $\gamma^2$ order, and become more approximate and qualitative beyond that.  However, bear in mind that a significant revival usually requires that the various revival times associated with the different entries of $\partial \bOmega$ should not differ too much on a scale $\sqrt{\heff}$.  In a generic system, this can be achieved by tuning external parameters of the Hamiltonian.  For the Bose-Hubbard system, this is obtained whenever $\gamma$ is not too far away from the perturbative regime, implying that for all the $\gamma$-domain of interest, [P3-P5] remains at least qualitatively true.

\section{The classical Bose-Hubbard dimer}
\label{sec:classical-dimer}

In the previous section, we provided an overview of the physical mechanisms at play in the observed revival phenomena. Now, we turn to a more complete discussion of the various ingredients necessary for the semiclassical description of revivals. This section begins with a study of the classical Bose-Hubbard dimer and continues in the following sections with details of the semiclassical treatment.

\subsection{Semiclassical scaling}
\label{sec:semiscal}

Following the general philosophy of  Sec.~\ref{sec:monomer}, the first step of our semiclassical analysis is to construct the classical analog of the quantum Hamiltonian, Eq.~\eqref{eq:cs.BHd}. The logic for the monomer terms is evidently unchanged, and dealing with the hopping term   $ -  J \left( \hat{a}_1^\dagger \hat{a}_2 + \hat{a}_2^\dagger \hat{a}_1 \right)$ is much simpler than with the cubic perturbation of Sec.~\ref{sec:monomer}, since no ordering problems arise.

Introducing $(s_j,\theta_j)$ the action-angle coordinates for the simple harmonic oscillator, namely $p_j = \sqrt{2 s_j} \sin \theta_j$ and $q_j = \sqrt{2 s_j} \cos \theta_j$, the classical symbol for the hopping in the dimer Hamiltonian Eq.~\eqref{eq:cs.BHd} can therefore be obtained directly through the substitution
\begin{equation} \label{eq:to-classic}
    \hat a_j  \rightarrow \sqrt{ \frac{s_j}{\heff}} e^{- i\theta_j} \qquad  \hat a^\dagger_j  \rightarrow \sqrt{ \frac{s_j}{\heff}} e^{ + i\theta_j} \; ,
\end{equation}
with $\heff$ defined by Eq.~\eqref{eq:heff}. In these coordinates, the classical Hamiltonian associated with the Bose-Hubbard dimer in Eq.~\eqref{eq:cs.BHd}  reads 
\begin{equation}  \label{eq:ham2}
    H(s_1,s_2,\theta_1,\theta_2) = \frac{U}{2\heff^2} \left[ s_1(s_1-2\heff) + s_2(s_2-2\heff) \right]  -  \frac{2 J}{\heff} \sqrt{s_1 s_2} \cos(\theta_1 - \theta_2) \quad .
\end{equation}

As for the monomer, a proper definition of the mean-field/semiclassical limits requires to perform the scaling
    \begin{equation} 
    \label{eq:rescaling}
        \tH := \left( \frac{\heff^2}{U} \right) H \; ,  
        \qquad \gamma := \frac{J \heff }{U},  \qquad {\rm and}
        \qquad \tt := \left( \frac{U}{\heff \hbar} \right) t \; 
     \quad .
    \end{equation}
In this way the scaled Hamiltonian is
\begin{equation}  \label{eq:ham2scaled}
    \tH(s_1,s_2,\theta_1,\theta_2) = \left[ s_1(s_1-2\heff) + s_2(s_2-2\heff) \right]  -  2 \gamma  \sqrt{s_1 s_2} \cos(\theta_1 - \theta_2) \quad ,
\end{equation}
which, beyond the $\heff$ sub-leading corrections,  has a well defined classical limit as $\heff \to 0$. 
The form Eqs.~\eqref{eq:ham2}-\eqref{eq:ham2scaled} for the classical Hamiltonian is convenient as it disentangles completely the classical properties, which are fixed by the parameter $\gamma$, and the phenomena associated with the variation of $\heff$.  In the rest of this section and the following one we keep the tilde to indicate scaled variables for the energies and time, but drop it for other quantities such as the angular frequencies $\omega$ and their derivatives as they are defined only for the scaled Hamiltonian, and thus no confusion is possible.

Starting from Eq.~\eqref{eq:ham2scaled}, there are two coordinate systems of interest. The first one, the ``relative coordinates'' is akin to a Mercator/cylindrical projection of the dynamics, and is useful to compute some quantities (action, frequencies and their derivatives, etc..) needed as input for the semiclassical expressions.  The second one, which is a kind of spherical representation on the other hand, provides a clearer picture of the dynamics.

\subsection{Dynamics in the relative-coordinates Mercator plane}
\label{sec:sth-dyn}

Introducing the relative coordinates given by the canonical transformation 
\begin{equation} \label{eq:relcoord}
    (s_1,s_2,\theta_1,\theta_2) \longmapsto ( S,s,\Theta,\theta) = \left(  s_1 + s_2, \frac{s_1 - s_2}{2} , \frac{\theta_1 + \theta_2}{2},  \theta_1 - \theta_2  \right) \quad , 
\end{equation}
the Hamiltonian \eqref{eq:ham2} now becomes
\begin{equation} \label{eq:Htilde}
   \tH (S,s,\Theta,\theta) =  H_S + H_s \; .
\end{equation}
with
\begin{equation} 
    H_S =  \frac{ S(S-4\heff)}{4}  \qquad H_s = s^2 -  2 \gamma \sqrt{(S/2)^2 - s^2} \cos \theta 
    \quad . \label{eq:H0Hs}
\end{equation} 
In Eqs.~\eqref{eq:Htilde}-\eqref{eq:H0Hs}, the fact that $\tH$ does not depend on $\Theta$ expresses the conservation of the total number of particles $[\ntot]^W = S/\heff-1$.  The system is thus integrable, and the nontrivial part of the dynamics is the one of the variables $(s,\theta)$ governed by $H_s$.  Note, though, that the dynamics of the variable $\Theta$ do affect the quantities considered in the next section.

\begin{figure}
    \centering
    \includegraphics[width=\linewidth]{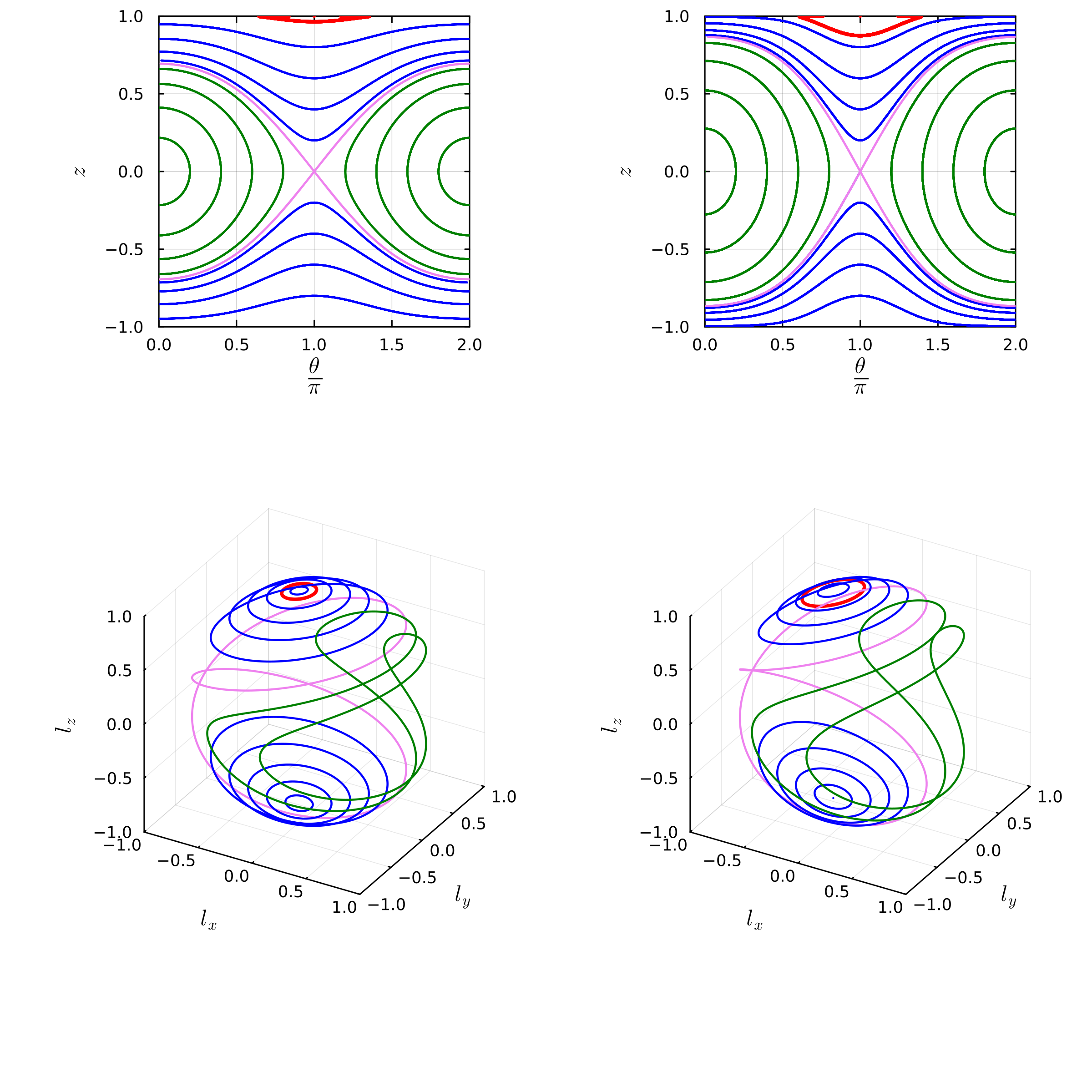}
    \caption{Phase portrait of the classical Bose-Hubbard dimer. Top row : Mercator representation $(\theta,z)$ (cf Eqs.~\eqref{eq:Hs}-\eqref{eq:zthdyn}). Bottom row: spherical representation $(l_x,l_y,l_z)$ (cf Eqs~.\eqref{eq:spherical}-\eqref{eq:toupie}).
    Left column : $\gamma = 0.14$.  Right column : $\gamma = 0.25$ ($S = \Sa = 2$ in both cases).  The red curve corresponds to the north-pole trajectory.}
    \label{fig:level_curves}
\end{figure}

For a fixed value of the constant of motion $S$, it is useful to note one further scaling property of $H_s$, namely that 
\begin{align}
H_s(s,\theta) & = \frac{S^2}{4} h_z(z,\theta)  \qquad z= \frac{2s}{S} \in[-1,1] \label{eq:Hs}\\
h_z(z,\theta) &=  {z^2} -  4 {\gs } \sqrt{1 - z^2}\cos \theta  \; , \label{eq:hz}  \\
\mbox{with} \qquad     \gs & :=  \frac{\gamma }{S}  \; . \label{eq:gammas}
\end{align}
Note, though, that $(z,\theta)$ is not a canonical system of coordinates, so one should always have in mind the original  $(s,\theta)$ variables when computing actions.  The  equations of motion in  the $(z,\theta)$  variables read
\begin{equation} \label{eq:zthdyn}
    \left\{
    \begin{aligned}
     \dot \theta & =   \{\theta,H_s\} =   + \frac{S}{2} \frac{\partial h_z}{\partial z} =  S z \left(1 + 2 \gs \frac{\cos \theta}{\sqrt{1-z^2}} \right) \\
        \dot z & =   \{ z,H_s\} =   - \frac{S}{2} \frac{\partial h_z}{\partial \theta} = -  2 S \gs \sqrt{1-z^2} \sin \theta 
    \end{aligned}
    \right. \; .
\end{equation}
Eqs.~\eqref{eq:Hs}-\eqref{eq:zthdyn} however show that up to a scaling factor, the dynamics at fixed $S$ is entirely fixed by $\gs$, but that this parameter depends on $S$.   In Fig.~\ref{fig:level_curves} the level curves of $h_z$, Eq.~\eqref{eq:hz}, are plotted for different values of the perturbation parameter $\gs$.

The initial state, Eq.~\eqref{eq:cs.BEC},  $|\boldsymbol{\zee}\rangle = |{\sqrt{\Nav}, 0}\rangle$  (with $\Nav  = \Sa/\heff $) is such that all particles are located in the first well.  The corresponding Wigner transform 
is thus centered at  $(S \smeq s_1 + s_2 \smeq \Sa, s \smeq \frac12(s_1-s_2) \smeq \Sa/2)$, which corresponds to the boundary $z=1$ of the $(z,\theta)$ Poincaré section.  In the semiclassical regime, the interest is therefore mainly in what is going on near the upper boundary of the $(z,\theta)$ phase space.

More specifically, note that $h_z(z \smeq 1,\theta) = 1$, for all $\theta$ and independently of $\gs$.  The classical trajectory at the center of the coherent state therefore evolves on the curve $h_z(z,\theta) = 1$, which appears as the union of the upper boundary of the phase space (the line $z=1$) and the curve $\Cstar$ (in red on Fig.~\ref{fig:level_curves}) separating the rotation-like motion with ``energy'' $h_z > 1$  around the fixed point $(z^*,\theta^*) = (\sqrt{1- 4 \gs^2},0)$  ($h_z^{\rm max} := h_z(z^*,\theta^*) = 1+ 4\gs^2$), and the libration-like motion further from the fixed point with $h_z < 1$.

We stress that this representation is slightly misleading as the full $z=1$ line corresponds to a single point in the original $(s_1,s_2,\theta_1,\theta_2)$ variables (in the sense that $s_2=0$ and therefore $\theta_2$ is undefined).  Below a spherical representation of the motion is introduced for which the curve $\Cstar$ has exactly the same topology as the neighboring $h_z(z,\theta) = {\rm const.}$ curves (see Sec.~\ref{sec:spherical-coordinates}).  $\Cstar$ is thus not properly speaking a separatrix, and we rather label it as the ``north-pole trajectory''. 

\subsection{Classical input for semiclassical expressions}
\label{eq:input}

From the discussion in Sec.~\ref{sec:EBKboot}, it is expected that the ``classical information'' that is needed for our semiclassical analysis is the action of the curve $\Ca = \Cstar(S \smeq\Sa)$, as well as the angular frequencies and their derivatives on that curve.
The equation for $\Ca$ derives simply from Eq.~\eqref{eq:hz} noting that on the north-pole trajectory, $h_z(z_{\Ca}(\theta),\theta) = 1$.   This implies $\cos \theta < 0$ (eg  $\theta \in [\pi/2, 3\pi/2]$), and  
\begin{equation} \label{eq:zstar}
    z_{\Ca}(\theta) = \sqrt{1 - (4 \gamma/\Sa)^2 \cos^2\theta}   \qquad \theta \in [\frac{\pi}{2},\frac{3\pi}{2} ] \; .
\end{equation}
The dynamics on $\Ca$ is then obtained directly from  Eqs.~\eqref{eq:zthdyn}, which given that $h_z(z_{\Ca}(\theta),\theta) = 1$ implies
\[
 \frac{2\gamma}{\Sa}\frac{\cos \theta}{\sqrt{1 - z_{\Ca}^2(\theta)}} = -  \frac12
\]%
gives
\begin{equation} \label{eq:thpt}
    \frac{d\theta}{d t} = \frac{\Sa}{2}z_{\Ca}(\theta) = \frac{\Sa}{2} \sqrt{1 - (4\gamma/\Sa)^2 \cos^2 \theta}  \; .
\end{equation}
(Note that with the definition of $z$, this means that on the north-pole trajectory, $\dot \theta = s$.)
Introducing the incomplete elliptic integral  of the first kind $\mathbb{F}(\phi;m) = \int_0^\phi {d\theta}/{\sqrt{1 - m \sin^2 \theta}}$, and $\mathbb{K}(m) = \int_0^{\pi/2} {d\theta}/{\sqrt{1 - m \sin^2 \theta} }$, its complete version, Eq.~\eqref{eq:thpt},  can be integrated as $t = ({2}/{\Sa}) \mathbb{F}[(\theta - \pi/2 ) ; (4\gamma/\Sa)^2 ]$
and the period of the motion is 
 \begin{equation} \label{eq:Tstar}
      T_m^\alpha = \frac{4}{\Sa} \mathbb{K}[ (4\gamma/\Sa)^2 ] \; . 
 \end{equation}
 Defining $\varphi_m =  \omega^\alpha_m t = - 2\pi t/ T_m^\alpha$, the angular coordinate on the north-pole trajectory (recall that $\omega^\alpha_m < 0$) is
\begin{equation} \label{eq:varphi}
    \varphi_m = - \pi \frac{\mathbb{F}[(\theta - \pi/2 ) ; (4\gamma/\Sa)^2 ] }{ \mathbb{K}[ (4\gamma/\Sa)^2 ] } \; .
\end{equation}
Finally, this function can be inverted in terms of the Jacobi amplitude $\mathbb{A}$, which is the inverse of the Jacobi incomplete elliptic integral, $\mathbb{F}$ .  One last technicality is that the angle $\theta$ is constrained to remain in the rangle $[\pi/2,3\pi/2]$, which implies that Eq.~\eqref{eq:varphi} is valid only for $t \in [0,T^\alpha_m]$ (i.e. $\varphi_m \in [-2\pi,0]$).   To extend the relation beyond this range, the fact that at each return to the north pole the angle $\theta$ picks up a $-\pi$ discontinuity must be taken into account; see the discussion in appendix~\ref{app:DynSeparatrix}.  This gives

\begin{align} 
      \ttheta(\varphi_m) &  = \tttheta(\varphi_m) \pmod{\pi} \; + \; \frac{\pi}{2} \qquad   \in [\pi/2, 3\pi/2] \label{eq:tilde-theta} \\
     \tttheta(\varphi_m) &  = \mathbb{A} \left[ -\left( \frac{ \varphi_m}{\pi} \right) \mathbb{K}\left(\frac{4 \gamma^2}{\Sa}\right) , \frac{4\gamma^2}{\Sa} \right] \label{eq:tilde-tilde-theta}
    \quad . 
\end{align}
where $\tttheta(\varphi_m)$ is a continuous function such that $\tttheta(\varphi_m \smeq 0) = 0$, and $\ttheta(\varphi_m)$ its discontinuous version mapped on $[\pi/2,3\pi/2]$.

An invariant torus of the perturbed system can be characterized by a pair of actions $(S,m)$ where the second action variable $m$ is chosen as the symplectic area in the $(s,\theta)$ plane {\em above} the invariant curve $\mathcal{C}$ corresponding to the torus under consideration
\begin{equation} \label{eq:action-m}
    \frac{S}{4\pi} \oint_\mathcal{C} (1 - z_{\mathcal{C}}(\theta)) d\theta \; ,
\end{equation}
so that in the uncoupled limit one has $\lim_{\gamma \to 0}m = s_2$ (we furthermore gain in this way that, with $S=\Sa$, $\lim_{h_z \to 1+} m = \lim_{h_z \to 1-} m = \ma $).   For the north-pole trajectory,
\begin{equation} \label{eq:ma}    
\ma = \frac{\Sa}{4\pi} \int_{\frac{\pi}{2}}^{\frac{3\pi}{2}} (1-z_{\Ca}(\theta)) d\theta  = \frac{\Sa}{4\pi} \left[ \pi - 2 \mathbb{E}((4\gamma/\Sa)^2) \right] 
=  \Sa {(\gamma/\Sa)}^2 + \mathcal{O}((\gamma/\Sa)^4)
\quad ,
\end{equation}
with $\mathbb{E}(m) = \int_0^{\pi/2} {d\theta}{\sqrt{1 - m \sin^2 \theta} }$ the complete elliptic integral of the second kind.  As shown in appendix~\ref{app:omegas}, the angular frequencies and their derivatives can be obtained analytically in essentially the same way.  On the level curve $\Ca$, noting $\zeta = (4 \gamma/\Sa)^2$ and, for second order derivatives,  neglecting the irrelevant $\heff$ corrections, one obtains
\begin{subequations} \label{eq:dxm}
\begin{align} 
    \dfrac{\partial m}{\partial E} \bigg\vert_{E_*,\Sa} 
            &= -\dfrac{2\mathbb{K}[\zeta]}{\pi \Sa} 
             \label{eq:dmE}    \\
    \dfrac{\partial m}{\partial S} \bigg\vert_{E_*,\Sa} &= \frac14 + 
        \dfrac{(6-8\heffS) \mathbb{K}\left[\zeta \right]}{4\pi} 
        \label{eq:dmS}\\
        \dfrac{\partial^2 m}{\partial E^2} \bigg\vert_{E_*,\Sa} 
            &= \frac{1}{2 \pi  \Sa \gamma ^2} \left[\frac{ \mathbb{E} 
            \left[ \zeta \right]}{1-\zeta} - \mathbb{K}\left[\zeta \right] \right]
        \label{eq:d2m2}\\
        \frac{\partial^2 m}{\partial S^2} \bigg\vert_{E_*,\Sa} 
            & = \frac{\Sa}{2 \pi  \gamma ^2}
             \left(\frac{1-\frac{7}{16} \zeta}{1-\zeta} \mathbb{E}[\zeta]
            - \left(1-\frac{3}{16} \zeta \right) \mathbb{K}[\zeta]\right)
        \label{eq:d2S2}\\
        \frac{\partial^2 m}{\partial S \partial E} \bigg\vert_{E_*,\Sa}  &  = 
        \frac{1}{8 \pi  \gamma^2 
        {\left(4  \mathbb{K}[\zeta] -\frac{4-\zeta}{1-\zeta} \mathbb{E}[\zeta] \right)} } \; . \label{eq:d2mS}
\end{align}
\end{subequations}
From these expressions, the angular frequencies $(\omega_m \smeq \partial_m \tH|_S,\omega_S \smeq \partial_S \widetilde H|_m )$ and their derivatives can be obtained as (again see appendix~\ref{app:omegas})
\begin{subequations} \label{eq:omegas-5}
\begin{align} 
    \omega_m^\alpha   & =[(\partial_E m)^\alpha]^{-1}\label{eq:omegam} \\
    \omega_S^\alpha &=  - \omega_m (\partial_S m)^\alpha \label{eq:omegaS} \\
    (\partial_m \omega_m)^\alpha  &= - {\omega^\alpha_m}^3 (\partial^2_{EE} m)^\alpha \label{eq:dm-omegam} \\
    (\partial_S \omega_m)^\alpha  &= -{\omega^\alpha_m}^2((\partial^2_{ES} m)^\alpha + \omega^\alpha_S (\partial^2_{EE} m)^\alpha) \label{eq:dS-omegam} \\
    (\partial_S \omega_S)^\alpha  &= -\omega^\alpha_m ((\partial^2_{SS} m)^\alpha + 2\omega^\alpha_S (\partial^2_{ES} m)^\alpha + ({\omega^\alpha_S})^2 (\partial^2_{EE} m)^\alpha) \; . \label{eq:dS-omegaS} 
\end{align}
\end{subequations}
Note that Eqs.~\eqref{eq:omegam}-\eqref{eq:dmE} imply $T^\alpha = 2\pi/|\omega_m^\alpha | = ({4/\Sa}) \mathbb{K}[\zeta]$, which indeed corresponds to Eq.~\eqref{eq:Tstar}.  Furthermore, from Eqs.~\eqref{eq:dmE}-\eqref{eq:dmS} and Eqs.~\eqref{eq:omegam}-\ref{eq:omegaS}, 
\begin{equation} 
\label{eq:omegaS-omegam}
    \omega^\alpha_S = \left(\frac{3}{4} \Sa -\heff \right) - \frac14 \omega^\alpha_m \;  ,
\end{equation}
whose physical meaning is given in appendix~\ref{app:DynSeparatrix}.

\subsection{Spherical representation}
\label{sec:spherical-coordinates}

\newcommand{\bL}{{\bf L}}

Even though the relative coordinates $(S,s,\Theta,\theta)$ used in the previous subsections are convenient to compute actions,  frequencies, and their derivatives, the fact that the initial state considered is centered on the north pole of the phase space makes the cylindrical projection somewhat non-transparent.  Following closely~\cite{Strzys08}, another, perhaps more transparent, representation of the Bose-Hubbard dimer (akin to a  spherical representation) is introduced.

Starting from the form, Eq.~\eqref{eq:ham2scaled}, of the Bose-Hubbard dimer Hamiltonian, introduce $ \mathbf{L} = (\ell_x,\ell_y,\ell_z)$ as
\begin{align}
    \ell_x =& \sqrt{s_1 s_2} \cos(\theta_1 - \theta_2) = \sqrt{(S/2)^2 - s^2} \cos \theta \notag \\
    \ell_y =& \sqrt{s_1 s_2} \sin(\theta_1 - \theta_2) = \sqrt{(S/2)^2 - s^2} \sin \theta  \qquad .
    \label{eq:spherical}\\  
    \ell_z =& \frac12 (s_1 - s_2) =  s\; \notag
\end{align}
One has  $\lVert \mathbf{\bL} \rVert^2 =  \ell_x^2+\ell_y^2 +\ell_z^2 = (S/2)^2$, and it can be checked easily that $(\ell_x,\ell_y,\ell_z)$ fulfill the Poisson bracket relation $\{\ell_x,\ell_y\} = \ell_z$, together with its cyclic permutations. $\bL$ can therefore be seen as the angular momentum vector of a classical rotor subject to a Hamiltonian  which, expressing Eq.~\eqref{eq:Htilde} in terms of the components of $\bL$ reads
\begin{align} 
    \tH  & =  \frac{ S(S-4\heff)}{4}  + \ell_z^2  - 2 \gamma \ell_x \label{eq:ham-toupie1} \\ 
    & = \frac{ S(S-2\heff)}{2} + \gamma^2 - \left( (\ell_x  + \gamma)^2 + \ell_y^2 \right) \; , \label{eq:ham-toupie2} 
\end{align}
(note that, as in Eq.~\eqref{eq:H0Hs}, there is actually an irrelevant $\frac34 \heff^2$ correction to $\tH $).    This leads to the equations of motion $\dot \ell_\alpha = \{\ell_\alpha,\tH \}$ for the angular momentum
\begin{align}
    \dot \ell_x & = - 2 \ell_z \ell_y     \notag  \\
    \dot \ell_y & = + 2 \ell_z (\ell_x  + \gamma)  \label{eq:toupie} \; .\\
    \dot \ell_z & = - 2 \gamma \ell_y \; \notag
\end{align}
Note that this works with both forms, Eqs.~\eqref{eq:ham-toupie1} and \eqref{eq:ham-toupie2}, of the Hamiltonian, and that the latter equation is equivalent to the second equation of Eq.~\eqref{eq:zthdyn}.  From Eq.~\eqref{eq:toupie} it is readily visible that as the coupling increases, the fixed point that was originally on the north pole $\bL = (0,0,S/2)$ moves to $\bL = (-\gamma,0,\sqrt{(S/2)^2 - \gamma^2})$ and increases its energy by $\gamma^2$.

The bottom row of Figure~\ref{fig:level_curves} gives a representation of the dynamics for the  on the sphere $||\bL||= S/2$, demonstrating that what appears separatrix-like for  the $(\theta,s)$ motion is actually an artifact of the Mercator projection.

\section{Semiclassical description of the revivals}
\label{sec:semirev}

Here the full details of the semiclassical theory are presented.  As mentioned in Sec.~\ref{sec:EBKboot}, they are essentially resulting from a EBK-bootstrapped TWA theory since TWA does not inherently contain the necessary quantum interference to describe revivals.  Detailed expressions for the various measures, i.e.~expectation values, occupancy oscillations, coherences, and autocorrelation functions, respectively, are derived.

\subsection{Warmup : revival time for the uncoupled dimer}
\label{sec:warmup}

The monomer Hamiltonian $\hat H_0^{\rm mono} = \frac{1}{2} U  \hat{n}_1 (\hat{n}_1 - 1)$ gives with the rescaling, Eq.~\eqref{eq:rescaling}, $\widetilde H_0^{\rm mono} = \frac12 s_1(s_1 -2\heff) $,  leading to ${\omega'}^{\rm mono} = 1 $, and thus  to the rescaled revival time $\ttrevo = 2 \pi /\heff \mapsto \trevo = 2 \pi \hbar / U$.   Furthermore, as discussed in Sec.~\ref{sec:coherence-return}, for the monomer the revival is exact and independent of the shape and location of the initial coherent state. In the absence of coupling, $J=0$, the two sites of the dimer evolve independently from each other and this leads naturally to a perfect revival at the rescaled time $\ttrevo$.  It will be useful for the following to reconstruct how this can be understood within the action angle variables $(S, m;\varphi_S,\varphi_m)$ used in the coupled case.

For $\gamma = 0$, the scaled Hamiltonian, Eq~\eqref{eq:Htilde}, reads $\tH_0(S,s;\Theta,\theta)  = \frac14 S(S-4\heff) +  s^2$, and the ``perturbed'' action variable reduces to $m = s_2 = \frac12 S - s$, so that 
\begin{equation} 
\label{eq:H0-pv}
    \tH_0(S,m)  =   \frac12 { S(S-2\heff)} +  m^2 - Sm \; .
\end{equation}
As the quantization conditions are simply $S_N =  \heff  (N+1)$ and $m_k =  \heff(k+\frac12)$, it turns out that 
\begin{align}
    \frac{\widetilde{E}^0_{Nk} \ttrevo}{\heff} & = 2\pi \left[\frac12 (N+1)(N-1) + (k+\frac12)^2 - (N+1)(k+\frac12)\right]   \\
    &= 2\pi \left[\frac12 N(N-1) - k (N-k) \right] -\frac{3}{4} = {\rm const.} \quad \text{[modulo $2\pi$]} 
    \; ,
\end{align}
so that a perfect revival is indeed obtained at $\ttrevo$. Note that these EBK energies differ from the exact ones (cf. Eq.~\eqref{eq.pt.Enk}) by an irrelevant $O(\heff^2)$ constant $(-3/4)\heff^2$.  In the following, the interest is in the neighborhood of a torus $\bIa^T = (\Sa,\ma)$ on which the initial coherent state  
is centered.  Here it is chosen such that   $\Sa/\heff = \Nav$ is an integer, but  $\la = (\ma/\heff - 1/2)$  can take on an arbitrary real value.  Note that $\delta \la = (\la - \round{\la})$ is the distance of $\ma$ to a quantized torus.  Thus Eq.~\eqref{eq:H0-pv}, up to an irrelevant constant, takes the form
\begin{align} \label{eq:H0-pv2}
    \tH_0(\delta \bI)  & =  \tH_0^{(1)}(\delta \bI) + \tH_0^{(2)}(\delta \bI)  \\
    \tH_0^{(1)}(\delta \bI) & = \bOmega^0  \cdot \delta \bI =   (\Sa - \ma - \heff) \delta S  +(2 \ma -\Sa) \delta m  \\
    \tH_0^{(2)}(\delta \bI) & =   \frac12  \delta \bI^T \cdot \partial \bOmega^0 \cdot \delta \bI 
    = \frac12 \delta S^2 + \delta m^2 - \delta S \delta m \; 
\end{align}
where $\bI^T = (S,m)$, $\delta \bI = (\bI-\bIa)$, and $\bOmega^0$ and $\partial \bOmega^0$ are given in Eq.~\eqref{eq:Omegas0}.  

Let us focus on the second order term $\tH_0^{(2)}(\delta \bI)$.    The EBK quantized torus $(S_N,m_k)$  is such that $(\delta S_N,\delta m_k ) = \heff (\delta N,\delta k + \delta \la)$, where  $(\delta N,\delta k) = (N+1-\Nav, k - \round{\la}) \in \mathbb{N}^2$.  
This leads to  $\widetilde{E}^{0,(2)}_{Nk} = \heff^2 \left(  \frac12 \delta N^2 + (\delta k + \delta \la)^2  - \delta N (\delta k + \delta \la) \right)$ for the energies, and thus    
\begin{align}
    \frac{\widetilde{E}^{0,(2)}_{Nl} \ttrevo}{\heff} & = 2\pi \left(\delta N \Delta_S +\delta l \Delta_m\right)  
    + {\rm const.} \quad \text{[modulo $2\pi$]} \; ,   \label{eq:E02} \\
    \bDa & = (\Delta_S ,\Delta_m) = (\pi + 2\pi  \delta \la, -4\pi \delta \la) \; .  \label{eq:Delta2d}
\end{align}  
The shift $\bDa$, Eq.~\eqref{eq:Delta2d}, is the two-dimensional equivalent of the phase, Eq.~\eqref{eq:DeltaTh}, introduced in the one dimensional case, and plays a role below when looking at the dynamics near the revival time.

\subsection{ Wigner function for the initial state}
\label{sec:Wigner}

As in Sec.~\ref{sec:EBKboot}, both of the classical and semiclassical approaches to the dimer are based on the Truncated Wigner Approximation (TWA), in which an initial state is evolved classically by means of its Wigner function. For an initial coherent state, the Wigner function is a Gaussian in quadratures:
\begin{equation} 
\label{eq:wigner-base}
    W_\alpha(p_1,p_2,q_1,q_2) = \frac{1}{(\pi \heff)^{2}} \exp \left\{ - \frac{1}{\heff} \left[ (p_1 - p_1^0)^2 + (p_2 - p_2^0)^2 + (q_1 - q_1^0)^2 + (q_2 - q_2^0)^2 \right] \right\} \quad ,
\end{equation}
where $(p_1^0,p_2^0,q_1^0,q_2^0)$ is the Wigner function's center.   
The coherent state, Eq.~\eqref{eq:PSI0}, considered here is such that all atoms start on site 1 (and thus  $p_2^0 = q_2^0 = 0$), and the phase on site 1 is such that $p^0_1=0$, implying $q_1^0 = \sqrt{2\Sa} $.

Moving to the unperturbed action-angle coordinates $(s_1,s_2,\theta_1,\theta_2)$, Eq.~\eqref{eq:to-classic}, gives
\begin{align}
    (p_1 - p_1^0)^2 + (q_1 - q_1^0)^2 &= 2s_1 + 2\Sa - \sqrt{2 s_1 \Sa} \cos \theta_1 \\
    p_2^2 +q_2^2 &= 2s_2 \quad .
\end{align}
Since $s_1 = S + s/2$ and $\theta_1 = \Theta + \theta/2$, the Wigner function in relative coordinates reads
\begin{align} \label{eq:W0-rel}
    W_\alpha(S,\Theta,s,\theta) &= \frac{1}{(\pi \heff)^{2}} \exp \left[ - \frac{2}{\heff} \left( S + \Sa \right) + \frac{4}{\heff} 
    \sqrt{\Sa \left( \frac{S}{2} + s \right)} \, \cos \left( \Theta + \frac{\theta}{2} \right) \right] \quad .
\end{align}
As shown in Appendix~\ref{app:wigner},  this can be expressed in (true) action-angle coordinates $\bI=(S,m)$, $\bPhi =(\varphi_S,\varphi_m)$ as Eq.~\eqref{eq:app-wigfull}, for which, in practice, only the Gaussian approximation is needed
\begin{equation}    
\label{eq:wigner-full}
    W_\alpha \left( \delta \bI,  \bPhi \right) = \frac{1}{(\pi \heff)^{2}} 
    \exp \left[ -\frac{1}{\heff} \left( 
    \frac{\delta m^2}{2\hma} + 2\hma \varphi_m^2
    + \frac{\delta S^2}{2 \Sa} + 2\Sa\left(\varphi_S + \frac{\varphi_m}{4}(1 +\frac{\Sa}{\omega^\alpha_m}\right)^2  \notag
    \right) \right] 
\end{equation}
with $\delta \bI = (S-\Sa, m-\ma)$, and (see Eq.~\eqref{eq:hma})
\begin{equation} \label{eq:hma-sec6}
    \hma \equiv \frac{\sqrt{S} \gamma}{\omega^\alpha_m}  = \ma + 0(\gamma^4) \; .
\end{equation}
Integrating this function over the angular variables $(\varphi_S,\varphi_m)$, gives the Gaussian approximation of the function $\Wr_\alpha (\bI) := \int d\varphi_S d\varphi_m W_\alpha (\bI,\bPhi)$ of Eq.~\eqref{eq:W0app-dmS} as 
\begin{equation} 
\label{eq:WaGauss}
 \Wr_\alpha( \delta\bI) = 
 \frac{1}{2 \pi \heff \sqrt{\det{\MMinv}}} 
 \exp\left(-\frac{1}{2\heff}\delta\bI^T \MMinv^{-1} \delta\bI \right) \, ,
 \end{equation}
with the matrix $\MMinv$  expressed as 
\begin{equation} 
\label{eq:M}
 \qquad \MMinv =\begin{pmatrix}
    \Sa   & 0\\ 0 &  {\hma}
    \end{pmatrix} \; .
 \end{equation}

\subsection{Expectation value of observables (general expressions)}
\label{sec:genobs}

Starting with a derivation of general expressions for the expectation of an observable, specific results are generated below for the particular cases of the number operator $\hat n_2$ and the coherence $\hat a_2$.   Consider an arbitrary operator $\hat B$, which as in Sec.~\ref{sec:semic-1d} is assumed classical in the sense that its Weyl symbol $\BW$ has no significant variation on the scale $\heff$.
The time evolution of the mean value of $\hat B$ can be obtained  by adapting the approach of Sec.~\ref{sec:EBKboot} to two dimensions with the main difference being that the damping of the revivals is much less due to third order terms and much more so to the mismatch of the frequencies and their derivatives for the different degrees of freedom.  
 
 \subsubsection{Truncated Wigner Approximation for short times}
 
 Near $t=0$, the generalization of Eq.~\eqref{eq:TWA-oneD-a} to two dimensions reads 
 \begin{equation} \label{eq:mean-twoD}
    \mean{B}(\tt) = \int {\rm d}^2 \delta\bI \, \Wr_\alpha( \delta\bI) B^\alpha(\bPhi(\delta\bI,\tt)) \; ,
\end{equation}
 with $\delta\bI =  (\bI - \bIa)$.    Here, $B^\alpha(\bPhi) := \BW(\bIa,\bPhi)$ is the Weyl symbol of the observable $\hat B$ {\em on the torus $\bIa$}, which would have to be computed specifically for each observable considered (here $\hat n_2$ and $\hat a_2$). $\Wr_\alpha( \delta\bI)$, given by Eq.~\eqref{eq:WaGauss}, is the Wigner function of the initial state expressed in action-angle variables and integrated over the angular variables (so as to keep $\Wr_\alpha$ normalized to unity).  Finally, the time evolution of the angles $\bPhi^T = (\varphi_S,\varphi_m)$ conjugate to the actions $\bI$ is given near $t=0$
by
\begin{equation} \label{eq:PhiTWA}
   \bPhi_{\rm TWA}(\delta \bI,\tt) = (\bOmega + \partial \bOmega \cdot \delta \bI) \, \tt \; .
\end{equation}
Expressing $B^\alpha(\bPhi) $ as a Fourier series
\begin{equation} \label{eq:fourier-2d}
     B^\alpha(\bPhi) =  \sum_{\bk = (k_S,k_m)}   b_\bk \exp(i(\bk \cdot \bPhi + \chi_\bk)) \; ,
\end{equation}
(with $(b_\bk,\chi_\bk) \in \mathbb{R}^2$, and, if $B^\alpha$ is real,  $b_{-\bk} =  b_{\bk}$ and  $\chi_{-\bk} = - \chi_{\bk}$), noting that because of normalization of the Wigner function, the integral of $b_{(0,0)}$ is just $b_{(0,0)}$  itself,  and performing the remaining  Gaussian integral then yields 

\begin{equation} \label{eq:mean-twoD-gen}
    \TWA{B}(t) =  b_{(0,0)} + \sum_{\bk = (k_S,k_m) \neq (0,0)}   b_\bk \exp(i(\bk \cdot \bOmega t + \chi_\bk)) 
    \exp \left[ - \frac{\heff}{2} t^2 \, \bk^T \cdot \GTWA  \cdot \bk \right]
    \; , 
\end{equation}
where $\GTWA := \partial \bOmega^T \cdot \MMinv \cdot \partial \bOmega $, with $\MMinv$ given by Eq.~\eqref{eq:M}, is expressed as
\begin{equation} \label{eq:LambdaTWA}
    \GTWA
    = \begin{pmatrix}
    \Sa(\partial_S \omega^\alpha_S)^2 + \hma (\partial_S \omega^\alpha_m)^2  & 
        (\partial_S \omega^\alpha_m) ( \Sa (\partial_S \omega^\alpha_S) + \hma  (\partial_m \omega^\alpha_m))\\
        (\partial_S \omega^\alpha_m) ( \Sa(\partial_S \omega^\alpha_S) + \hma (\partial_m \omega^\alpha_m)) &
            \Sa (\partial_m \omega^\alpha_S)^2 + \hma  (\partial_m \omega^\alpha_m)^2  
    \end{pmatrix} \; .
\end{equation}

 \subsubsection{EBK-Embedded WTA Approximation near revival time}

Considering now the neighborhood of the revival time $\trevo$, the same approach as in Sec.~\ref{sec:OneDtrev} can be followed.  Defining $\delta \tt \equiv \tt-\ttrev$, $\delta H = H - \tH_0^{(2)}$ and using Eq.~\eqref{eq:E02} gives 
\begin{align}
  \hat U(\tt) =& \exp \left[- \frac{i}{\hbar} \tH_0^{(2)} \ttrev \right] \exp \left[- \frac{i}{\hbar} \delta H \ttrev \right]
 \exp \left[- \frac{i}{\hbar} \hat H \delta \tt \right]
 \notag \\
=&  \hat R(\bDa)
\exp \left[- \frac{i}{\hbar} \delta \hat H \ttrev \right] \exp \left[- \frac{i}{\hbar} \hat H \delta \tt \right] \; ,
\end{align}
where $\hat R (\Delta \varphi_S,\Delta \varphi_m)$ is the operator performing a rotation of angle $\Delta \varphi_S$ on $\varphi_S$ and of angle $\Delta \varphi_m$ on $\varphi_m$.  In Eq.~\eqref{eq:mean-twoD}, this amounts to using the angular time dependence
\begin{equation} 
\label{eq:PhiBoot}
   \bPhi_{\rm emb} (\delta \bI,t) =  (\bOmega + \partial \bOmega \cdot \delta \bI) \tt 
   + \bDa - (\partial \bOmega^0 \cdot \delta \bI)\ttrevo \; ,
\end{equation}
with 
\begin{equation} \label{eq:Deltas}
    \bDa^T  \equiv (\Delta_S, \Delta_m) = (\pi  (1 + 2 \delta \la) , -4 \pi \delta \la ) \; , 
\end{equation}
and
\begin{equation}   \label{eq:la}
 \qquad \delta \la := \cmodun{\ma/\heff -1/2}
\end{equation}
being, up to the factor $1/\heff$, the distance of $\ma$ to the nearest quantizing torus.  In this way near $\trevo$
\begin{align} 
    \mean{B}_{\rm emb}(\tt) & = \sum_{\bk = (k_S,k_m)}   b_\bk \exp(i(\bk \cdot (\bOmega t + \bDa) + \chi_\bk)) 
    \exp \left[ - \frac{\heff}{2}  \, \bk^T \cdot \GBoot \cdot \bk \right]
    \; \label{eq:boot-twoD-gen} \\ 
    \text{with} \quad \GBoot  & :=  (\partial \bOmega \tt- \partial \bOmega^0 \ttrevo)\cdot  \MMinv \cdot (\partial \bOmega \tt- \partial \bOmega^0 \ttrevo)  \; \label{eq:LambdaBoot}
\end{align}
(note the time dependence is included in $\GBoot$, whereas it was left out of $\GTWA$).

\subsection{Number oscillations}
\label{sec:n2dimer}
 
 Next, consider the application of the general formalism of Sec.~\ref{sec:genobs} to the observables considered in Sec.~\ref{sec:RevInBHdimer}.  
 Let us begin with the time evolution of the second-well occupancy $\mean{n_2}$,
where the average is taken over the evolved state from the initial coherent state, Eq.~\eqref{eq:PSI0}.  The Weyl symbol $\ntwoW$ of the second-well particle number is, up to the factor $\heff$ and a subleading $\heff$-correction, the action variable $s_2$ (cf Eq.~\eqref{eq:nclass}).  Thus, in leading order in $\heff$
\begin{equation}
\ntwoW (S,s,\Theta,\theta) \simeq \frac{s_2}{\heff} = \frac{1}{\heff}(({S}/2) -s) \; .
\end{equation}
The Wigner transform of the evolved state is centered on the north-pole torus $(\Sa,\ma)$, for which, from Eq.~\eqref{eq:relcoord},  $s^\alpha(\theta)$ is given by Eqs.~\eqref{eq:Hs}-\eqref{eq:zstar} as
\begin{equation} \label{eq:stheta}
    s^\alpha(\theta) = \frac{\Sa}{2} \sqrt{1 - (4 \gamma/\Sa)^2 \cos^2\theta}   \qquad \theta \in [-\frac{\pi}{2},+ \frac{\pi}{2} ] \; .
\end{equation}
It just remains to express the angle $\theta$ (which is the angular variable conjugate to $s$  in the relative coordinates, Eq.~\eqref{eq:relcoord}) in terms of the angles of the action-angle variables  $(S,m,\varphi_S,\varphi_m)$.  On the north-pole torus $(\Sa,\ma)$,  $\theta$ remarkably depends only on the second component of $\bPhi^T = (\varphi_S,\varphi_m)$, and 
thus $\theta = \ttheta (\varphi_m)$, with $\ttheta$ the function, Eq.~\eqref{eq:tilde-theta}, introduced in Sec.~\ref{sec:sth-dyn}.

\subsubsection{Near \texorpdfstring{$t=0$}{t = 0}}

Near $t=0$, the number oscillation is therefore obtained within the TWA approximation from Eq.~\eqref{eq:mean-twoD} with $\hat n_2$ as the operator $\hat B$, and thus with $B^\alpha$ as $n^\alpha_2 := \ntwoW(\bIa)$ expressed as
\begin{equation} \label{eq:n2alpha}
    n_2^\alpha(\varphi_m)  = \left(\frac{2\gamma^2}{\Sa \heff}\right)  \tf(\varphi_m) \;  , 
\end{equation}
where 
\begin{equation}  \label{eq:tf_of_phi}
    \tf(\varphi_m) := \left(\frac{\Sa}{2\gamma} \right)^2 \left(1-\sqrt{1 - \left(\frac{4 \gamma}{\Sa}\right)^2 \cos^2 \widetilde \theta \left(  \varphi_m \right) } \right)
    \quad  . 
 \end{equation}
The scaling in the r.h.s.\ of Eq.~\eqref{eq:n2alpha} is chosen such that in the perturbative regime
\begin{equation}  \label{eq:tf_of_phi_pert}
    \tf(\varphi_m) =  (1 - \cos \varphi_m) + O(\gamma^2) 
    \quad  . 
 \end{equation}
That gives\footnote{To the attentive reader, it may seem that the subleading $\heff$ correction associated with the $1/2$ shift in Eq.~\eqref{eq:nclass} is being neglected.  However, as can be seen in Figs.~\ref{fig:number-t0}-\ref{fig:number-trev}, and discussed in Appendix~\ref{sec:shift}, this correction is actually canceled out by a term associated with the $\heff^{1/2}$ extension of $W^\alpha(\bI,\bPhi)$ around $\bIa$.  This also leads to an $O(\heff)$ correction term that exactly compensates for the term due to the 1/2 shift in Eq.~\eqref{eq:nclass}.}
\begin{equation}
\label{eq:TWAn2-base}
\TWA{n_2}(\tt) = \left(\frac{2 \gamma^2}{\Sa \heff} \right)
    \int {\rm d} S {\rm d}m \, \Wr_\alpha( S,m) \tf((\omega^\alpha_m + (\partial_m \omega^\alpha_m ) \delta m) \tt) \; ,
\end{equation}
with $\delta m = (m - m^\alpha)$ and  $\Wr_\alpha(S,m)$ given by Eq.~\eqref{eq:WaGauss}.  

A more explicit expression for  $\braket{n_2}(t)$  is then obtained by decomposing the (periodic) function  $f(\varphi_m) $  as a Fourier series:
\begin{equation}
     \tf(\varphi_m ) = \sum_{k=-\infty}^{+\infty} \tf_k \exp \left( + {i k \varphi_m} \right) \, \quad \tf_k = \frac{1}{2\pi} \int_0^{2\pi} d \varphi_m
      \tf(\varphi_m) \exp \left( -  i k \varphi_m \right) \quad , \label{eq:fourier}
\end{equation}
with  $\tf_k$  real and $\tf_{-k} = \tf_k$  since $f(\varphi_m)$ is an even real function, and furthermore, with the scaling used in Eq.~\eqref{eq:tf_of_phi}, $\lim_{\gamma \to 0} \tf_0 = 1$ and $\lim_{\gamma \to 0} \tf_{\pm 1} = -1/2$.   Thus, in the case of $\hat n_2$, the general form, Eq.~\eqref{eq:fourier-2d}, reduces to a single sum on $k_m =k$ since only the $k_S=0$ terms are non zero.  This implies in practice that only the entry $\GTWA_{22}$  of the matrix  Eq.~\eqref{eq:LambdaTWA} plays a role, and this gives
\begin{align} 
\label{eq:n2t0-secVI}
  & \TWA{n_2}(\tt) = \left( \frac{2\gamma^2}{\Sa\heff} \right) \times \nonumber \\
  & \left(\tf_0
    +  
    2 \sum_k \tf_k \cos \left[   \omega_m^\alpha  (k \tt) \right]
    \exp \left[ -\frac{\heff}{2} \,(k \tt)^2 
    \left(  
      \Sa   (\partial_S \omega^\alpha_m)^2 +
    \hma (\partial_m \omega_m^\alpha )^2  
    \right)
     \right]\right) \; .
   \end{align}
 As can be seen in Fig.~\ref{fig:number-t0}, the expression Eq.~\eqref{eq:n2t0-f0f1}, which only keeps the constant term $f_0$ and the first harmonic $f_1$ of the Fourier series,  provides a very good account of the exact quantum result. Adding further Fourier components, or numerically integrating Eq.~\eqref{eq:TWAn2-base} has no effect on the curves for small $\gamma$'s, and only a very slight effect  otherwise.
 
 As the Fourier series, Eq.~\eqref{eq:fourier}, is dominated by its constant term $f_0$ and its first component $f_1$, the behavior near $t=0$ is characterized by oscillations of frequency $\omega^\alpha_m$ and amplitude $(2\gamma^2/\Sa \heff)  \tf_1$  that, neglecting $\hma$ in front of $\Sa$, decay on a time scale $\tdec \simeq 2/ (\partial_S \omega^\alpha_m \sqrt{\Sa \heff})$, and saturate to a constant value $(2\gamma^2/\Sa \heff) \tf_0$.

\subsubsection{Near \texorpdfstring{$\trev$}{trev}}

Considering times near the revival time $\trev$, the main difference with the analysis above is that in Eq.~\eqref{eq:mean-twoD}, the angle dynamics, Eq.~\eqref{eq:PhiTWA}, should be replaced by Eq.~\eqref{eq:PhiBoot}, leading to 
\begin{equation} 
\label{eq:BOOTn2-base}
\Boot{n_2}(\tt) = \left( \frac{2\gamma^2}{\Sa\heff} \right)
    \int {\rm d} S {\rm d}m \, \Wr_\alpha( S,m) \tf[(\omega_m^\alpha + (\partial_m \omega_m^\alpha ) \delta m) \,\tt     -4 \pi \delta \la  - (\partial_m \omega_m^\alpha)_0 \delta m \, \ttrevo ] \; .
\end{equation}
Expanding $f(\varphi_m)$ in a Fourier series, Eq.~\eqref{eq:boot-twoD-gen} (with just the component $\GBoot_{22}$ of the matrix  Eq.~\eqref{eq:LambdaBoot} involved since the only non-zero components for the Fourier series have $k_S=0$), gives 
\begin{align} 
    \Boot{n_2}(\tt)   = & \left(\frac{2\gamma^2}{\Sa\heff} \right)\left(\tf_0   +  
    2 \sum_{k\geq 1} \tf _k \cos \left[   k(\omega_m^\alpha  \tt  - 4\pi \dna)\right] 
    \exp \left[ -\frac{\heff}{2} k^2 \GBoot_{22}(\tt) \right] \right)   \label{eq:n2trev-SecVI}
    \\
     \mbox{with} \quad \GBoot_{22}(\tt) = &
     \Sa \left( (\partial_S \omega^\alpha_m)^2 \delta \tt + \trevo (\partial_S \omega^\alpha_m - \partial_S \omega^{\alpha,0}_m ) \right)^2  
      +  \notag \\
   & \qquad \hma \left( (\partial_m \omega^\alpha_m)^2 \delta \tt + \trevo (\partial_m \omega^\alpha_m -\partial_m \omega^{\alpha,0}_m ) \right)^2    \notag 
\end{align}
with $\tt = \ttrevo + \delta \tt$.  Keeping only the Fourier components $k=0,1$ in Eq.~\eqref{eq:n2trev-SecVI}  yields Eq.~\eqref{eq:n2trev-f0f1}, which, as seen in Fig.~\ref{fig:number-trev}, also gives excellent agreement with the quantum data.  Again, adding further Fourier components or numerically integrating Eq.~\eqref{eq:BOOTn2-base} only marginally modifies the curves.
 
 From Eq.~\eqref{eq:n2trev-SecVI}, the shift between $\ttrevo$ and the time corresponding to the maximum amplitude of the revived $\mean{n_2}$ is 
\begin{equation}
    \delta \trev^{(n_2)} = -
    \frac{\Sa (\partial_S \omega^\alpha_{m}-\partial_S \omega^{\alpha,0}_{m})
    + \hma  (\partial_m \omega^\alpha_{m}-\partial_m \omega^{\alpha,0}_{m})}
    {\Sa \partial_S \omega^\alpha_{m} + \hma  \partial_m \omega^\alpha_{m}}
    \; ,
\end{equation}
which, for small $\gamma$'s is dominated by the behavior of $\partial_S \omega^\alpha_m$,  and the amplitude of the revival is reduced for the harmonic $k$ by a factor $\exp[-(\heff k^2 /2)] \GBoot_{mm}(\ttrevo + \delta \trev^{(n_2)} ) $.

\subsection{Coherence}
\label{sec:a2dimer}

The calculation for the coherence follows the same route as for the particle number.  The only difference is that the phase space function $n_2 = s_2/\heff$  should be replaced by $a_2 = \sqrt{s_2/\heff} \exp(-i\theta_2)$.  In the relative variables, $s_2 = \frac12 S - s$ and $\theta_2 = \Theta - \frac12 \theta $, and thus on the north-pole torus $(\Sa,\ma)$, in term of the angle variables $(\varphi_S,\varphi_m)$,  $s_2 = \frac12 S - s^\alpha(\ttheta (\varphi_m))$ with  $s^\alpha(\theta)$ given by Eq.~\eqref{eq:stheta}, and the functions $\ttheta(\varphi_m)$ and $\tttheta(\varphi_m)$ by Eqs.~\eqref{eq:tilde-theta}-\eqref{eq:tilde-tilde-theta}. Furthermore in appendix~\ref{app:DynSeparatrix}, it is shown that on the north-pole torus (see Eq.~\eqref{eq:theta2-phis-app})
\begin{equation} \label{eq:theta2-phis}
        \theta^\alpha_2(\varphi_S,\varphi_m)  = \varphi_S + \frac14(\varphi_m  - 2 \tttheta(\varphi_m) ) +  \pi \floor{-\varphi_m/2\pi} - \frac{\pi}{2}\; ,
\end{equation}
 with the understanding that  $\frac14 (\varphi_m - 2 \tttheta(\varphi_m) )$ is  a smooth monotonously increasing function,  and that the discontinuities imposed by $\pi \lfloor- \varphi_m/2\pi \rfloor$, with $\lfloor x \rfloor$ the integer part of $x$, is just what is necessary to make $\theta_2(\varphi_S,\varphi_m)$ $2\pi$-periodic in $\varphi_m$.  
 The coherence is thus obtained from Eq.~\eqref{eq:mean-twoD} with $B^\alpha(\bPhi)$ taking the form
\begin{equation} \label{eq:a2-explicit}
    a_2(\varphi_S,\varphi_m) = i \sqrt{\frac{\frac12 \Sa - s^\alpha(\ttheta(\varphi_m))}{\heff}}   
    (-1)^{\floor{-\varphi_m/2\pi} }
    \exp \left[ -\frac{i}{4} \left(\varphi_m - 2 \tttheta(\varphi_m) \right) \right] 
    \exp \Bigg[- i \varphi_S \Bigg] \; ,
\end{equation}
and $\bPhi= \bPhi_{\rm TWA}$, Eq.~\eqref{eq:PhiTWA}, near $t=0$ and $\bPhi = \bPhi_{\rm emb}$, Eq.~\eqref{eq:PhiBoot}, near $\trev$.

To get further insight, decompose $a_2$ into its Fourier components.  Writing 
\[
a^\alpha_2 (\varphi_S,\varphi_m) = \sqrt{\frac{2 \gamma^2}{\Sa \heff}}  \tg(\varphi_m)  \exp (-i \varphi_S) 
\]
with 
\[
\tg(\varphi_m) := i (-1)^{\floor{-\varphi_m/2\pi} } \sqrt{\tf(\varphi_m)}  
    \exp \left[ -\frac{i}{4} \left(\varphi_m - 2 \tttheta(\varphi_m) \right) \right] \; ,
\]
it turns out that given the trivial $\varphi_S$ dependence of $a_2$, the only non zero Fourier coefficient are of the form $\bk =(-1,k_m)$.  Furthermore, it can be shown that in the limit $\gamma \to 0$, $\tg(\varphi_m)  =  \left(e^{-i\varphi_m} - 1\right) + O(\gamma^3)$, so that $\tg_{-1} \to 1$ , $\tg_{0} \to -1$, and all the other Fourier components vanish.
As a consequence, even for finite $\gamma$, $\tg_0$ and $\tg_{-1}$ largely dominate the Fourier expansion of $g$.  Keeping only these two terms of the Fourier series near $t=0$ gives
\begin{equation}  \label{eq:a2t0-secVI}
\begin{aligned}
    \TWA{a_2}(t) & =   \sqrt{\frac{2 \gamma^2}{\heff \Sa}}
    \Bigg\{ 
    \tg_0 \exp(-i\omega^\alpha_S t) \exp \left[ - \frac{\heff}{2} t^2 \, \GTWA_{11}  \right] \\
& +
        \tg_{-1} \exp(-i(\omega^\alpha_S + \omega^\alpha_m) t) \exp \left[ - \frac{\heff}{2} t^2 \, \left(\GTWA_{11}  +  2 \GTWA_{12} + \GTWA_{22}\right) \right]
    \Bigg\}
    \end{aligned}
    \qquad  , 
\end{equation}%
which with
\[ 
\GTWA_{11}  +  2 \GTWA_{12} + \GTWA_{22} = \Sa (\partial_S(\omega^\alpha_S+\omega^\alpha_m))^2  + \hma (\partial_m(\omega^\alpha_S+\omega^\alpha_m))^2 )
\]
gives Eq.~\eqref{eq:mean-a2t0-s4}.  

In the same way, near $\trev$,
\begin{equation}  \label{eq:a2trev-secVI}
\begin{aligned}
    {\Boot{a_2}(\tt)} & =  \sqrt{\frac{2 \gamma^2}{\heff \Sa}}
    \Bigg\{ 
    \tg_0 \exp[-i\omega^\alpha_S \tt + i\pi  (1 + 2 \delta \la)] \exp \left[ - \frac{\heff}{2} \, \GBoot_{11}  \right] \\
& +
        \tg_{-1} \exp[-i(\omega^\alpha_S + \omega^\alpha_m) \tt + i\pi  (1 - 2 \delta \la)]  \exp \left[ - \frac{\heff}{2} \, \left(\GBoot_{11}  +  2 \GBoot_{12} + \GBoot_{22}\right) \right]
    \Bigg\}
    \end{aligned}
    \qquad  .
\end{equation}
As shown in Figs.~\ref{fig:coherence-t0}-\ref{fig:coherence-trev}, Eqs.~\eqref{eq:a2t0-secVI}-\eqref{eq:a2trev-secVI} provide an excellent approximation to the exact quantum result.  They also allow us to link the  atypical behavior of $\mean{a_2}$ with the classical precession associated with the fact that $(\omega^\alpha_S + \omega^\alpha_m)  = O(\gamma^2)$.

\subsection{Autocorrelation function}
\label{sec:autocor}

The autocorrelation function $\autocor(t) =  |\langle \Psi_0 | \Psi (t) \rangle|^2 $ is somewhat simpler to compute than the mean value of observables since, beyond the initial Wigner function, only the dynamics within the action-angle variables are required.  A slight extra-complication, however, is that it is then necessary to take into account the full Wigner function and not its angle-integrated version.  

The autocorrelation function can be expressed as  
\begin{equation} 
\label{eq:autocor}
    \autocor(t)  = 2 \pi^2 \heff^2 \iiiint  d^2\delta \bI   d^2\bPhi\,  W_\alpha(\delta \bI,  \bPhi)  
    W^t_\alpha(\delta \bI,  \bPhi) \; 
\end{equation}
where  the initial Wigner function $W_\alpha(\delta \bI, \bPhi)$ Eq.~\eqref{eq:wigner-full}  can be expressed as 
\begin{equation}
    W_\alpha \left( \delta \bI,  \bPhi \right) 
     = \exp \left[ -\frac{1}{\heff} \left( \frac12
\delta \bI^T \MMinv^{-1} \, \bI + 2 \bPhi^T \NN^T  \MMinv  \NN  \,\bPhi
     \right) \right]
   \label{eq:wigner-full-bis}
\end{equation}with $\MMinv$ given by Eq.~\eqref{eq:M} and 
\begin{equation} \label{eq:NN}
    \NN
    = \begin{pmatrix}
    1  & \frac{1}{4} \left(1 + \frac{\Sa}{\omega^\alpha_m}  \right)        \\
    0   & 1  
    \end{pmatrix} \; .
\end{equation}
The time evolved Wigner function $W^\tt_\alpha(\delta \bI,  \bPhi)$ is obtained as 
\begin{equation}
    \begin{aligned}
    W^t_\alpha(\delta \bI, \bPhi) & = 
    W_{\alpha} (\delta \bI, \bPhi-\bPhi_{\rm TWA}(\delta \bI,\tt)) \qquad \text{for } \tt \simeq 0 \\ 
    W^t_\alpha(\delta \bI, \bPhi) & = 
    W_{\alpha} (\delta \bI, \bPhi-\bPhi_{\rm emb}(\delta \bI,\tt)) \qquad \text{for } \tt \simeq \ttrev 
\end{aligned}
\; ,
\end{equation}
with $\bPhi_{\rm TWA}(\delta \bI,\tt)$ and  $\bPhi_{\rm emb}(\delta \bI,\tt)$ given by  Eq.~\eqref{eq:PhiTWA} and Eq.~\eqref{eq:PhiBoot} respectively.

Equation~\eqref{eq:autocor} then takes the form of a Gaussian integral, which near $t=0$ results in 
\begin{equation}
    \autocor(t) = \frac{1}{\sqrt{\det(\MMinv \cdot  \AAt)}}
    \exp \left[
    \frac{1}{\heff }\left(\BBt^T \AAt^{-1} \, \BBt- \CCt \right)
    \right]
\end{equation}
with 
\begin{subequations} \label{eq:auto-t0}
\begin{align}
    \AAt &= \MMinv^{-1} + (\partial \bOmega \cdot \NN^T \cdot \MMinv \cdot \NN \cdot \partial \bOmega )\, \tt^2 \\
    \BBt^T & = (\GG(\tt)^T \cdot \MMinv \cdot \NN \cdot \partial \bOmega) \tt \\
    \CCt &= \GG(\tt)^T \cdot \MMinv \cdot \GG_0(\tt) \\
    \GG_0(\tt) &= \NN \cdot \begin{pmatrix}
    \cmod{\omega^\alpha_S \, \tt}  \\
    \cmod{\omega^\alpha_m \, \tt}  
    \end{pmatrix}     \; ,
\end{align}
\end{subequations}
(with $\cmod{\varphi}$ the angle $\varphi$ the range brought back to the range $[-\pi,+\pi]$),  and near $\trev$ in 
\begin{equation} 
\label{eq:autocor_trev}
    \autocor(t) = \frac{1}{\sqrt{\det(\MMinv \cdot \AAb)}}
    \exp \left[
    \frac{1}{\heff }\left(\BBb^T \AAb^{-1} \, \BBb - \CCb \right)
    \right]
\end{equation}
with 
\begin{subequations} \label{eq:auto-trev}
\begin{align}
\AAb &= \MMinv^{-1} + \KKb^T(t) \cdot \MMinv \cdot \KKb(t) \\
    \BBb^T & = \DGr(t)^T \cdot\MMinv \cdot\KKb(t) \\
    \CCb &= \DGr(t)^T \cdot \MMinv \cdot \DGr(t)   \label{eq:auto-trevC}
    \end{align}
and
    \begin{align}
    \DGr(\tt) &= 
    \NN \cdot \left(\cmod{\Da + \bOmega \tt}\right)      \\
\KKb &= \NN \cdot[\partial \bOmega \tt -\partial \bOmega^0 \ttrevo ] \; .
\end{align}
\end{subequations}

Note that in the $1$D analog of Eq.~\eqref{eq:auto-trev} all the matrices become scalars and thus commute, and Eq.~\eqref{eq:autocor_trev} reads
\begin{equation} 
\label{eq:autocor_trev_1d}
    \autocor(t) = \sqrt{\frac{M^2}{M^2 + \KK^2}}
    \exp \left[-
    \frac{1}{\heff } \frac{ (M \, \DG)^2}{M^2 + \KK^2}
    \right] \; ,
\end{equation}
with  $M$ being a pure number describing the squeezing of the initial coherent state in the action-angle basis and $N \equiv 1$.   In the absence of squeezing (i.e.\ for $M=1$), this equation is the equivalent of Eq.~\eqref{eq:Previval}.

As seen in Fig.~\ref{fig:autocorr-trev},  Eq.~\eqref{eq:autocor_trev} reproduces very precisely the exact quantum revival of the autocorrelation function beyond the perturbative regime, and in the perturbative regime for values of $\gamma$ near the peaks of $\autocormax(\gamma)$.  However, although the global structure is correct, it under-evaluates significantly the global amplitude in the perturbative regime near dips of $\autocormax(\gamma)$.  As already discussed in Sec.~\ref{sec:syncres}, the origin of this behavior can be related to the smallness of the action $\ma = O(\gamma^2)$ of the north-pole trajectory.  Indeed, even for $\Nav = 200$, $\ma/\heff$ can be not much larger, or even smaller, than unity for the smaller $\gamma$-range.  

To see why this is important, consider Eq.~\eqref{eq:auto-trevC}.  As $\lim_{\gamma \to 0} \omega^\alpha_m = - \Sa $,  the off-diagonal term of the matrix $\NN$ Eq.~\eqref{eq:NN} is $O(\gamma^2)$ and actually does not modify significantly the end result, and for this discussion we make the approximation $\NN \simeq \mathbb{1}$.  In that case, $\CCb \simeq \left(\Sa \GGSe(t)^2 + \hma \GGme(t)^2 \right)$,  with 
$$(\GGSe(t),\GGme(t)) = \left(\cmod{\Delta_S + \Omega_S \tt},\cmod{\Delta_m + \Omega_m \tt}\right)$$ 
the center of the evolved coherent state. 
Focusing on the second term, it can be understood as resulting from the Gaussian integral
\begin{align}
   \mathcal{I}_m(t)  & =  \frac{2}{\pi \heff} \iint d\delta m \, d\varphi_m \, G(\delta m , \varphi_m ) \,G(\delta m , \varphi_m  - \GGSe(t) ) \label{eq:Im} \\
   & = \exp \left(- \frac{\hma}{\heff} [\GGSe(t)]^2 \right) \notag \\
   & G(\delta m , \varphi_m ) :=
   \exp \left(-\frac{1}{\heff} \left( \frac{\delta m^2}{2 \hma} + (2 \hma) \varphi^2 \right) \right) \; ,
\end{align}
which corresponds to the overlap between two ``frozen'' coherent states, one centered at the origin, $\varphi_m = 0$, and the second centered on $\GGme(t)$.   In the deep semiclassical limit, it is expected that $\hma / \heff \gg 1$.  Thus, only times for which $\GGme(t) \simeq 0$ are expected to give any significant contribution, and for such times a Gaussian approximation is fine.  However, if $(\hma/\heff) \lesssim 1$,  we get significant contributions even for $\GGme \simeq \pi$, where only the tails of the Wigner transform of the evolved coherent state are overlapping.  It is thus not such a big surprise that in that case the Gaussian approximation to the Wigner transform, which adequately describes only its center, does not necessarily provide a decent approximation.

One simple way out of this difficulty is to perform the Gaussian integral, Eq.~\eqref{eq:Im}, not in the action-angle $(m, \varphi_m)$ variables, but in the associated quadrature variables $(q_m,p_m) = \sqrt{2m }(\cos \varphi_m, \sin \varphi_m)$.  In that case (cf.~Eqs.~\eqref{eq:W0-aaa}-\eqref{eq:W0-aa}), the Gaussian approximation to the frozen  coherent state takes the form
\begin{equation}
\tilde G(q_m,p_m;t) = \frac{1}{\pi\hbar} \exp \left[ - \frac{1}{\hbar}\left((p _m- \Delta p_m(t))^2 +  (q_m - \Delta q_m(t))^2\right) \right] \; 
\end{equation}
with 
\begin{align}
\Delta q_m(t) & = \sqrt{2\ma }\cos \GGme(t) \nonumber \\
\Delta p_m(t) & = \sqrt{2\ma } \sin \GGme(t)\; .
\end{align}
for which the Gaussian approximation is expected to be much more precise.

Performing the Gaussian integral, Eq.~\eqref{eq:Im}, in the quadrature representation leads to replacing $\mathcal{I}_m$ by 
\begin{align}
    \tilde{\mathcal{I}}_m(t)=& \frac{2}{\pi \heff} \iint d q_ m \, d p_m \, \tilde G(q_ m , p_m ; 0) \, \tilde G(q_ m , p_m ; t) \label{eq:Imt} \\
    =& \exp \left( - \frac{2}{\heff} (1- \cos \GGme(t))\right) \; .
\end{align}
As expected, the integrals $\tilde{\mathcal{I}}_m(t)$ and ${\mathcal{I}}_m(t)$  correspond for $\GGme(t) \ll \pi$, but as shown in Figs.~\ref{fig:autocorr-trev} and \ref{fig:autocorr-max}, the former provides a much better approximation to the exact quantum results than the latter. A full uniform semiclassical approximation, valid in the peaks of $\autocormax(\gamma)$ for small $\gamma$'s as well as for larger $\gamma$'s and a bit further away from $\trev$ where the contribution of $\BBb^T \AAb^{-1} \, \BBb $ in the exponent of Eq.~\eqref{eq:autocor_trev} cannot be neglected, can be obtained by performing the whole integral, Eq.~\eqref{eq:autocor}, using the quadrature variables $(q_m,p_m)$.

\section{Conclusion}
\label{sec:conc}

In summary, our study sheds light on intricate collapse and revival phenomena that can emerge in a many-body bosonic system, owing to the intrinsic multidimensionality of the dynamics and to the variety of observables that are offered by the context of ultracold quantum gases.
Various revival phenomena, manifested in different observables such as the return probability to the initial state, the site occupancies, and the site coherences, were revealed for the Bose-Hubbard dimer and analyzed using semiclassical and quantum perturbative methods.   These two different types of analytical techniques yield complementary insights and enable a comprehensive quantitative picture of the physical mechanisms that are at work here.  
The semiclassical approach has the additional asset that it allows one to identify a classical (mean-field)  ``backbone'' for this bosonic many-body system, given in terms of the underlying phase-space structure, which in a semiclassical regime, characterized here by a large dimer population $\Nav \gg 1$, ultimately governs all sorts of quantum dynamical phenomena occurring in this system, including those that rely on genuinely nonclassical (interference or tunneling) effects as it is the case for revivals.

The diversity of observables offered by the physical context of ultra-cold quantum gases, as well as the increased number of degrees of freedom,  led to a much richer revival phenomenology than in the traditional 1D Schrödinger context.  This is evidenced by the extensive list of rather surprising phenomena that are discussed in Sec.~\ref{sec:explanation}, some of which  have already been identified in~\cite{Schlagheck22}. In particular we have highlighted 5 puzzles: the existence of more that one revival time [P1]; a very unexpected behavior of the smooth background for the site-2 coherence [P2]; the disappearance and resurgence of the coherent-state revival with increasing dimensionless inter-site hopping strength $\gamma$ [P3], correlated with the presence of a minima or maxima of the site-2 occupation $n_2(t)$ at $\trevauto$ [P4], and with the integer or half integer value of the time-averaged $\bar n_2$ [P5]. Our semiclassical approach provides a rather complete understanding of the revival mechanism's underlying physics, allowing us to fully understand the origin of these various puzzles.

Many of the findings reported here, especially those concerning the mean site occupancies and coherences, can be readily observed in state-of-the-art quantum simulation setups involving bosonic atoms in optical lattices.
Coherences can be probed using an adaptation of the collapse and revival experiment of Greiner et al.~\cite{Greiner02b} to dimer superlattices, possibly combined with band mapping techniques in order to distinguish between the two sites of the dimer~\cite{Folling07}, and the second-site occupancy can be detected with quantum gas microscopes~\cite{Bakr09,Sherson10}.
Let us note, furthermore, that the phenomenology discussed here is not restricted to the physical context of ultracold bosonic quantum gases, but can be observed in other systems as well.
Owing to the formal Lipkin-Meshkov-Glick equivalence between a Bose-Hubbard dimer and a spin with high angular momentum~\cite{Lipkin65, Meshkov65, Strzys08}, we expect that resurgent revivals can be encountered also with rotating molecules and nanorotors~\cite{Stickler18, Rieser, Wachter}.
More generally, qualitatively similar phenomena are expected to occur in generic integrable systems with two of degrees of freedom, including wave mechanical systems such as two-dimensional integrable (e.g., rectangular, circular, or elliptic) microwave billiards~\cite{StoeckmannStein,StockmannBook,Dembowski00} as well as  optical fibers \cite{Doya01,Doya02}.

It remains an open question what happens with revivals in the presence of a perturbation that breaks integrability. 
Owing to the Kolmogorov-Arnold-Moser theorem, most invariant tori of the integrable classical dynamics still prevail in presence of the perturbation, so no substantial changes are generically expected to occur up to the collapse time scale. 
On longer time scales, however, tunneling effects induced by nonlinear resonances \cite{Brodier01,Brodier02} might come into play and compete with the interference dynamics giving rise to the revival that could have a drastic impact on the resurgence of the latter despite the weakness of the perturbation.
As near-integrable systems are encountered much more frequently than strictly integrable ones, this particular research question is of practical relevance and merits further investigations.

As revivals are highly reliant on the preservation of global coherence (or, perhaps more precisely, on the absence of decoherence) during the time evolution process, their probing can generally be used to certify the functionality of quantum devices whose working principle is based on (single- or many-body) interference.
More specifically, revival phenomena in the Bose-Hubbard dimer can, in principle, be exploited for the purpose of quantum sensing, in particular concerning the second, initially unpopulated, site, and again with special emphasis on the preservation of coherence and interference.
More research into this direction is certainly needed in order to explore the full metrological and sensing potential that this particular revival setting can offer.

\appendix

\section{Quantum perturbation theory in the Bose-Hubbard dimer}
\label{app:qpt}

Quantum perturbative expressions are developed here for measures such as occupancies, coherences, and return probabilities for the Bose-Hubbard dimer.  For the interesting case of a maximally imbalanced initial state occupancy, Eq.~\eqref{eq:cs.BEC} with $z_2=0$, the occupancies of the second site mixed by the perturbation are in the neighborhood of zero, thus they can be considered small and far less than the total number of particles, which is taken to be large; i.e., the second site occupancies are effectively smaller by nearly a factor $\heff$.  The perturbation contains only real non-zero matrix elements on the two off-diagonals of a tridiagonal matrix and excluding a single diagonal matrix element for odd-$N$ cases and a single off-diagonal element for even-$N$ cases (largely irrelevant for our purposes),  they are essentially identical for symmetry decomposed spaces.  This tridiagonal structure has a very strong imprint on the quantum perturbation theory. In the absence of intersite hopping, i.e.~$J=0$, any product state in a lattice revives since the revival time scale is the same for all sites regardless of occupancy.

Initially the hopping strength $J$ represents the perturbation strength parameter, however ultimately the perturbation series for the frequencies becomes an expansion in $\gamma$, as it must.  For vanishing $\gamma$, the quantum Hamiltonian has the special feature that the spectrum is a purely quadratic function of the $\{\hat n_j\}$ operators.  Hence, the expansion of the on-site eigenenergies to quadratic order, Eq.~\eqref{eq:expans-sec2}, is exact separately for each site.  This property leads to some special spectral features.

The two key effects of a perturbation are the shifts of the eigenvalues and the rotation of the eigenvectors.  It turns out, as shown ahead, that to describe all of the revival phenomena observed for the dimer it is sufficient to stop at second order with the exception of return probabilities.  For that measure, additional terms related to eigenfunction rotations are required.  Second, although there are degeneracies due to symmetries in the unperturbed spectrum of the interaction (Kerr) Hamiltonian, the perturbation respects these symmetries and therefore does not split these degeneracies. In a basis fully accounting for the symmetries of fixed particle number and interchange of site index, there are no further degeneracies to take into account, and in fact, there are no small energy denominators either.

\subsection{General expression for eigenvalues and eigenvectors}

The eigenstates of the Bose-Hubbard dimer Hamiltonian, Eq.~\eqref{eq:cs.BHd}, are either symmetric or antisymmetric with respect to the interchange of site index.
It is thus convenient to develop the perturbative approach in a properly symmetrized basis.  The latter is constituted by the states
\begin{equation} 
\ket{N;k;\pm} = \frac{1}{\sqrt{2}} \left( \ket{N-k,k} \pm \ket{k,N-k} \right)
\end{equation}
for all integer $k$ satisfying $0 \leq k \leq \left[\frac{N-1}{2}\right]$ as well as the additional case for $n$ even,
\begin{equation} 
\ket{N; N/2; +} = \ket{N/2,N/2} \,.
\end{equation}
The Hamiltonian matrix is real symmetric tridiagonal within this basis.  Defining
\begin{align}
\mathcal{E}_{N;k} & =  \frac{U}{2}\left[N(N-1) - 2 k(N-k)\right] \,, 
\label{eq.pt.Enk} \\
\mathcal{J}_{N;k} & = J \sqrt{(k+1)(N-k)} \,, 
\label{eq:pt.Jnk}
\end{align}
its diagonal matrix elements $\mathcal{E}_{N;k}^\pm = \bra{N;k;\pm} \hat{H} \ket{N;k;\pm}$ are given by 
\begin{equation}
\mathcal{E}_{N;k}^\pm = \mathcal{E}_{N;k}
\label{eq:enk0}
\end{equation}
except for the odd $N$ case of $k = (N - 1)/2$, which is 
\begin{equation}
\mathcal{E}_{N;(N-1)/2}^\pm = \mathcal{E}_{N;(N-1)/2} \mp \mathcal{J}_{N;(N-1)/2} \,. 
\label{eq:qpt.Epm}
\end{equation}
The lower off-diagonal matrix elements $ \bra{N;k+1;\pm} \hat{H} \ket{N;k;\pm} = - \mathcal{J}_{N;k}^\pm$ are given by \begin{equation}
\mathcal{J}_{N;k}^\pm = \mathcal{J}_{N;k}
\end{equation}
with the one exceptional case of
\begin{equation} 
\mathcal{J}_{N;N/2 - 1}^+ = \sqrt{2} \mathcal{J}_{N;N/2 - 1} \ .
\label{eq:qpt.Jp}
\end{equation}

It is worth noting here that there is a consistency, as there must be, between the unperturbed classical Hamiltonian, Eq.~\eqref{eq:H0-pv}, rescaled using Eq.~\eqref{eq:rescaling}, and the unperturbed quantum spectrum, Eqs.~\eqref{eq.pt.Enk} and \eqref{eq:enk0}.  
Through semiclassical quantization, the action, $S$, is related to the total number of particles, $N$, and the action, $m$, is related to the second site occupancy, i.e.~quantum number $k$.  
The action derivatives of the classical Hamiltonian generate two frequencies, $\omega_S = S -m -\heff$ and $\omega_m = -S+2m$, which, after accounting for the scaling and $\hbar$, are perfectly consistent with the pair of energy differences, $\mathcal{E}_{N;k}^\pm - \mathcal{E}_{N-1;k}^\pm = U(N - k -1)$ (corresponding to $\omega_S$), and $(\mathcal{E}_{N;k+1}^\pm - \mathcal{E}_{N;k-1}^\pm)/2 = -U(N - 2k)$ (corresponding to $\omega_m$).  The particular energy differences chosen here are determined by the effects of phase shifts associated with caustics, i.e.~Maslov indices. Similarly, the frequency derivatives given in Eq.~\eqref{eq:Omegas0} follow from the appropriate second energy differences.

Denote the expansion coefficients of an eigenstate $\ket{\Psi}$ with respect to this symmetrized basis by $\psi_{N;k}^{\pm} = \braket{N;k;\pm|\Psi}$.  The eigenstate equation $\hat{H}\ket{\Psi} = E\ket{\Psi}$ can be written as
\begin{equation}
E \psi_{N;k}^\pm = \mathcal{E}_{N;k}^\pm \psi_{N;k}^\pm - \mathcal{J}_{N;k-1}^\pm\psi_{N;k-1}^\pm - \mathcal{J}_{N;k}^\pm\psi_{N;k+1}^\pm 
\label{eq:pt.pk}
\end{equation}
for $0 < k < k_{\rm max}^\pm(N)$, combined with
\begin{eqnarray}
E \psi_{N;0}^\pm & = & \mathcal{E}_{N;0}^\pm \psi_{N;0}^\pm - \mathcal{J}_{N;0}^\pm \psi_{N;1}^\pm \,, 
\label{eq:pt.p1} \\
E \psi_{N;k_{\rm max}^\pm(N)}^\pm & = & \mathcal{E}_{N;k_{\rm max}^\pm(N)}^\pm\psi_{N;k_{\rm max}^\pm(N)}^\pm - \mathcal{J}_{N;k_{\rm max}^\pm(N)-1}^\pm \psi_{N;k_{\rm max}^\pm(N)-1}^\pm \,. 
\label{eq:pt.pn}
\end{eqnarray}
Equation \eqref{eq:pt.p1} generates $\psi_{N;0}^\pm = \alpha_{N;1}^\pm \psi_{N;1}^\pm$ with
\begin{equation}
\alpha_{N;1}^\pm = \frac{\mathcal{J}_{N;0}^\pm}{\mathcal{E}_{N;0}^\pm - E} \,. 
\label{eq:pt.a1}
\end{equation}
Making the ansatz $\psi_{N;k}^\pm = \alpha_{N;k+1}^\pm \psi_{N;k+1}^\pm$ for $k > 0$ and inserting it into Eq.~\eqref{eq:pt.pk} yields the recursion relation
\begin{equation}
\alpha_{N;k+1}^\pm = \frac{\mathcal{J}_{N:k}^\pm}{\mathcal{E}_{N;k}^\pm - E - \mathcal{J}_{N;k-1}^\pm\alpha_{N;k}^\pm}  \ .
\label{eq:pt.aj}
\end{equation}
Conversely, starting on the opposite end of the chain with the notation $\psi_{N;k_{\rm max}^\pm(N)}^\pm = \beta_{N;k_{\rm max}^\pm(N) - 1}^\pm \psi_{N;k_{\rm max}^\pm(N)-1}^\pm$ leads to
\begin{equation}
\beta_{N;k_{\rm max}^\pm(N) - 1}^\pm = \frac{\mathcal{J}_{N;k_{\rm max} - 1}^\pm}{\mathcal{E}_{N;k_{\rm max}^\pm(N)}^\pm - E}
\label{eq:pt.bn}
\end{equation}
and the ansatz $\psi_{N;k}^\pm= \beta_{N;k-1}^\pm \psi_{N;k-1}^\pm$ gives the recursion relation
\begin{equation}
\beta_{N;k - 1}^\pm = \frac{\mathcal{J}_{N;k-1}^\pm}{\mathcal{E}_{N;k}^\pm - E - \mathcal{J}_{N;k}^\pm \beta_{N;k}^\pm} 
\label{eq:pt.bj}
\end{equation}
for $k > 0$.

Equation \eqref{eq:pt.pk} can be rewritten as
\begin{equation}
\left( E - \mathcal{E}_{N;k}^\pm + \mathcal{J}_{N;k-1}^\pm \alpha_{N;k}^\pm + \mathcal{J}_{N;k}^\pm \beta_{N;k}^\pm\right) \psi_{N;k}^\pm = 0 \,.
\end{equation}
Excluding the trivial unphysical solution in which all the coefficients vanish, 
$E - \mathcal{E}_{N;k}^\pm + \mathcal{J}_{N;k-1}^\pm \alpha_{N;k}^\pm + \mathcal{J}_{N;k}^\pm \beta_{N;k}^\pm = 0$, yielding with Eqs.~\eqref{eq:pt.a1}--\eqref{eq:pt.bj}, 
\begin{equation}
    E = \mathcal{E}_{N;k}^\pm - \frac{(\mathcal{J}_{N;k - 1}^\pm)^2}{\mathcal{E}_{N;k-1}^\pm - E - \frac{(\mathcal{J}_{N;k-2}^\pm)^2}{ \mathcal{E}_{N;k-2}^\pm - E - \frac{(\mathcal{J}_{N;k-3}^\pm)^2}{ \cdots \frac{\cdots}{\mathcal{E}_{N;0}^\pm - E}}}} - \frac{(\mathcal{J}_{N;k}^\pm)^2}{\mathcal{E}_{N;k+1}^\pm - E - \frac{\mathcal{J}_{N;k + 1}^\pm)^2}{ \mathcal{E}_{N;k+2}^\pm - E - \frac{(\mathcal{J}_{N;k+2}^\pm)^2}{\cdots \frac{\cdots}{\mathcal{E}_{N;k_{\rm max}^\pm(N)}^\pm - E}}}} \,. \label{eq:pt.Ek}
\end{equation}
This equation exhibits $k_{\rm max}^\pm(N) + 1$ distinct solutions for even and odd parity, respectively, which together yield the $N + 1$ eigenvalues of the Hamiltonian.

For perturbatively small $\mathcal{J}_{N;k}^\pm$, it is possible to obtain from Eq.~\eqref{eq:pt.Ek} an approximate expression for the eigenvalue $E_{N;k}^\pm$ that is close to $\mathcal{E}_{N;k}^\pm$ in a rather straightforward manner, namely by consistently replacing occurrences of $E$ on the right-hand side of Eq.~\eqref{eq:pt.Ek} with the entire right-hand side of this equation and by performing then a Taylor series expansion in all $(\mathcal{J}_{N;k'}^\pm)^2$ up to the desired order.  Linear order in these coefficients (and thus second-order perturbation theory) gives the expected coefficients from an ordinary Rayleigh-Schr\"odinger perturbation theory aproach, i.e.,
\begin{eqnarray}
    E_{N;k}^\pm &\simeq &\mathcal{E}_{N;k}^\pm + \frac{(\mathcal{J}_{N;k-1}^\pm)^2}{\displaystyle \mathcal{E}_{N;k}^\pm - \mathcal{E}_{N;k-1}^\pm} + \frac{(\mathcal{J}_{N;k}^\pm)^2}{\mathcal{E}_{N;k}^\pm - \mathcal{E}_{N;k+1}^\pm}\nonumber \\
   & = & \frac{U}{2}\left( N(N-1)-2(N-k)k  + \frac{2J^2}{U^2}\left[1 + \frac{(k+1)^2}{N-2k-1}-\frac{k^2}{N-2k+1}\right] \right)  \,.
    \label{eq:pt.E2}
\end{eqnarray}
In the last form, the unity constant in square brackets has been separated out that cannot contribute to anything more than an irrelevant global phase in the quantum dynamics.  Recalling the definition of $\heff$, Eq.~\eqref{eq:heff}, the remaining correction terms turn out to be lower order, $O(\heff\gamma^2)$, in the sense that for $\heff \rightarrow 0$ the values of $N$ are clustered around the very large value of $\Nav$ and the values of $\{j,k\}$ ahead end up being the integers from zero up to some rather limited value, depending on $\gamma$ and $\Nav$, due to the tridiagonal nature of the perturbation.  For the analysis of all the revival measures, it turns out to be unnecessary to go further in the series for the eigenvalue corrections.  In addition, the perturbation corrections to the energy differences mentioned above, e.g.~$\mathcal{E}_{N;k}^\pm - \mathcal{E}_{N-1;k}^\pm\approx -U(2k+1)\gamma^2/4$, imply consistent results for the perturbative shifts in the classical frequencies $\omega_S$ and $\omega_m$.  The second energy difference is similar, but weaker by a factor $\heff$, $\approx -U(2k+1)\heff\gamma^2/4$, which implies small shifts in features near the revival time.

On the other hand, these expressions are extremely useful for generating higher order corrections to the eigenstates where needed.  The corresponding eigenvector is given by $(\psi_{N;0}^{(k)\pm},\ldots,\psi_{N;k_{\rm max}^\pm(N)}^{(k)\pm})$ with 
\begin{equation}
    \psi_{N;j}^{(k)\pm} = \psi_{N;k}^{(k)\pm} \left\{ \begin{array}{c@{\quad : \quad}l} \alpha_{N;j+1}^\pm \alpha_{N;j+2}^\pm \cdots \alpha_{N;k}^\pm & j < k \\ \beta_{N;j-1}^\pm \beta_{N;j-2}^\pm \cdots \beta_{N;k}^\pm & j > k \end{array} \right. \,,
    \label{eq:exactev}
\end{equation}
noting that $E$ is to be replaced by the eigenvalue $E_{N;k}^\pm$ when evaluating those coefficients via recursive applications of  Eqs.~\eqref{eq:pt.a1}--\eqref{eq:pt.bj}.
Proper normalization of the eigenvector yields
\begin{equation}
\psi_{N;k}^{(k)\pm} = \left( 1 + \sum_{j=0}^{k-1} \prod_{i=j+1}^{k} (\alpha_{N;i}^\pm)^2 + \sum_{j=k+1}^{k_{\rm max}^\pm(N)} \prod_{i=k}^{j-1} (\beta_{N;i}^\pm)^2 \right)^{-1/2} 
\label{eq:pt.renorm}
\end{equation}
up to an unimportant global phase factor.  The leading nonvanishing order of the individual coefficients in $J$ is
\begin{equation}
    \alpha_{N;j+1}^\pm \simeq \frac{\mathcal{J}_{N;j}^\pm}{\mathcal{E}_{N;j}^\pm - \mathcal{E}_{N;k}^\pm} = \frac{J\sqrt{(j+1)(N-j)}}{U(N-k-j)(k-j)}
    \label{eq:pt.alpha}
\end{equation}
for all $j \in\{ 1,\ldots,k\}$ (the final form excludes the exceptional cases requiring Eqs.~\eqref{eq:qpt.Epm} or \eqref{eq:qpt.Jp}).  For small $(j,k)$, the coefficient picks up an order $N^{-1/2}$ in magnitude. Similarly, 
\begin{equation}
    \beta_{N;j-1}^\pm \simeq \frac{\mathcal{J}_{N;j-1}^\pm}{\mathcal{E}_{N;j}^\pm - \mathcal{E}_{N;k}^\pm} = \frac{J\sqrt{j(N-j+1)}}{U(N-k-j)(k-j)}
    \label{eq:pt.beta}
\end{equation}
for all $j \in \{k,\ldots,k_{\rm max}^\pm(N)-1\}$ (with the same exclusion). This yields
\begin{equation}
\psi_{N;j}^{(k)\pm} \simeq \left\{ \begin{array}{c@{\quad : \quad}l} \displaystyle \frac{\mathcal{J}_{N;j}^\pm}{\mathcal{E}_{N;j}^\pm - \mathcal{E}_{N;k}^\pm}\frac{\mathcal{J}_{N;j+1}^\pm}{\mathcal{E}_{N;j+1}^\pm - \mathcal{E}_{N;k}^\pm} \cdots \frac{\mathcal{J}_{N;k-1}^\pm}{\mathcal{E}_{N;k-1}^\pm - \mathcal{E}_{N;k}^\pm} & j < k \\ \displaystyle 1 & j = k \\ \displaystyle \frac{\mathcal{J}_{N;j-1}^\pm}{\mathcal{E}_{N;j}^\pm - \mathcal{E}_{N;k}^\pm}\frac{\mathcal{J}_{N;j-2}^\pm}{\mathcal{E}_{N;j-1}^\pm - \mathcal{E}_{N;k}^\pm} \cdots \frac{\mathcal{J}_{N;k}^\pm}{\mathcal{E}_{N;k+1}^\pm - \mathcal{E}_{N;k}^\pm} & j > k \end{array} \right. 
\label{eq:pt.admix}
\end{equation}
in leading order.  More compactly,
\begin{equation}
\psi_{N;j}^{(k)} \simeq \left\{ \begin{array}{c@{\quad : \quad}l} \displaystyle \left(\frac{J}{U}\right)^{k-j}\frac{\sqrt{\binom{k}{j}\binom{N-j}{k-j}}}{(k-j)! \binom{N-k-j}{k-j}} \sim \left(\frac{J}{\sqrt{N}U}\right)^{k-j}\sqrt{\frac{1}{(k-j)!}\binom{k}{j}} & j < k \\ 1 & j=k \\ \displaystyle \left(\frac{-J}{U}\right)^{j-k} \frac{\sqrt{\binom{j}{k}\binom{N-k}{j-k}}}{(j-k)! \binom{N-2k-1}{j-k}} \sim \left(\frac{-J}{\sqrt{N}U}\right)^{j-k}\sqrt{\frac{1}{(j-k)!}\binom{j}{k}}  & j > k \end{array} \right. 
\label{eq:pt.admix2}
\end{equation}
The asymptotic form has an error of $O(1/N)$ multiplied by quadratic and linear terms of $(j,k)$.  

Note that using these  individual leading order coefficients leaves the eigenvectors unnormalized.   Similar to the leading normalization multiplication in Eq.~\eqref{eq:exactev}, except not using the exact normalizing factor of Eq.~\eqref{eq:pt.renorm}, in this case multiplying the coefficients by
\begin{equation}
\psi_{N;k}^{(k)} = 
\left[\left(1+\exp\left[\frac{J^2}{NU^2}\right]\right) 
L_k\left(-\frac{J^2}{NU^2}\right)-1\right]^{-1/2} \genfrac{}{}{0pt}{2}{\displaystyle \Longrightarrow}{N\rightarrow \Nav}
\left[\left(\frac{\gamma^2}{2\heff}\right)^k \frac{{\rm e}^{\frac{\gamma^2}{2\heff}}}{k!}\right]^{-1/2}
\label{eq:norm1}
\end{equation}
restores their normalizations to $O(1/N)$.  The asymptotic form above is for large argument.  The $L_k(x)$ are the ordinary Laguerre polynomials, but not in the usual domain of positive argument.

\subsection{Expectation values: occupancies}

The calculations just presented do not need to be carried out to unlimited order for our purposes, and thus it is possible to safely neglect the extremely subtle differences between the symmetric and antisymmetric subystems, which are solely induced by the Hamiltonian matrix elements that concern states with equal or almost equal site populations (namely Eq.~\eqref{eq:qpt.Epm} and Eq.~\eqref{eq:qpt.Jp} for odd and even $n$, respectively). The leading order expressions given above for the two subspaces generate essentially identical results to the needed orders here and the $\pm$ indicator can be dropped in the following.  

Consider first expectation values of one-body observables.  Of particular interest is the time-dependent site-2 occupancy of a propagated initial Glauber coherent state, Eqs.~\eqref{eq:alpha} and \eqref{eq:Glauber}, which displays the characteristic revival feature that is seen in Fig.~\ref{fig:hop}(a-c); the site-1 occupancy is just $n_1(t) = \Nav - n_2(t)$.  Given the particularly simple evaluation of the matrix element in the unsymmetrized Fock basis,
\begin{equation}
\bra{N-k;k}\hat{n}_2\ket{N'-k';k'} = k \times \delta_{k k'} \delta_{N N'} \,,
\end{equation}
and the near perfect equivalence of the results for the symmetric and antisymmetric representations, 
\begin{eqnarray}
    E_{N;k}^\pm & = & E_{N;k} \nonumber \\
    \psi_{N;k}^{\pm} & = & \psi_{N;k} = \frac{1}{\sqrt{2}}\braket{N-k,k|\Psi} \ ,
\end{eqnarray}
a little algebra leads to
\begin{eqnarray}
\label{eq:n2q}
n_2(t) &=& \bra{\alpha(t)}\hat n_2 \ket{\alpha(t)} \nonumber \\
&=& \sum_{N=0}^\infty |c_N|^2 \sum_{k=0} k \left| \sum_{k'=0} \psi_{N;0}^{(k')} \psi_{N;k}^{(k')}  e^{-i E_{N;k'}t / \hbar}\right|^2 \ .
\end{eqnarray}
From the above eigenvector rotation expressions, Eq.~\eqref{eq:pt.admix2}, the coefficients scale as
\begin{equation}
\psi_{N;k}^{(k')} \sim \left(\frac{J^2}{NU^2}\right)^{|k-k'|/2} \ ,\label{eq:pt.scaling2}
\end{equation}
which implies that the lowest order contribution to the $k^{th}$ term in Eq.~\eqref{eq:n2q} is this factor to the $k^{th}$ power with the $k'$ summation cutoff at $k$ ($k'>k$ terms are smaller since they are of higher order).  Inserting the explicit expressions gives
\begin{align}
&\braket{\hat n_2(t)} = {\rm e}^{-\Nav}\sum_{N=1}^\infty \frac{\Nav^N}{N!} \sum_{k=1}^{N} \left(\frac{J}{U}\right)^{2k} \frac{k}{k!^2 } \binom{N}{k}\left| \sum_{k^\prime=0}^{k} \frac{(-1)^{k'}\binom{k}{k^\prime}^2}{\binom{N-k^\prime}{k^\prime}\binom{N-2k^\prime-1}{k-k^\prime}}   {\rm e}^{-iE_{N,k^\prime} t/\hbar} \right|^2 
\label{eq:pt.n2}
\end{align} 
except that the appropriate eigenfunction normalizations, e.g.~\eqref{eq:pt.renorm}, must multiply this equation.  For the $k=0$ terms, it can be shown that the $k'>k$ terms are exactly cancelled by the result of the post-normalization, but they do not contribute due to the factor $k$ from the number operator.  Curiously, there are two seemingly nontrivial combinatorial identities implied this result and another similar one.  For all integer $N\ge k \ge 0$,
\begin{equation}
\delta_{k0}=\sum_{k^\prime=0}^{k} \frac{(-1)^{k'}\binom{k}{k^\prime}^2}{\binom{N-k^\prime}{k^\prime}\binom{N-2k^\prime-1}{k-k^\prime}} = \sum_{k^\prime=0}^{k} \frac{(-1)^{k'}}{N-k'} \binom{N-k^\prime}{k} \binom{k}{k^\prime} \ .
\end{equation}

The more global leading order evaluation of Eq.~\eqref{eq:n2q} restricts the summation to $k=1$.  This gives
\begin{equation}
\braket{\hat n_2(t)} = {\rm e}^{-\Nav}\sum_{N}^\infty \frac{\Nav^N}{N!} \left(\frac{2J}{U}\right)^{2} \frac{N}{(N-1)^2}   \sin^2\left(\frac{[E_{N;0} - E_{N;1}]t}{2\hbar}\right) \ .
\label{eq:pt.n3}
\end{equation} 
With Eq.~\eqref{eq:pt.E2}, the energy difference is found to be approximately
\begin{equation}
E_{N;0} - E_{N;1} \simeq (N - 1) U\left(1 - \frac{2J^2}{U^2} \frac{N+1}{(N-1)^2(N-3)} \right)\quad (N>3)\ .
\label{eq:enerdiff}
\end{equation}
In this form, the decomposition of the occupancy into a mean part (infinite temporal average), $\bar{n}_2$, and an oscillatory part $n_2^{\rm osc}(t)$ is straightforward, i.e., $\braket{\hat{n}_2(t)} = \bar{n}_2 + n_2^{\rm osc}(t)$.  By making a large-$\Nav$  expansion of the square of the relation
\begin{equation}
\frac{1}{N}=\frac{1}{\Nav\left(1+\frac{N-\Nav}{\Nav}\right)}
\label{eq:expnav}
\end{equation}
to quadratic order, the mean turns out to be
\begin{equation}
\bar{n}_2 \approx \frac{\gamma^2}{\heff}\left(1+\frac{3}{2}\heff\right)
\end{equation}
in terms of the perturbation parameter $\gamma$.  

There are a couple of straightforward ways to approximate the summation in Eq.~\eqref{eq:pt.n3}.  One can make use of the general identity
\begin{equation}
    \sum_{N=0}^\infty \frac{\Nav^N}{N!} \cos (N x + \varphi) = e^{\Nav \cos x} \cos(\Nav \sin x + \varphi)
    \label{eq:identity}
\end{equation}
which is valid for all $x,\Nav,\varphi\in\mathbb{R}$.  Or, one can replace the factorials by Stirling's approximation, convert the sum to an integral, and apply a saddle point approximation.  This leads to 3 saddles, which have simple forms as long as the $N^{-5/2}$ factor is not brought up into the exponential argument, thus left as a multiplying function out front, and Eq.~\eqref{eq:expnav} is only used to linear order in expanding Eq.~\eqref{eq:enerdiff}.  In this case, the solutions are
\begin{equation}
N=\Nav,\ \Nav{\rm e}^{ix}, \ \Nav{\rm e}^{-ix} \qquad x=\frac{Ut}{\hbar}\left(1+\frac{\gamma^2}{2}\right) \qquad \varphi = -2\gamma^2\frac{Ut}{\hbar}\left(\frac{1}{\heff}+1\right) \ .
\label{eq:saddle}
\end{equation}
Either approach finally leads approximately to
\begin{eqnarray}
\langle\hat n_2(t) \rangle &\approx& \frac{\gamma^2}{\heff}\left\{1- \exp\left[ -\frac{4}{\heff}\sin^2\left(\pi\tau\left[1+\frac{\gamma^2}{2}\right]\right)\right]\right. \nonumber \\ && \times \left. \cos\left[ \frac{2}{\heff}\sin\left(2\pi \tau\left[1+\frac{\gamma^2}{2}\right]\right) - 4\pi\gamma^2\tau\left(\frac{1}{\heff}+1\right) \right] \right\}
\label{eq:pt.n2pert}
\end{eqnarray}
with $\tau = t / \trevo = U t / (2\pi\hbar)$. This expression contains the essential behaviors seen in the oscillations seen in the occupancies near the revival time, as well as the leading dependence of the long time average.

Higher-order corrections to Eq.~\eqref{eq:pt.n2pert} are rather tedious to calculate for the parameters $\gamma$ and $\heff$ considered here.  As was already pointed out above, the perturbation series effectively features $\gamma^2/\heff$ and not $\gamma^2$ as a small parameter, even though perturbative corrections of the order $\nu$ are bound to scale as $\gamma^{2\nu}$ and not as $(\gamma^2/\heff)^\nu$, as can be inferred from the semiclassical analysis of this problem.
Precise bookkeeping of terms belonging to a given order $\nu$ is thus required for a proper  evaluation of such corrections.
Carrying out such a calculation in order to obtain the next-to-leading order correction to the temporal average of the site-2 population yields
\begin{equation}
\bar{n}_2  \simeq \frac{1}{\heff} \left(\gamma^2 + \frac{5}{4} \gamma^4 + O(\gamma^6) \right)
\end{equation}
in agreement with the classical evaluation of this quantity. 

\subsection{Coherence}

Whereas the coherence on site-1 is only marginally affected by the presence of the second site (see Fig.~\ref{fig:coh}), the site-2 coherence features similar oscillations as the site-2 population, with a background that appears to be subject to temporal damping as is seen in Fig.~\ref{fig:hop}(e).
Using
\begin{equation}
\bra{N';k';\sigma'}\hat{a}_2\ket{N;k;\sigma} = \frac12 \left( \sqrt{k} \,\delta_{k',k-1} + \sigma \sigma' \sqrt{N-k}\,\delta_{k',k}\right)\delta_{N',N-1}
\end{equation}
for all $k < N/2$ and $k' < N'/2$ as well as
\begin{eqnarray}
\bra{N;N/2;+}\hat{a}_2\ket{N';k;\sigma} & = & \sqrt{\frac{k+1}{2}} \sigma\delta_{N',N+1}\delta_{N,2k} \,, \\
\bra{N';k;\sigma}\hat{a}_2\ket{N;N/2;+} & = & \sqrt{\frac{k+1}{2}}\delta_{N',N-1}\delta_{N,2k+2}
\end{eqnarray}
for even $N$ and $k < N'/2$, the coherence amplitude $a_2(t) = \bra{\alpha_t}\hat{a}_2\ket{\alpha_t}$ on site-2 is exactly obtained as
\begin{align}
a_2(t) & =\sum_{N=0}^\infty c_{N-1}^* c_N \sum_{\sigma,\sigma'=\pm}\sum_{k,k''=0}^{k_{\rm max}^\sigma(N)}\sum_{k',k'''=0}^{k_{\rm max}^{\sigma'}(N-1)} \psi_{N;k}^{(k'')\sigma}\psi_{N;0}^{(k'')\sigma}\psi_{N-1;k'}^{(k''')\sigma'}\psi_{N-1;0}^{(k''')\sigma'} \nonumber \\
& \times\left[\frac{1}{4}\left(\sqrt{k}\,\delta_{k',k-1}+\sigma \sigma' \sqrt{N-k}\,\delta_{k',k}\right)(1-\delta_{N,2k})(1-\delta_{N,2k'+1})  \right. \nonumber \\
& \left. +\sqrt{\frac{k'+1}{8}}\left(\sigma\delta_{N,2k+1}\delta_{k,k'}\delta_{\sigma'+} + \delta_{k,k'+1}\delta_{N,2k}\delta_{\sigma+}\right)\right] e^{-i t (E_{N;k''}^{(\sigma)}-E_{N-1;k'''}^{(\sigma')})/\hbar}\,. 
\label{eq:exacta}
\end{align}
Neglecting the extremely tiny difference between even and odd eigenvalues, Eq.~\eqref{eq:exacta} can be simplified yielding
\begin{equation}
a_2(t) \simeq \sum_{N=1}^\infty c_{N-1}^* c_N \sum_{k,k',k''=0}^{[N/2]} \sqrt{k}\,\psi_{N;k}^{(k')}\psi_{N;0}^{(k')}\psi_{N-1;k-1}^{(k'')}\psi_{N-1;0}^{(k'')}e^{-i t (E_{N;k'}-E_{N-1;k''})/\hbar} \,. 
\label{eq:pt.a2}
\end{equation}
Similarly to Eq.~\eqref{eq:pt.n2}, the lowest order contributions to the $k$-summation terms comes from $k\ge k'$ and $k-1\ge k''$.  As before for the site-2 occupancy, here inserting the explicit expressions from Eqs.~\eqref{eq:pt.admix2} and \eqref{eq:Glauber} yields
\begin{align}
a_2(t) = &{\rm e}^{-\Nav}\sqrt{\Nav}\sum_{N=4}^\infty \frac{\Nav^{N-1}}{(N-1)!} \sum_{k=1}^{[N/2]} \left(\frac{J}{U}\right)^{2k-1} \frac{1}{k!(k-1)! } \binom{N-1}{k-1} \nonumber \\
& \times\sum_{k^\prime=0}^{k} \frac{(-1)^{k'}\binom{k}{k^\prime}^2}{\binom{N-k^\prime}{k^\prime}\binom{N-2k^\prime-1}{k-k^\prime}}    \sum_{k''=0}^{k-1} \frac{(-1)^{k''}\binom{k-1}{k''}^2}{\binom{N-1-k''}{k''}\binom{N-2k''-2}{k-1-k''}}   {\rm e}^{-it\left(E_{N,k'} - E_{N-1,k''}\right)/\hbar} 
\label{eq:pt.a2low}
\end{align} 
From this expression, the lowest order term arises for $k=1$, which considerably simplifies the expression to
\begin{equation}
a_2(t) \simeq \frac{J}{U}{\rm e}^{-\Nav}\sqrt{\Nav}\sum_{N=4}^\infty \frac{\Nav^{N-1}}{(N-1)!}\left(\frac{1}{N-1}\right)\left(e^{-i t (E_{N;0}-E_{N-1;0})/\hbar} - e^{-i t (E_{N;1}-E_{N-1;0})/\hbar}\right)\,. \label{eq:pt.a2lin}
\end{equation}
From Eq.~\eqref{eq:pt.E2} it follows that ($N\gg 1$)
\begin{eqnarray}
E_{N;1} - E_{N-1:0} \simeq \frac{J^2}{U}\left( \frac{2N+1}{(N-1)(N-3)}\right) \nonumber  \\
E_{N;0} - E_{N-1;0} \simeq U(N-1) - \frac{J^2}{U} \left( \frac{1}{(N-1)(N-2)}\right)\ .
\label{eq:ediffk}
\end{eqnarray}
Using the same techniques (saddle point) and relations (e.g.~Eqs.~\eqref{eq:expnav} and \eqref{eq:identity}), the sum can be expressed as
\begin{align}
& a_2(t) \simeq \frac{\gamma}{\sqrt{2\heff}}\times \nonumber \\
&\left\{ \exp\left[-\frac{4}{\heff}\sin^2 \left(\pi\tau \left[1+\frac{\gamma^2\heff}{4}\right]\right) - i\left(\frac{2}{\heff}- \frac{3}{2}\right)\sin \left(2\pi\tau \left[1+\frac{\gamma^2\heff}{4}\right]\right) + \frac{3}{2}\pi i \gamma^2 \tau\right] \right. \nonumber \\
&\left. -\exp\left[-\frac{4}{\heff}\sin^2 \left(\frac{\pi\gamma^2\tau}{2} \right) + i\frac{2}{\heff}\sin \left(\pi\gamma^2\tau\right) - i \pi \tau\gamma^2 \left(\frac{4}{\heff}+ 5\right)  \right]\right\}
\label{eq:cohamp}
\end{align}
The absolute square of this relation gives the coherence.  There are essentially three terms.  The first exponential term generates the rapid decay equal to 
\begin{equation}
\frac{\gamma^2}{2\heff}\exp\left(-\frac{8}{\heff}\sin^2 \left[\pi\tau \left(1+\frac{\gamma^2\heff}{4}\right)\right]\right)
\label{eq:decay1}
\end{equation}
whereas the other decay term,
\begin{equation}
\frac{\gamma^2}{2\heff}\exp\left(-\frac{8}{\heff}\sin^2 \left[\frac{\pi\gamma^2\tau}{2} \right]\right)
\label{eq:decay2}
\end{equation}
is much slower due to the presence of the $\gamma^2$ in the argument of the $\sin$-function.  The cross terms interfere to generate a $\cos$-function oscillation damped by the square root of the product of these two decay functions.

Although the absolute square of Eq.~\eqref{eq:cohamp} captures much of the coherence behavior rather faithfully, it turns out that the decay described in Eq.~\eqref{eq:decay2} is much too slow to account properly for the long decay to extremely small values seen in Fig.~\ref{fig:hop} (e).  Recalling Eq.~\eqref{eq:ediffk}, it is the absence of the leading $N$-dependence, $U(N-1)$, in the first energy difference that lengthens this decay.  In fact, there is a series of terms in Eq.~\eqref{eq:pt.a2low} that are missing a leading energy difference dependence in $N$.  There is a separation of terms contributing to the coherence amplitude, those with and those without a leading $U(N-1)$ energy difference dependence.  Thus, the coherence amplitude divides into its fast and slow components,
\begin{equation}
a_2(t) = a_{2,s}(t) + a_{2,f}(t) \ ,
\label{eq:pt.a2lin2}
\end{equation}
by grouping the terms into two summations.  Consider only $a_{2,s}(t)$, its contributions arise from the $k'' = k' - 1$ terms in Eq.~\eqref{eq:pt.a2low}.  From Eq.~\eqref{eq:pt.E2} the differences after a bit of algebra are
\begin{equation}
E_{N;k'} - E_{N-1;k'-1} \simeq U(k' - 1) + \frac{J^2}{U} \left( \frac{2}{N} + \frac{10 k'}{N^2}+O(N^{-3})\right)
\end{equation}
By including this whole set of terms, it is shown next that the dephasing due to the presence of $U(k'-1)$ speeds up the slow decay relative to the simple result of Eq.~\eqref{eq:decay2}, which is then more in line with the behavior seen in Fig.~\ref{fig:hop} (e). 

The amplitudes associated with these terms in Eq.~\eqref{eq:pt.a2} scale as $\sim (\frac{\gamma^2}{2\heff})^k$ for $N \simeq \Nav$ and are thus, strictly speaking, of higher perturbative order.  However, for the parameter sets considered here $\frac{\gamma^2}{2\heff}$ is of $O(1)$, which implies that these terms nevertheless matter.  This also requires a proper renormalization.  Before accounting for this renormalization, taking the large $\Nav$ limit (dropping terms of relative $O(N^{-1})$ in the magnitude), and performing some algebra and a stationary phase approximation, the slow part of Eq.~\eqref{eq:pt.a2low}, i.e.~the terms with the restriction $k''=k'-1$, can be reduced to
\begin{equation}
a_{2,s}(t) \approx \sqrt{\frac{\gamma^2}{2\heff}} \sum_{k=0}^{\infty}\frac{1}{k!} \left(\frac{\gamma^2}{2\heff} \right)^k \sum_{k'=0}^k \binom{k}{k'}\binom{k+1}{k'+1}\exp\left\{-\frac{itU}{\hbar}\left[\left(1+\frac{5\gamma^2}{2}\right) k'+\frac{\gamma^2}{\heff}\right]\right\}
\end{equation}
Using an integral form of a Kronecker $\delta$-function
\begin{equation}
\delta_{jk} = \frac{1}{2\pi}\int_{-\pi}^{\pi}{\rm d}\theta e^{i\theta(j-k)}
\end{equation}
and renormalizing appropriately gives an expression for the slow decay that depends on $\gamma^2/\heff$ instead of $\gamma^4/\heff$.  Keeping just the more significant terms, defining $\tilde{\tau} = (1 + 5\gamma^2/2)\tau$, and dropping pure phases leads to
\begin{equation} \label{eq:A55}
a_{2,s}(t)  \sim \sqrt{\frac{\gamma^2}{2\heff}} \left[\frac{I_0\left(\frac{\gamma^2}{\heff} e^{-i \pi \tilde{\tau}}\right) + I_1\left(\frac{\gamma^2}{\heff}e^{-i\pi\tilde{\tau}}\right)e^{i\pi\tilde{\tau}}}{I_0\left(\frac{\gamma^2}{\heff}\right) + I_1\left(\frac{\gamma^2}{\heff}\right)}\right]  e^{- \frac{\gamma^2}{\heff} \sin^2\pi\tilde{\tau}} \,,
\end{equation}
where $I_k(z)$ is the modified Bessel function of the first kind of order $k$ continued into the complex plane.  Using the asymptotic behavior
\begin{equation}
I_k(z) \simeq \frac{1}{\sqrt{2\pi\sqrt{z^2}}}\exp\left[\sqrt{z^2} - i k \pi (1 - \mathrm{sgn} \mathrm{Re}z )\right]
\end{equation}
valid for large $|z|\gg 1$, the leading decay becomes
\begin{eqnarray}
a_{2,s}(t) & \sim & \sqrt{\frac{\gamma^2}{2\heff}} \exp\left[- \frac{\gamma^2}{\heff} \left(\sin^2{ \pi \tilde{\tau}} + 2 \sin^2 \frac{\pi \tilde{\tau}}{2}\right)\right] 
\end{eqnarray}
for $\frac{\gamma^2}{\heff} > 1$.  

\subsection{Return probability}

\label{sec:pt.rp}

The above expressions for the eigenvalues and eigenvectors and techniques can be used to calculate the time evolution of the system's return amplitude $\braket{\alpha|\alpha_t}$.  Neglecting the extremely subtle differences between the even- and odd-parity energies and eigenvector coefficients as throughout this appendix,  the autocorrelation function can be expressed as
\begin{eqnarray}
\braket{\alpha|\alpha_t} &\simeq& \sum_{N=0}^\infty |c_N|^2 \sum_{k=0}^N (\psi_{N;0}^{(k)})^2 e^{-i E_{N;k} t/\hbar}
 \nonumber \\
&\simeq& {\rm e}^{-\Nav-\frac{\gamma^2}{2\heff}} \sum_{N=0}^\infty \frac{\Nav^N}{N!}\sum_{k=0}^N \left(\frac{J^2}{U^2}\right)^k\frac{\binom{N}{k}{\rm e}^{-iE_{N;k} t/\hbar}}{k!^2\binom{N-k}{k}^2}  \nonumber \\
&\simeq& {\rm e}^{-\Nav-\frac{\gamma^2}{2\heff}} \sum_{N=0}^\infty \frac{\Nav^N}{N!}\sum_{k=0}^N \frac{1}{k!} \left(\frac{J^2}{NU^2}\right)^k f(N,k){\rm e}^{-iE_{N;k} t/\hbar} \,. \\
f(N,k)&\simeq & \left(1+\frac{5k^2-k/2}{N}\right) \approx 1
\label{eq:pt.auto}
\end{eqnarray}
where the issue of renormalizing the eigenstate intensities is accounted for using Eq.~\eqref{eq:norm1} for $k=0$.  Inserting this expression into Eq.~\eqref{eq:pt.auto} (and neglecting the $N$-dependence of the upper bound of the $k$ summation, which can safely be carried out to infinity if it is supposed to converge quickly in a perturbative regime) generates the approximate expression for large $N$
\begin{equation}
|c_N|^2 (\psi_{N;0}^{(k)})^2 \simeq \frac{1}{\sqrt{2\pi\Nav}} \left(\frac{J^2}{\Nav U^2}\right)^{k} {\rm e}^{-\frac{\gamma^2}{2\heff}} f(N, k) e^{-(N - \Nav + k + 1/2)^2/(2 \Nav)} \,, \label{eq:pt.gauss}
\end{equation}
which implies
\begin{eqnarray}
\sum_{N=0}^\infty|c_N|^2 (\psi_{N;0}^{(k)})^2 e^{-i E_{N;k}t/\hbar} &\simeq& \frac{1}{\sqrt{2\pi\Nav}} \left(\frac{J^2}{\Nav U^2}\right)^{k} {\rm e}^{-\frac{\gamma^2}{2\heff}} f(\Nav - k - 1/2, k) \nonumber \\ 
&& \times \sum_{N=0}^\infty e^{-(N - \Nav + k + 1/2)^2/(2 \Nav)} e^{-i E_{N;k} t/\hbar} \,.
\label{eq:pt.gauss1}
\end{eqnarray}

Theoretically, the perturbation theory developed here properly converges for $1 \gg J^2 / (\Nav U^2) = \gamma^2 / ( 2 \heff)$.
Note, however, that this latter inequality is not necessarily fulfilled if $\gamma^2 \ll 1$ and $\heff \ll 1$ as well, with these two parameters being independently tuned. 
In particular, many of the examples discussed in Sec.~\ref{sec:RevInBHdimer} were studied for $\Nav = 200$ and $\gamma = 0.14$, thus implying $J^2 / (\Nav U^2) \simeq 1$.
Applying perturbation theory in order to calculate observables for such a parameter set is still possible but requires a meticulous book-keeping of all contributing terms $\sim (J/U)^{2k}$ of a given order $k$, with terms of higher power in $(J/U)^2$ being consequently removed from the calculation.
Mutual cancellations among these contributing terms then yields an effective scaling $\sim \gamma^{2k}$, and not $\sim [\gamma^2/(2\heff)]^k$, of the $k$th order term of the observable under consideration.  Instead a viable alternative to this procedure of a simple renormalization of the coefficients $(\psi_{N;0}^{(k)})^2$, Eq.~\eqref{eq:norm1}, was applied in arriving at Eq.~\eqref{eq:pt.auto} that ensures proper normalization of the autocorrelation function.

In leading order, differences of eigenenergies are not concerned by the above complication since they feature perturbative corrections that effectively scale as $\gamma^2$ and not as $\gamma^2/(2 \heff)$.  It is thus sufficient to replace the energies by the second-order expansion Eq.~\eqref{eq:pt.E2} given by $E_{N;k} \simeq \mathcal{E}_{N;k} + \delta E_{N;k}$ with
\begin{equation}
\delta E_{N;k} = \frac{J^2}{U}\left( 1 + \frac{(k+1)^2}{N - 2 k - 1} - \frac{k^2}{N - 2 k + 1} \right) \,. \label{eq:pt.dEnk}
\end{equation}
Using this approximation, the expression Eq.~\eqref{eq:pt.auto} allows one to reproduce the exact return probability in a remarkably faithful manner, as can be seen in Fig.~\ref{fig:max}.

Anticipating that $N$ is to be evaluated at $\Nav - k - 1/2$ according to Eqs.~\eqref{eq:pt.gauss} and \eqref{eq:pt.gauss1}, a quadratic expansion of $\delta E_{n;k}$ can be performed about $N=\Nav - k - 1/2$ yielding
\begin{eqnarray}
\delta E_{N;k} & \simeq & \frac{J^2}{U}\left[\left( 1 + \frac{(k+1)^2}{\Nav - 3 k - 3/2} - \frac{k^2}{\Nav - 3 k + 1/2}\right) \right.\nonumber \\ && - (N - \Nav + k + 1/2) \left( \frac{(k+1)^2}{(\Nav - 3 k - 3/2)^2} - \frac{k^2}{(\Nav - 3k + 1/2)^2} \right) \nonumber \\ && \left. + (N - \Nav + k + 1/2)^2 \left(\frac{(k+1)^2}{(\Nav - 3 k - 3/2)^3} - \frac{k^ 2}{(\Nav - 3 k + 1/2)^3} \right) \right] \,. \label{eq:dEexp}
 \end{eqnarray}
Focusing on the most dominant $k=0$ contribution, the perturbatively modified recurrence and revival times can be inferred.  By analogy with Eq.~\eqref{eq:expans-sec2} one finds
\begin{equation}
  E_{N;0} \simeq \Ea + \frac{2\pi \hbar}{\tosc} (N - \Nav + 1/2) + \frac{\pi\hbar}{\trev}(N - \Nav + 1/2)^2
\end{equation}
with 
\begin{eqnarray}
\frac{2\pi\hbar}{\tosc} & \simeq & (\Nav -1)U - \frac{J^2}{(\Nav - 3/2)^2U} \stackrel{\Nav \gg 1}{\simeq} (\Nav - 1 - \gamma^2/4)U\,, \\
\frac{2\pi\hbar}{\trev} & \simeq & U + \frac{2 J^2}{(\Nav - 3/2)^3 U} \stackrel{\Nav \gg 1}{\simeq} (1 + \heff \gamma^2 / 4) U\,.
\end{eqnarray}
The number of recurrence oscillations that the system carries out until the revival time is then approximately evaluated as
\begin{equation}
\trev / \tosc \simeq \Nav - 1 - 3 \gamma^2 / 4 + \heff \gamma^2 / 4
\end{equation}
in leading order in $\gamma^2$.
The hopping-induced shift of this ratio is thus of the order of $\gamma^2$ and can therefore be neglected for the parameters considered here.

In order to understand the origin of the oscillations of the revival peak height as a function of $\gamma$, as displayed in Fig.~\ref{fig:max}, recall that the peak shift is $O(\gamma^2\heff)$, which is quite small and can be neglected for the purposes here.  The maximal return probability is thus nearly attained at $t = \trevo = 2\pi\hbar/U$, where $e^{- i E_{N;k} t / \hbar} = e^{-2\pi i \delta E_{N;k}/U}$.  By neglecting the $k$-dependence in the denominators, using $k \ll \Nav$, this yields 
\begin{equation}
\delta E_{N;k} \simeq \frac{J^2}{U}\left( 1 + \frac{2 k + 1}{\Nav} \right) \,. 
\label{eq:pt.dEnk0}
\end{equation}
Inserting this approximation into Eq.~\eqref{eq:pt.gauss1} generates an $N$-summation, which can be evaluated via Gaussian integration, and the $k$-summation becomes an exponential series. This gives
\begin{equation}
|\langle\alpha|\alpha_{\trev}\rangle|^2 \simeq \exp\left( - \frac{2 \gamma^2}{\heff} \sin^2 \frac{\pi \gamma^2}{\heff}\right).
\end{equation}
As is clearly confirmed in Fig.~\ref{fig:max}, the revival peak height thus attains maxima near integer values of $\gamma^2 / \heff$ and minima near half-integer values of $\gamma^2 / \heff$, with the first minimum exhibiting $e^{-1}$ as peak height. 
Its oscillation as a function of $\gamma^2$ is then attributed to hopping-induced mismatches of the phases $E_{N;k} \trev / \hbar$ (mod $2\pi$) associated with the eigenstates, which in leading perturbative order scale linearly with $k$.  Thus, $\delta E_{N;k} \trev / \hbar \sim 4\pi k  J^2 / (\Nav U^2)$, and are in phase again if $2 J^2 / (\Nav U)$ is an integer value.

Higher-order corrections with respect to these latter approximations yield, in addition, a globally decaying envelope of the revival peak height, implying that the peak height maxima do not attain unity in reality.  The result, Eq.~\eqref{eq:dEexp}, inserted into the expression, Eq.~\eqref{eq:pt.gauss1}, for the autocorrelation function and evaluated at a time $t$ that is very close to the unperturbed revival time $\trevo$ is sufficient.
Defining 
\begin{equation}
\delta\tau = 1 - t / \trevo
\end{equation}
with $|\delta\tau| \ll 1$ as well as
\begin{subequations} \label{eq:angles}
\begin{align}
\frac{\varphi_k}{2\pi} & = \frac{1}{2} (\Nav - k - \frac12)(\Nav - k - \frac32) \delta \tau
+ \frac{\gamma^2}{4}  \left(\Nav^2 + \Nav (2k+1)\right)(1+\delta \tau) \,, 
\label{[eq:pt.phik} \\
\frac{\Omega_k}{2\pi} & =  (\Nav - 2 k - 1) \delta\tau -  \frac{\gamma^2}{4} (2 k + 1)(1+\delta\tau) \,, \label{eq:pt.Omk} \\
\frac{\Delta_k}{2\pi} & = \delta \tau + \frac{\gamma^2}{2\Nav} (2k + 1) (1+\delta\tau) \,, \label{eq:pt.Dek}
\end{align}
\end{subequations}
Eq.~\eqref{eq:pt.gauss1} is rewritten as
\begin{eqnarray}
\braket{\alpha|\alpha_t} & \simeq & \frac{e^{-\gamma^2/(2 \heff)}}{\sqrt{2\pi \Nav}} \sum_{k=0}^\infty \frac{1}{k!} \left(\frac{\gamma^2}{2 \heff}\right)^{k} e^{-i \varphi_k} \nonumber \\ && \times \sum_{N=-k}^\infty \exp\left[-\frac{1}{2}\left(\frac{1}{\Nav} + i \Delta_k\right)\left(N - \Nav + \frac{1}{2}\right)^2 - i \Omega_k\left(N - \Nav + \frac{1}{2}\right)\right]
\end{eqnarray}
and can be evaluated via Poisson summation yielding
\begin{eqnarray}
\braket{\alpha|\alpha_t} & \simeq & e^{-\gamma^2/(2 \heff)} \sum_{l=-\infty}^\infty e^{4\pi i l(1 - \heff/4)/\heff} \nonumber \\ && \times \sum_{k=0}^\infty \frac{1}{k!} \left(\frac{\gamma^2}{2 \heff}\right)^{k} \frac{e^{-i \varphi_k}}{\sqrt{1 + 2i \Delta_k/\heff}} \exp\left(- \frac{(\Omega_k - 2\pi l)^2}{\heff + 2i \Delta_k}\right) \,. 
\label{eq:pt.autopoisson} 
\end{eqnarray}

The maximal revival peak time is approximately defined as the time where $\Omega_0 = 0$, yielding, with Eq.~\eqref{eq:pt.Omk}, the shift
\begin{equation}
\delta\tau \simeq \frac{\gamma^2\heff}{8} \,.
\end{equation}
At that time Eqs.~\eqref{[eq:pt.phik}--\eqref{eq:pt.Dek} are evaluated as
\begin{subequations}
    \begin{eqnarray}
    \varphi_k/(2\pi) & \simeq & \frac{\gamma^2}{\heff^2} \left(1 + \heff k  
    + \frac{\heff^2k^2}{8}\right) \,, \\
    \Omega_k/(2\pi) & \simeq & - k \gamma^2/2 \,, \\
    \Delta_k/(2\pi) & \simeq & ( k + 3/4) \frac{\gamma^2\heff}{2} \,,
    \end{eqnarray}
\end{subequations}

and in the autocorrelation function, Eq.~\eqref{eq:pt.autopoisson}, the denominator can be simplified to
\begin{equation}
\heff+2i\Delta_k = \heff\left(1+4\pi i\gamma^2 [k+3/4]\right) \simeq \heff\ .
\end{equation}
Restricting the $l$-summation to the central-peak term $l=0$, Eq.~\eqref{eq:pt.autopoisson} is then evaluated at the maximal revival peak time $t_{\rm peak}$ as
\begin{equation}
\braket{\alpha|\alpha_{t_{\rm peak}}} \simeq e^{-\gamma^2/(2 \heff) - 2 \pi i \gamma^2 / \heff^2} \sum_{k=0}^\infty \frac{1}{k!} \left(\frac{\gamma^2}{2 \heff} e^{-2\pi i \gamma^2 / \heff} \right)^{k}e^{-k^2(\pi^2 \gamma^4/\heff+i\pi\gamma^2/2)} \,,
\end{equation}
which yields the return probability
\begin{equation}
|\braket{\alpha|\alpha_{t_{\rm peak}}}|^2 \simeq e^{-\gamma^2/\heff} \left| \sum_{k=0}^\infty \frac{1}{k!} \left(\frac{\gamma^2}{2 \heff} e^{-2\pi i \gamma^2 / \heff} \right)^{k}e^{-k^2(\pi^2 \gamma^4/\heff+i\pi\gamma^2/2)}\right|^2 \,.
\end{equation}
An approximate evaluation of this expression can be obtained using the general relation
\begin{equation}
    \sum_{k=0}^\infty \frac{\xi^k}{k!}e^{-\eta k^2} = \frac{1}{\sqrt{\pi}}\int_{-\infty}^\infty \exp\left(-x^2 + \xi e^{2 i \sqrt{\eta} x}\right) dx \simeq \frac{e^{\xi - \eta \xi^2}}{\sqrt{1 + 2 \eta \xi}} \,,
\end{equation}
for $\xi,\eta\in\mathbb{C}$ with $|\eta| \ll 1$ and $\mathrm{Re}(\eta) > 0$.
This yields
\begin{equation}
|\braket{\alpha|\alpha_{t_{\rm peak}}}|^2 \simeq \frac{\exp\left( - \frac{2 \gamma^2}{\heff} \sin^2 \frac{\pi \gamma^2}{\heff}- \frac{\pi^2 \gamma^8}{2\heff^3} \cos4\pi\frac{\gamma^2}{\heff}\right)}{\sqrt{1 + \frac{2\pi^2 \gamma^6}{\heff^2}\cos2\pi\frac{\gamma^2}{\heff} + \frac{\pi^4\gamma^{12}}{\heff^4}}} \,,
\end{equation}
valid for $(\pi^2 \gamma^4/\heff)^2 + (\pi \gamma^2/2)^2 \ll 1$, and clearly displays an attenuation of the revival peak height oscillations as a function of $\gamma$.

\section{Quantum to classical frequencies}
\label{app:qcf}

It is possible to use certain quantum mechanical energy differences and second differences to derive various frequencies, i.e.~$\omega=\Delta E/\hbar$, and the shearing rates, respectively, which can then, in principle, be compared after the appropriate scaling directly with the classical (mean field) frequencies and shearing rates.  For example, the scaling between the classical frequencies, $\Omega$, and the quantum frequencies, $\Omega^{(q)}$, is given by
\begin{equation}
\label{eq:cqscale}
\Omega = \frac{\hbar\heff}{U}\Omega^{(q)}\ .
\end{equation}
The eigenenergies given in Appendix~\ref{app:qpt} are labeled $E_{N;k}$ and their expression as a function of the two quantum numbers $(N,k)$ is given in Eq.~\eqref{eq:pt.E2}to second order in perturbation theory.  With $S=s_1+s_2$ and $m=s_2$, the desired discrete partial derivatives centered at $E_{N;k}$ are given by
\begin{align}
\label{eq:quantfreq}
\left(\Delta E_{N;k}\right)_S  
= &\frac{E_{N+1;k}-E_{N-1;k}}{2}\nonumber \\ 
= & U\left(N-k-\frac{1}{2}-\frac{J^2}{U^2(N-2k)}\left[\frac{(k+1)^2}{N-2k-2}-\frac{k^2}{N-2k+2}\right]\right) \nonumber \\
\left(\Delta E_{N;k}\right)_m = & \frac{E_{N;k+1}-E_{N;k-1}}{2} \nonumber \\ =&-U\left(N-2k +
\frac{J^2}{2U^2}\left[\frac{(k+2)^2}{N-2k-3} - \frac{(k+1)^2}{N-2k-1} - \frac{k^2}{N-2k+1} + \frac{(k-1)^2}{N-2k+3}\right]\right) 
\ .
\end{align}
Applying the scaling of Eq.~\eqref{eq:cqscale} and the quantization conditions, $S_N =  \heff  (N+1)$, $m_k =  \heff(k+\frac12)$, and dropping terms of $O(\heff^2)$ gives the relations 
\begin{equation} 
\label{eq:Omegas00}
 \Omega_{N,k} =\begin{pmatrix}
        S_N - m_k - \heff - \frac{2\gamma^2m_k}{\left(S_N-2m_k\right)^2}\left[1+\frac{2m_k}{S_N-2m_k}\right]\\
        - \left[S_N - 2m_k\right]+\frac{2\gamma^2}{S_N-2m_k}\left[1+\frac{4m_k}{S_N-2m_k}+\frac{4m_k^2}{\left(S_N-2m_k\right)^2}\right]
    \end{pmatrix}
\end{equation}
Proceding in a similar way, the second differences with respect to $S_N$ and $m_k$ dropping $O(\heff^2)$ terms are given by
\begin{align} 
\label{eq:dOmegas00}
&\partial \Omega_{N,k} = \nonumber\\
&\begin{pmatrix}
1 +\frac{4\gamma^2m_k}{\left(S_N-2m_k\right)^3}\left[1+\frac{3m_k}{S_N-2m_k}\right] & -1 - \frac{2\gamma^2}{\left(S_N-2m_k\right)^2}\left[1+\frac{8m_k}{S_N-2m_k}+\frac{12m_k^2}{\left(S_N-m_k\right)^2}\right] \\
-1 - \frac{2\gamma^2}{\left(S_N-2m_k\right)^2}\left[1+\frac{8m_k}{S_N-2m_k}+\frac{12m_k^2}{\left(S_N-m_k\right)^2}\right] & 2 + \frac{12\gamma^2}{\left(S_N-2m_k\right)^2} \left[1+\frac{4m_k}{S_N-2m_k}+\frac{4m_k^2}{\left(S_N-2m_k\right)^2}\right]
\end{pmatrix} .
\end{align}
At this point, dropping the $N,k$ subscripts gives frequencies and shearing rates that must correspond to an expansion of the classical relations, i.e.~$\Omega_{N,k}\to \Omega(S,m)$ and $\delta\Omega_{N,k}\to \delta\Omega(S,m)$.

In Sec.~\ref{sec:freqshear}, the frequencies and shearing rates are given for the north-pole trajectory expanded to include only $\gamma^2$ corrections, i.e.~Eqs.~\eqref{eq:Omegas}-\eqref{eq:Omegas1}.  These same relations follow from Eqs.~\eqref{eq:Omegas00} and \eqref{eq:dOmegas00} by dropping the $N,K$ subscripts and substituting $S=\Sa$ and $m=\gamma^2/\Sa$, expanding and dropping all terms of order greater than $\gamma^2$.  The consistency of the classical frequencies and derivatives with those derived from the quantum eigenvalues assures a consistency between semiclassical and quantum perturbative approaches to the order being evaluated.

\section{Wigner function of a general wave packet}
\label{app:wigner-function}

For a generalized version $|\alpha\rangle$ of Eq.~\eqref{eq:cohquad}, the Wigner transform is given by
\begin{equation}
\label{eq:wtgauss-gen}
W_\alpha(\mathbf{q},\mathbf{p})= \left(\pi \hbar \right)^{-D} \exp \left[ - \left(\vec p - \vec p_\alpha, \vec q - \vec q_\alpha \right) \cdot \frac{{\bf A}_\alpha}{\hbar} \cdot \left(\vec p - \vec p_\alpha, \vec q - \vec q_\alpha \right) \right] 
\end{equation}
where ${\bf A}_\alpha$ is
\begin{equation}
\label{mvg}
{\bf A}_\alpha = \left(\begin{array}{cc}
{\bf  c^{-1}} & {\bf  c}^{-1} \cdot {\bf  d}  \\
 {\bf  d} \cdot {\bf c}^{-1} & {\bf c} + {\bf  d} \cdot {\bf c}^{-1} \cdot {\bf  d} \end{array}  
\right)   \qquad {\rm Det}\left[ {\bf A}_\alpha \right] =1 \ ,
\end{equation}
and whose matrix elements are determined by a wave packet/coherent state covariance matrix,
\begin{equation}
\label{mvgwf}
{\bf b}_\alpha = {\bf c} + i {\bf d} \ ,
\end{equation}
which contains all the shape information and which also must be positive definite and symmetric.  Without squeezing, in the coherent state context $\mathbf{A_\alpha}=\mathbb{1}$, and we recover the simple form Eq.~\eqref{eq:wtgauss}.
The $2D \times 2D$ dimensional matrix ${\bf A}_\alpha$ is real and symmetric.  If $\bf b_\alpha$ is real, there are no covariances between $\mathbf{p}$ and $\mathbf{q}$ (${\bf d}$ vanishes); i.e.~the wave packet is not chirped.  The off-diagonal blocks of the matrix ${\bf A}_\alpha$ disappear. 

\section{Quadratures ordering}
\label{app:ordering}
 
This appendix provides some detailed discussion about the $\heff$ corrections  that should be included in the {\em classical symbol} associated with a quantum operator and, in particular, in the classical symbol $H_{\rm cl}(q,p)$ associated with a quantum Hamiltonian $\hat H$ (the discussion here is performed in the context of the Bose-Hubbard problem, but would remain unchanged, up to the transmutation  $\heff \to \hbar$, for a traditional quantum mechanical problem).

 We first stress that it is necessary to be able to evolve initial states for times of the order of $\ttrev = O(\heff^{-1})$.   Given an eigenstate $\ket{n}$ of energy $E_n$, for the phase $(2\pi /\heff) E_n \ttrev$ to be computed up $O(\heff^0)$, corrections up to $O(\heff)$ in $E_n$ must be included, which implies that the details of EBK quantization (including Maslov indices, etc..) are taken into account, and also that the classical symbol $H_{\rm cl}(q,p)$ includes the first order corrections in $\heff$.

As mentioned in Sec.~\ref{sec:monomer}, the derivation of the WKB approximation~\cite{Maslov81} (and thus of the EBK quantization conditions) assumes that the classical symbol $H_{\rm cl}(q,p)$ associated with a quantum Hamiltonian $\hat H$ is of the form
\begin{equation} \label{eq:app-classical-symbol}
  \hat H = \frac12 \left( H_{cl} (\overleftarrow{\hat q},\overrightarrow{\hat p}) 
   + H_{cl} (\overrightarrow{\hat q},\overleftarrow{\hat p})  \right) \; . 
\end{equation}
In other words, the quantum $\hat H$ should be the half-sum of the quantized $H$ for which all $\hat q$'s have been pushed to the left and thus all $\hat p$'s have been pushed to the right, and vice versa.  

\subsection{Ordering for the perturbed monomer}

 Consider first the Hamiltonian for the perturbed Bose-Hubbard monomer, Eq.~\eqref{eq:cubicBHmonomer}, which reads $\widehat{H}_{\rm mono} = (U/2) \widehat{H}^{(2)}_{\rm mono} + (\epsilon U/6) \widehat{H}^{(3)}_{\rm mono}$ 
 with 
 \begin{equation*}
    \widehat{H}^{(2)}_{\rm mono} \equiv      {\widehat{n}(\widehat{n}-1)}
    \qquad  \widehat{H}^{(3)}_{\rm mono}  =  \left( \widehat{n} + \frac12 \right)^3 \,  \, .
\end{equation*}
Using the definition, Eq.~\eqref{eq:quadratures}, of the quadratures for the quadratic term, and permuting the operators to reach the form of Eq.~\eqref{eq:app-classical-symbol}, this reads
\[
\hat H^{(2)}_{\rm mono} = \frac{1}{  \heff^2} \left[ 
\frac14 (\hat q^4 + \hat p^4 + \hat q^2 \hat p^2 + \hat p^2 \hat q^2 ] ) - \heff (\hat q^2 +\hat p^2 ) + \frac34 \heff^2 \right] \; . 
\]
Dropping the irrelevant constant $\frac34 \heff^2 $, and denoting the simple harmonic oscillator action variable as $s(q,p) \equiv \frac12 (q^2+p^2)$, the appropriate classical symbol for $\hat H^{(2)}_{\rm mono}$ is thus
\[
H^{(2)}_{\rm mono} = \frac{1}{2 \heff^2} \left[ s(q,p)( s(q,p) - 2 \heff )  \right] \; .
\]

Moving to the cubic perturbation term, it is expressed in terms of the quadratures as
\begin{equation}\label{eq:n3}
    \hat H^{(3)}_{\rm mono} = \frac{1}{8 \heff^3}   \left[\left( \widehat{p}^6 + \widehat{p}^4 \widehat{q}^2 + \widehat{p}^2 \widehat{q}^4 \right) + \left( \widehat{q}^6 + \widehat{q}^4 \widehat{p}^2 + \widehat{q}^2 \widehat{p}^4 \right) + \left[ \widehat{p}^2 \widehat{q}^2 \widehat{p}^2 + \widehat{q}^2 \widehat{p}^2 \widehat{q}^2 \right]\right] \; .
\end{equation}
Terms within parenthesis in the above fulfill the requirement for EBK quantization, but the ones within brackets do not. To symmetrize them, move either position or momentum to the right or left by employing the relations 
\begin{align}
    \widehat{q}^2 \widehat{p}^2 = \widehat{p}^2 \widehat{q}^2 + 4 i \heff \widehat{p} \widehat{q} - 2 \heff^2 \quad \Longleftrightarrow \quad \widehat{p}^2 \widehat{q}^2 = \widehat{q}^2 \widehat{p}^2 - 4 i \heff \widehat{q} \widehat{p} - 2 \heff^2 \; ,
\end{align}
which were obtained through five consecutive substitutions of the canonical commutation relations. Thus, an EBK-friendly form for the first term in the brackets of Eq.~\eqref{eq:n3} is
\begin{align}
    \widehat{p}^2 \widehat{q}^2 \widehat{p}^2 = \frac{1}{2} \left[ \left(  \widehat{p}^4 \widehat{q}^2 + \widehat{q}^2 \widehat{p}^4 \right)  + 4 i \heff \left(  \widehat{p}^3 \widehat{q} - \widehat{q} \widehat{p}^3 \right) - 2\heff \widehat{p}^2 \right] \; .
\end{align}
It is straightforward to derive a similar identity for $\widehat{q}^2 \widehat{p}^2 \widehat{q}^2$. Plugging the resulting expressions in Eq.~\ref{eq:n3} gives a form that is ready to be mapped to its classical symbol:
\begin{equation*}
    \hat H^{(3)}_{\rm mono} \longmapsto  H^{(3)}_{\rm mono} = \frac{1}{\heff^3} \left( s^3 - \frac{1}{4}\heff^2 \, s\right) \; ,
\end{equation*}
so that the classical Hamiltonian corresponding to \eqref{eq:cubicBHmonomer} is
\begin{equation}
    H(s) = \frac{U}{\heff^2} \left\{ \frac{s(s-2\heff)}{2} + \frac{\varepsilon}{\heff} \left[ \frac{s(s^2 - 4 \heff^2)}{6} \right] \, \right\} \, .
\end{equation}
Performing the scaling of Eq.~\eqref{eq:rescaling-Et} generates the classical Hamiltonian, Eq.~\eqref{eq:monomer_classical}, for which the corrections of $O(\heff)$ are explicit.

\subsection{Ordering for the dimer}

The Hamiltonian for the Bose-Hubbard dimer reads 
\[ \widehat{H}_{\rm di} = (U/2) \left(\widehat{H}^{(2)}_{\rm mono}(\hat n_1)  +  \widehat{H}^{(2)}_{\rm mono}(\hat n_2) \right) 
- J \widehat{H}^{(J)}_{\rm di} 
 \]
with
\[
\widehat{H}^{(J)}_{\rm di} =  \left( \hat{a}_1^\dagger \hat{a}_2 + \hat{a}_2^\dagger \hat{a}_1 \right) \; .
\]
The ordering for the $\widehat{H}^{(2)}_{\rm mono}$ terms is performed in exactly the same way as for the monomer, and since the $(\hat{a}_1^\dagger, \hat{a}_1)$ commute with the $(\hat{a}_2^\dagger, \hat{a}_2)$, there are no ordering corrections for $\widehat{H}^{(J)}_{\rm di}$.
Therefore,
\begin{equation} 
\label{eq:app-ham2}
    H_{\rm di} = \frac{U}{2\heff^2} \left[ s_1(s_1-2\heff) + s_2(s_2-2\heff) \right] + \frac{2 J}{\heff} \sqrt{s_1 s_2} \cos(\theta_1 - \theta_2) \quad . 
\end{equation}

\subsection{Ordering for the observables}

The issue of quadrature operator ordering appears not only for the Hamiltonian, but also for the observables (here for instance $\hat n_2$ and $\hat a_2$) whose averages need to be computed.  From a general point of view, for some observable $\hat B (\hat \bp, \hat \bq)$, the quantum expectation value $\mean{B} = \bra{\psi} \hat B \ket{\psi}$  is expressed exactly in term of the Wigner function $W_\psi(\bq,\bp)$ and the Weyl symbol $\BW(\bq,\bp)$  as (see Eq.~\eqref{eq:w-meanB})
\[
\mean{B} = \int d\bq \, d\bp \, W_\psi(\bq,\bp) \BW(\bq,\bp) \; .
\]
Since this equation forms the basis of the semiclassical approach, it is natural to define the classical symbol associated with the observable $\hat B$ as its Weyl symbol $\BW(\bq,\bp)$, which in practice amounts to replacing the quantum $(\hat q,\hat p)$ with their classical symbol $(q,p)$ in the form of $\hat B (\hat q,\hat p)$ which is fully symmetrized. This is slightly different from the prescription of Eq.~\eqref{eq:app-classical-symbol}.  However, for the case considered here, $\hat B$ is either $ \hat q$ or $\hat a_2 = (1/\sqrt{2\heff}) (\hat q_2 + i \hat p_2)$, which are linear in the quadratures, and no ordering issues arise, or $\hat B = \hat n_2$ is quadratic and both procedures result in exactly the same classical symbol, Eq.~\eqref{eq:nop}.  For higher order observables, the two procedures might give rise to $0(\heff^2)$ differences, which however are anticipated to be negligible.

\section{Discussion of the -1/2 shift}
\label{sec:shift}

As mentioned in Sec.~\ref{sec:rev-1d-quant}, the expansion, Eq.~\eqref{eq:expans-sec2}, is traditionally done around an integer $\na = n_0$~\cite{Nauenberg90,Bluhm96,Robinett04}.  Truncated to second order that gives 
\begin{equation}
\label{eq:expans-usual}
  E_n = E_{n_0} +  \frac{2\pi \hbar}{\tosc^{n_0}} (n - n_0) + \frac{2 \pi\hbar}{\trev^{n_0}}\frac{(n - n_0)^2}{2}   + O[(n - n_0)^3] \ ,
\end{equation}
where ${2\pi}/{\tosc^{n_0}}  =(E_{n_0+1}-E_{n_0-1})/2\hbar$, and ${2\pi}/{\tosc^{n_0}} = (E_{n_0+1}+E_{n_0-1}-2E_N)/\hbar$, (the discreet derivatives being exactly equal to the true ones if terms beyond the second order are neglected).  However, applying the semiclassical picture of Sec.~\ref{sec:semic-1d} to the Bose-Hubbard monomer in Sec.~\ref{sec:monomer} for an initial state $\ket{z_\alpha}$ with an integer average number of particle $\Nav = |z_\alpha|^2 \in \mathbb{N}$ naturally leads to an expansion near $\na = \Nav -1/2$, which is half-integer.  Here, three legitimate questions are answered:  i) what is the origin of the $-1/2$ integer shift?;  ii) is this expansion at integer values of $\na$ or not irrelevant for the final results?; and, iii) if it gives the same final answer, why not stick to the usual practice of choosing $\na$ as an integer?

\subsection{Origin of the -1/2 shift}

The origin of this $-1/2$ shift is connected to distinctions between the semiclassical pictures provided by Weyl-Wigner formalism and EKB quantization.  First, it is quite natural to expand the Hamiltonian around the action $s_\alpha = (q_\alpha^2 + p_\alpha^2)/2 = \heff \Nav$ where the Wigner transform of the initial state is maximal.  However, because the Maslov indices are equal to 2, the EBK quantization condition reads 
\begin{equation}
E_n = H(s_n) \; , \qquad s_n = \heff(n+1/2) \; .
\end{equation}
If $\Nav$ is an integer, $s_\alpha$ falls right in the middle of two quantizing actions, which translated into the variable $n$ leads to $\na$ half integer.

Another way to understand the origin of this shift is to remember that  the Weyl symbol of the operator $\hat n$ is given by  Eq.~\eqref{eq:nclass}
\begin{equation}
\nW (q,p) = \frac{(q^2 + p^2)}{2\heff} - \frac12 = \frac{s(q,p)}{\heff} - \frac12 \; .
\end{equation}
Therefore $\nW(s_\alpha) = \Nav - \frac12$, when because of the Maslov index $\nW(s_n) = n$.   The EBK-quantized tori are located where $\nW$ is quantized to integer values, but a coherent state with an integer average number of particle $\Nav$ is centered at a point of phase space where $\nW$ is $-\frac12$ smaller.  Bearing in mind that the state $\ket{\zee_\alpha}$ is such that $(\mean{q} \smeq  q_\alpha, \mean{p} \smeq  p_\alpha)$, and that by definition of the variance $\mean{q^2} = \mean{q}^2 + {\rm var}(q)$ (idem for $\hat p$), $\nW(s_\alpha)$ has to be slightly below $\Nav$, with the small shift $-1/2$ being seen as the consequence of the fact that coherent states are the quantum states with the smallest possible variance of $\hat q$ and $\hat p$ (minimum uncertainty).  Starting from the  wave packet/coherent state Wigner function, Eq.~\eqref{eq:W0-aa}, we indeed get (with $s(q,p) = \frac12 (q^2+p^2)$, $s_\alpha = s(q_\alpha,p_\alpha)$)
\begin{equation}
\braket{s}_\alpha = 
     \frac{1}{\pi\heff} \int dq dp \, s(q,p) \, 
    \exp \left[ - \frac{1}{\heff}\left((q - q_\alpha)^2 + (p - p_\alpha)^2 \right) \right]  
    = s_\alpha + \frac{\heff}{2}
\end{equation}
This mechanism also justifies that one should not take into account the 1/2 shift in Eq.~\eqref{eq:nclass} for Eqs.~\eqref{eq:TWAn2-base}-\eqref{eq:BOOTn2-base}.

\subsection{Checking consistency of results}

Assuming that a coherent state contains an integer average number of particles, $\Nav = n_0 \in \mathbb{N}$, it is not, of course, expected that the decision to expand near $n_0 \in \mathbb{N}$ or near $\na = n_0 -1/2$ can change the final results.  So let's check this.  Neglecting terms of order three and higher in the expansions Eqs.~\eqref{eq:expans-sec2} and \eqref{eq:expans-usual} implies that their second order coefficients $ 2\pi\heff /\trev$ are  identical, and thus $\trev^\alpha = \trev^{n_0}$.  On the other hand, the first order coefficients are slightly shifted leading to 
\begin{equation}
    \frac{1}{\tosc^{n_0}} -   \frac{1}{\tosc^{\alpha}}  =   \frac{\delta \na}{\trev}  \; ,
\end{equation}
with $\delta \na = \na - n_0 = -1/2$.  This shift of  $1/\tosc$ is $O(\heff)$ and therefore $t/\tosc$ will not be affected for short times.  However, if $t \sim \trev = O(\heff^{-1})$, the difference between $t/\tosc^{n_0}$ and $t/\tosc^\alpha$ would be of order one.
This difference is however exactly what is taken into account by the extra phase $\Da$, Eq.~\eqref{eq:DeltaTh}.  It can be checked that Eq.~\eqref{eq:Etrev} remains unchanged if $\tosc^{n_0}$ with $\delta \na=0$ or $\tosc^\alpha$ with the actual $\na$ is used.  Any other differences induced by $\tosc^{n_0}$ being slightly different from $\tosc^\alpha$ are negligibly small in $\heff$.

\subsection{Why bother with the -1/2?}

So the question is: why not do what physicists have always done, i.e.~assume $\na \in \mathbb{N}$ and not bother with the extra phase $\Da$ Eq.~\eqref{eq:DeltaTh}?  If one starts with, for instance, a one dimensional problem for which the function $E(n)$ is analytically known (as for the hydrogen atom for which $E(n) = -R/n^2$, or the Bose-Hubbard monomer for which $E(n) = (U/2) n (n-1)$, it makes indeed little sense to expand away from an integer (irrelevant to whether $\Nav$ is an integer or not).  On the other hand, if the information about the system is essentially classical and one gets to the eigenstates through EBK quantization, it is much more effective to clearly split between the classical information (torus on which the initial wave packet is centered, etc...) and to have the consequences of the EBK quantization neatly confined within $\Da$.   For instance, here the wave packet is centered on the north pole, and it is actually the torus going through the north pole that makes the most sense to be used as a reference, and not its quantized neighbors.

\section{Actions and frequencies for the north-pole trajectory}
\label{app:omegas}

\subsection{The action and its derivatives}

For a given value of $S$, the upper north-pole trajectory can be parameterized in terms of $\theta$ as
\begin{equation}\label{eq:separatrix}
     s^*(\theta) = 
    \begin{cases}
        \dfrac{S}{2} \sqrt{1 - (4 \gamma/S)^2  \cos^2 \theta} \quad , &\quad \dfrac{\pi}{2} \leq \theta \leq \dfrac{3\pi}{2} \\
        \dfrac{S}{2} \quad , &\quad  \theta \in [0,\dfrac{\pi}{2} ] \cup [\dfrac{3\pi}{2},2\pi ]
    \end{cases} \quad .  
\end{equation}
The action $m$ is defined as the area between this curve and the top of the Poincaré section $s=S/2$.  With $E^* = \tH(S,s\smeq S/2) = S(S-2\heff)/2$ the energy of the north-pole trajectory,
\begin{equation} \label{eq:mstar-append}
    m(S,E^*) =: m^*(S) = \frac{S}{4\pi} \int_{\pi/2}^{3 \pi/2}  d\theta \, \left( 1 - \sqrt{1 -  (4 \gamma/S)^2 \cos^2 \theta} \right) 
    = 
     \frac{S}{4\pi} \left( \pi - 2 \mathbb{E}\left[\left(\frac{4 \gamma}{S}\right)^2\right] \right) 
\end{equation}
and $\ma := m^*(\Sa)$.

The expression, Eq.~\eqref{eq:separatrix}, assumes the action $S$ and the energy $E\smeq E^*(S)$ are fixed.  To work out the frequencies $\omega_{S,m} := \partial_{S,m} \tilde H$, which involve derivations, we need to relax this constraint. Writing the equation $\tH(S,s) = E$ as  $H_S(S) + s^2 -E =  \gamma \sqrt{S^2 - 4s^2} \cos \theta$, with $H_S = S(S-4\heff)/4$, (see Eq.~\eqref{eq:H0Hs}) and squaring both sides of the equality  gives the fourth order equation 
\begin{equation}
    s^4 +  2s^2 \left[ H_S(S) + 2 \left(  \gamma  \cos \theta \right)^2  -E \right] 
     + \left[ \left( H_S(S) - E \right)^2 - \left(  \gamma S \cos \theta \right)^2 \right] = 0 \quad .
\end{equation}
This equation has four solutions, but only the two positive ones near the top of the Poincaré section are of interest.  Noting $\delta E = (E-E^*) = (E- H_S(S) -S^2/4)$, these two solutions can be expressed as 
\begin{equation}
    s_\pm(\theta;E,S) 
    = \left[\left(  \frac{S^2}{4} + \delta E -  2 \gamma^2\cos^2 \theta \right)  \pm 2 \gamma \cos \theta \sqrt{-\delta E  +   \gamma^2 \cos^2 \theta } \right]^{1/2}   \geq 0\; , 
\end{equation}
(note $  s_-(\theta) =   s_+(\theta+\pi)$).  One can check that for $\delta E < 0$, the curve describing the libration-like motion is $s_+(\theta)$, $\theta \in [-\pi,+\pi]$, and for $\delta E > 0$, the curve describing the rotation-like motion is made of both sheets $s_+(\theta)$ and $s_-(\theta)$ for $\theta \in [\pi -\theta_M,\pi +\theta_M]$, with $\theta_M = \arccos \left( \sqrt{\delta  E }/\gamma\right)$ .

Keeping in mind that for the libration-like motion $\delta E<0$, the action $m$ is the area between the curve $s_-(\theta)$ and the top of the Poincaré section $s=S/2$, it turns out that
\begin{align}
    m(S,\delta E \leq 0) & = \frac{1}{2\pi}  \int_{-\pi}^{+\pi} d\theta \,\left(\frac{S}{2} - s_+(\theta) \right)
    = \frac{1}{2\pi} \int_{-\pi/2}^{+\pi/2} d\theta \, \left(S -(  s_+(\theta) +   s_-(\theta))  \right) \label{eq:mEinf} \\
    m(S,  \delta E \geq 0) &= \frac{1}{2\pi} \int_{\pi-\theta_M}^{\pi +\theta_M} d\theta \,  (  s_-(\theta) -   s_+(\theta)) = \frac{1}{2\pi} \int_{-\theta_M}^{+\theta_M} d\theta \,  (  s_+(\theta) -   s_-(\theta))  \label{eq:mEsup} 
\end{align}
where on the right hand side of both equations $s_-(\theta) =  s_+(\theta+\pi)$ has been used.  Noting that for $\theta \in [\pi/2,+3\pi/2]$, $s_-(\theta;\delta E \smeq 0) = S/2$ and  $ s_+(\theta;\delta E \smeq 0) = (S/2)\sqrt{1 - (4\gamma/S)^2 \cos^2\theta}$, it can be verified that both expressions indeed recover Eqs.~\eqref{eq:mstar-append} and \eqref{eq:ma} in the limit $(E,S) \to (E^*,\Sa)$.

Differentiating either expression above with respect to the energy $E$ or S, generates the derivatives of the north-pole trajectory action.  Focusing on the form, Eq.~\eqref{eq:mEinf} ($\delta E \leq 0$), which is somewhat simpler to handle (but starting from Eq.~\eqref{eq:mEsup} would give the same result),  one finds 
\begin{equation}\label{eq:pre_E_derivs}
    \frac{\partial^{k_1+k_2} \, m}{\partial E^{k_1} \partial S^{k_2}} ( E^*,\Sa) = \frac{1}{2\pi} \int_{-\pi/2}^{\pi/2} d\theta \left[ S - \left(\frac{\partial^{k_1+k_2} s_+(\theta;E,S)}{\partial E^{k_1} \partial S^{k_2}} + \frac{\partial^{k_1+k_2} s_-(\theta;E,S)}{\partial E^{k_1} \partial S^{k_2}} \right) 
    \right]_{\begin{array}{cc} E= E^* \\ S= \Sa \end{array}} \quad .
\end{equation}
In this way, with $\zeta = (4 \gamma / \Sa)^2$ and including  $\heff$ corrections even for second derivatives, the following relations are obtained,
\begin{subequations} \label{eqs:DmDse}
\begin{align}
    \dfrac{\partial m}{\partial E} \bigg\vert_{ E^*,\Sa} 
            &= -\dfrac{2\mathbb{K}[\zeta]}{\pi \Sa} 
                \\
    \dfrac{\partial m}{\partial S} \bigg\vert_{ E^*,\Sa} &= \frac14 + 
        \dfrac{(6-8\heffS) \mathbb{K}\left[\zeta \right]}{4\pi} 
        \\
        \dfrac{\partial^2 m}{\partial E^2} \bigg\vert_{ E^*,\Sa} 
            &= \frac{1}{2 \pi  \Sa \gamma ^2} \left[\frac{ \mathbb{E} 
            \left[ \zeta \right]}{1-\zeta} - \mathbb{K}\left[\zeta \right] \right]
        \\
        \dfrac{\partial^2 m}{\partial S^2} \bigg\vert_{ E^*,\Sa} 
            &=  \frac{\Sa}{2 \pi \gamma^2} \left[\frac{
         (\left(1-\heffS \right)^2 - \left(7 - 8 \heffS\right)\frac{\zeta}{16})}{1 - \zeta} \mathbb{E}\left[\zeta \right] 
         - ( \left(1-\heffS \right)^2- \frac{3}{16} \zeta) \mathbb{K}\left[\zeta\right] \right]
        \\
        \dfrac{\partial^2 m}{\partial S \partial E} \bigg\vert_{ E^*,\Sa}  & = 
        \frac{1}{2 \pi\gamma ^2
\left[
\left(1-\heffS\right)( \mathbb{K}[{\zeta}]-2 \mathbb{E}[{\zeta}])+
\frac{\left(1-\heffS \right)-\left(\frac{7}{4} - 2\heffS \right)\zeta}{1-{\zeta}} \mathbb{E}[{\zeta}]
\right] \; .
   }
\end{align}
\end{subequations}
where $\mathbb{K}$ and $\mathbb{E}$ are Jacobi's complete elliptic integrals of first and second kinds, respectively.  

\subsection{The frequency and its derivatives}

Consider the calculation of the angular frequencies $(\omega_m \smeq {\partial_m\tH}\vert_{S}, \omega_S \smeq {\partial_S \tH} \vert_{m})$ and their derivatives in terms of the various derivatives of $m(E,S)$.  For $\omega_m$,  
\begin{equation}
    \omega_m  = \frac{\partial \tH}{\partial m} \bigg\vert_{S} = \left( \frac{\partial m}{\partial E}\bigg\vert_{S} \right)^{-1} \quad .
\end{equation}
Thus,
\begin{equation}
    \omega_{m}^\alpha = - \frac{\pi \Sa}{2\mathbb{K}(\zeta)} < 0
    \; . \label{eq:omega-app}
\end{equation}
For the first derivative with respect to action apply the chain rule in $\omega$:
\begin{equation}
    \frac{\partial \omega_m}{\partial m}\bigg|_{S} 
    = \frac{\partial}{\partial m} \left( \dfrac{\partial E}{\partial m} \right)\bigg|_{S} 
    = \dfrac{\partial E}{\partial m}\bigg|_{S}  \frac{\partial}{\partial E} \left( \dfrac{\partial m}{\partial E}\bigg|_{S}  \right)^{-1}
    = -  (\omega_m)^3 \, \dfrac{\partial^2 m}{\partial E^2}\bigg|_{S} \quad .
\end{equation}
For the derivatives with respect to $S$, and further derivatives  with respect to $m$, note that for a generic function $f(E,S)$, $f(m,S) := f(E(m,S),S)$ and thus 
\begin{align}
(\partial_S f)\Big\vert_m & = (\partial_S f)\Big\vert_E + (\partial_E f)\Big\vert_S (\partial_S E)\Big\vert_m =  (\partial_S f)\Big\vert_E + \omega_S (\partial_E f)\Big\vert_S  \\
(\partial_m f)\Big\vert_S & = (\partial_E f)\Big\vert_S  (\partial_m E)\Big\vert_S = \omega_m (\partial_E f)\Big\vert_S \; .
\end{align}
This gives 
\begin{subequations} \label{eqs:DomegaDs}
\begin{align}
    \frac{\partial \omega_{m}}{\partial S}\bigg|_{m} & = 
        - \omega_m^2 \frac{\partial^2  m }{\partial E \partial S}
         -\frac{\partial  \omega_{m} }{\partial m}\bigg|_{S} 
         \frac{\partial m}{\partial S}\bigg|_{E} \\
    \omega_S   = 
        \frac{\partial \tH}{\partial S} \bigg\vert_{m} 
        & = -\left( \frac{\partial m}{\partial E}\bigg\vert_{S} \right)^{-1} 
        \left( \frac{\partial m}{\partial S}\bigg\vert_{E} \right) 
        = - \omega_m  \frac{\partial m}{\partial S}\bigg\vert_{E} \\
    \frac {\partial \omega_S}{\partial S}\bigg\vert_{m}  &    =
        - \frac{\partial \omega_{m}}{\partial S}\bigg|_{m}  
        \frac{\partial m}{\partial S}\bigg\vert_{E} 
        - \omega_m  \frac{\partial^2 m}{\partial S^2}\bigg\vert_{E} 
        - \omega_m \omega_S \frac{\partial^2 \, m}{\partial S \partial E} \; .
\end{align}
\end{subequations}
Furthermore, 
\begin{equation}
    \frac {\partial \omega_S}{\partial m}\bigg\vert_{S}  
    =
    - \frac{\partial \omega_{m}}{\partial m}\bigg|_{S}  \frac{\partial m}{\partial S}\bigg\vert_{E} 
    - \omega_m  \frac{\partial }{\partial m}\bigg\vert_{S} 
   \frac{\partial m}{\partial S }\bigg\vert_{E}  
   = - \frac{\partial \omega_{m}}{\partial m}\bigg|_{S}  \frac{\partial m}{\partial S}\bigg\vert_{E} 
    - \omega_m^2  
   \frac{\partial^2 m}{\partial S \partial E}   \; .
\end{equation}
where  $(\partial_S \omega_{m}) = (\partial_m \omega_{S}) = (\partial^2_{mS} \widetilde H)$ as expected.

This information is summarized in Table~\ref{table:omegas}, where the derivatives of $m(E,S)$, namely $(\partial_E m)$, $(\partial_S m)$, $(\partial^2_{EE} m)$, $(\partial^2_{SS} m)$ , $(\partial^2_{ES} m)$, are evaluated at $( E^*,\Sa)$.
\begin{table}[!h]
    \centering
    \begin{tabular}{|l|}
    \hline
         $\omega_m^\alpha   = ((\partial_E m)^\alpha)^{-1}$\\
        $\omega_S ^\alpha =   - \omega^\alpha_m (\partial_S m)^\alpha$\\
        $(\partial_m \omega_m)^\alpha  = - {\omega^\alpha_m}^3 (\partial^2_{EE} m)^\alpha$\\
        $  (\partial_S \omega_m)^\alpha  =  -{\omega^\alpha_m}^2((\partial^2_{ES} m)^\alpha + \omega^\alpha_S (\partial^2_{EE} m)^\alpha)$\\
        $(\partial_S \omega_S)^\alpha  = -\omega^\alpha_m ((\partial^2_{SS} m)^\alpha+2 \omega^\alpha_S(\partial^2_{ES} m)^\alpha + {\omega^\alpha_S}^2 (\partial^2_{EE} m)^\alpha)$\\
        \hline
    \end{tabular}
    \caption{Recap of the expressions of the frequencies $(\omega_m,\omega_S)$ and of their derivative (as functions of $(m,S)$). }   \label{table:omegas}
\end{table}

\section{Wigner function of the initial state}
\label{app:wigner}

Starting from Eq.~\eqref{eq:W0-rel} 
\begin{equation} \label{eq:W0app-rel}
    W_{\alpha}(S,s,\Theta,\theta) = \frac{1}{(\pi \heff)^{2}} \exp \left\{ - \frac{2}{\heff} \left[ S + \mathcal{S}_\alpha 
    - 2 \sqrt{\mathcal{S}_\alpha \left( \frac{S}{2} + s \right)} \, \cos  \theta_1\right] \right\} \quad , 
\end{equation}
with $\theta_1 = \Theta + {\theta}/{2} $.  The goal here is to derive the form of the initial state Wigner transform in the action angle coordinates $(\bI,\bPhi)$ (with $\bI=(S,m)$ and $\bPhi = (\varphi_S,\varphi_m)$).  To proceed, introduce  $\delta s \equiv S/2 - s$, which represents the unperturbed action $m$ at $\gamma = 0$.   Since the initial state is localized at the top of the $(s,\theta)$ plane, $s = S/2 - \delta s$ can be substituted in Eq.~\eqref{eq:W0app-rel} and expanded up to $\mathcal{O}(\delta s)$: 
\begin{equation} \label{eq:W0app-ds}
     W_\alpha \left( S,\delta s,\Theta,\theta \right) = \frac{1}{(\pi \heff)^{2}} \exp \left[ -\frac{2( S + \mathcal{S}_\alpha)}{\heff} \right] \exp \left[ \frac{4}{\heff} \sqrt{ \frac{\mathcal{S}_\alpha}{S}} \left( 
    S - \frac{\delta s}{2} \right) \cos \theta_1  \right] \quad . 
\end{equation}
Equation~\eqref{eq:st-to-mp} of Appendix~\ref{app:top-dyn} gives 
\begin{equation}
    \delta s = {\hma \varphi_m^2} + \frac14 \dfrac{\delta m^2}{\hma} 
\end{equation}
(with $\hma \simeq \ma$ given by Eq.~\eqref{eq:hma}), and thus
\begin{align}
         W_\alpha \left(  S,\delta m,\Theta,\varphi_m \right) 
         &= \frac{1}{(\pi \heff)^{2}} \exp \left[ -\frac{2( S + \Sa)}{\heff} \right] 
         \exp \left\{ \frac{4}{\heff} \sqrt{ \frac{\Sa}{S}} 
         \left( S - \frac12  \hma \varphi_m^2 - \frac18 \dfrac{ \delta m^2}{\hma} \right) \right. \notag \\
    &\qquad \qquad \times\left. \cos \left( \Theta + \frac{1}{2} \theta(\varphi_m, \delta m) \right)  \right\} \quad , \label{eq:app-wigfullbis}
\end{align}
where the precise form of $\theta(\varphi_m, \delta m)$ given by Eq.~\eqref{eq:theta-of-deltamphim} is, for now, unimportant.
For the mean value of the observable discussed in Sec.~\ref{sec:semirev} only the action dependence of the Wigner function is needed.  This can be obtained directly by integrating Eq.~\eqref{eq:app-wigfullbis} on the angular variable $(\Theta,\varphi_m)$.  Thus, to within the stationary phase approximation (and noting that in the integral over $\Theta$, the term $+ \frac{1}{2} \theta(\varphi_m, \delta m)$ can be eliminated by a change of variable),
\begin{equation} \label{eq:W0app-dmS}
    \Wr_\alpha(\delta m, S) = \frac{1}{2 \pi \heff} \left( \frac{1}{\mathcal{S}_\alpha \hma} \right)^{1/2} \exp \left\{ -\frac{2}{\heff} \left[ S + \mathcal{S}_\alpha - 2 \sqrt{ \frac{\mathcal{S}_\alpha}{S}} \left( 
    S - \frac18 \dfrac{\delta m^2}{\hma } \right) \right] \right\} \; .
\end{equation}
This form keeps the full $S$ dependence in the exponent, but simplifies to the stationary values in the prefactor.  The form, Eq.~\eqref{eq:WaGauss}, given in Sec.~\ref{sec:Wigner} corresponds to a Gaussian approximation (for the $S$ variable) of this expression.

When computing the autocorrelation function in Sec.~\ref{sec:autocor}, the full angle dependence of the initial Wigner function is required, and thus $\theta_1$ must be expressed in terms of the action angle coordinates.  However, the convention used for the origin of the angles $(\varphi_S,\varphi_m)$ is that $\theta_1(S,\delta m,\varphi_S \smeq 0,\varphi_m \smeq 0) = 0$ for all values of $S$ and $\delta m$ so that the $\cos \theta_1$ in the Wigner function exponent is indeed a maximum for $(\varphi_S \smeq 0,\varphi_m \smeq 0)$.  Therefore, on the north pole $\partial_m \theta_1 = \partial_S \theta_1 = 0$, and only the $(\varphi_S,\varphi_m)$ dependence of $\theta_1$ is required for the Gaussian approximation of the Wigner function, which can be computed on the north-pole trajectory (and close to the north pole).  In this way (cf Eq.~\eqref{eq:theta1-Wigner})
$\theta_1 = \varphi_S + \frac{\varphi_m}{4}  \left( 1 + \frac{S}{\omega^\alpha_m} \right)$, 
\begin{align}
         W_\alpha \left( S,\delta m,\varphi_S,\varphi_m \right) 
         &= \frac{1}{(\pi \heff)^{2}} \exp \left[ -\frac{2( S + \mathcal{S}_\alpha)}{\heff} \right] 
         \exp \left\{ \frac{4}{\heff} \sqrt{ \frac{\mathcal{S}_\alpha}{S}} 
         \left( S - \frac12 {\hma \varphi_m^2} - \frac18  \dfrac{ \delta m^2}{\hma} \right) \right. \notag \\
    &\qquad \qquad \times \left. \cos \left[ \varphi_S + \frac{\varphi_m}{4}  \left( 1 + \frac{S}{\omega^\alpha_m} \right) \right] \right\} \quad . \label{eq:app-wigfull}
\end{align}
Eq.~\eqref{eq:wigner-full} is a Gaussian approximation (for the variable $S$), of this expression, and  integrating Eq.~\eqref{eq:app-wigfull} over $(\varphi_S,\varphi_m)$ recovers Eq.~\eqref{eq:W0app-dmS}.

\section{Dynamics near the top of the Poincaré section}
\label{app:top-dyn}

To study the linearized motion near the north pole, it is best to start from  the spherical representation Eqs.~\eqref{eq:ham-toupie2}-\eqref{eq:toupie}.  More specifically, it is necessary to describe the motion in the very small region of phase space in which the initial Wigner function Eq.~\eqref{eq:W0app-ds} is significant, which implies, in particular, that $\delta s = (S/2) -s = 0(\heff)$.  Keeping  terms of order $\heff^{1/2}$ and neglecting higher order terms, Eq.~\eqref{eq:spherical} reads
\begin{align}
    \ell_x \simeq & \sqrt{S \delta s} \cos \theta = O(\heff^{1/2}) \notag \\
    \ell_y \simeq &  \sqrt{S \delta s} \sin \theta  = O(\heff^{1/2}) \qquad .
    \label{eq:spherical-HO}\\  
    \ell_z \simeq & \frac{S}{2} = O(\heff^{0})\; \notag
\end{align}
In this approximation, the Poisson bracket $\{\ell_x, \ell_y \} = (S/2)$, which means that up to the scaling $\sqrt{S/2}$, the pair $(\ell_x,\ell_y)$ can be considered as a canonical pair.  For a given value of $S$, the Hamiltonian Eq.~\eqref{eq:ham-toupie2} reads
\begin{equation}
    \tH  (\ell_x,\ell_y) = \gamma^2 - ((\ell_x + \gamma)^2 + \ell_y^2) + \text{const.}
\end{equation}
and therefore corresponds to a harmonic oscillator centered in $(-\gamma,0)$, and associated with the equation of motion
\begin{align}
    \dot \ell_x & = \{\ell_x,\tH  \} = - S \ell_y     \notag  \\
    \dot \ell_y & = \{\ell_y,\tH  \}  =+ S(\ell_x  + \gamma)  \label{eq:toupie-HO} \; .
\end{align}

Introducing $Z = (\ell_x + \gamma) + i \ell_y$, Eq.~\eqref{eq:toupie-HO} reads $\dot Z = i S Z$, and is therefore integrated as $Z(t) = Z_0 \exp(i S t)$.  Therefore, for  trajectories starting on the line $\ell_y=0$ (which by convention corresponds to $\varphi_m=0$) 
\begin{equation}
    \begin{aligned} \label{eq:ho-lxly}
    \ell_x(t)  &=  (\ell_x^0  + \gamma) \cos(St)  - \gamma   \;  \\
   \ell_y(t)  & =  (\ell_x^0 + \gamma)  \sin(St) \; .
\end{aligned}
\end{equation}
We stress that this approximation is only valid near the north pole, and in general not on the entire trajectory started near the north pole.  When the trajectories wander away, for the most part the approximation Eq.~\eqref{eq:spherical-HO} is not valid.  In the perturbative regime $\gamma \ll 1$, however, the motion of trajectories starting near the north pole follow a harmonic oscillator motion of frequency $S$ centered in $(-\gamma,0)$, and the $\lim_{\gamma \to 0} \omega^\alpha_m = -S$.  The action along the north-pole trajectory is in this approximation 
\begin{equation} \label{eq:mho}
    \ma^{\rm h.o.} = \frac{2}{S} \frac{1}{2\pi}\oint_{\genfrac{}{}{0pt}{2}{\ell_x^0 = 0}{\ell_y^0 = 0}}  \ell_x  d\ell_y = \frac{\gamma^2 }{S} \; .
\end{equation}

The next step is to relate the coordinates $(\delta s, \theta)$ to the action-angle variables $(m,\varphi_m)$ in order to express the Wigner function in these latter variables.  For this, consider that the energy $\tH (S,m)$ is only a function of the action variable, and thus assuming $S$ fixed, and noting $\delta m = (m - \ma)$, $\delta E = \tH (S,m) - \tH (S,\ma)$, and since $\tH (S,\ma) = \tH (\ell_x \smeq 0, \ell_y \smeq 0)$,
\begin{equation*}
     \delta E   = \omega^\alpha_m \delta m  = - ((\ell_x + \gamma)^2 + \ell_y^2)   + \gamma^2 = 
      - 2 \gamma \ell_x - (\ell_x ^2 + \ell_y^2) = - 2 \gamma \sqrt{S \delta s}\cos\theta + 0(\heff) \; .
\end{equation*}
In the same way, $\varphi_m = \omega^\alpha_m t$, so that in leading $\heff$ order 
\[
\ell_y = \sqrt{S\delta s} \sin \theta = \gamma \sin (St)  = \gamma \sin \left( \frac{S \,\varphi_m}{\omega^\alpha_m} \right)\; .
\]
Let us introduce the parameter 
\begin{equation} \label{eq:hma}
\hma \equiv   \frac{ S}{(\omega^\alpha_m)^2  } \gamma^2 \; .
\end{equation}
Since $\lim_{\gamma \to 0} \omega^\alpha_m = -S$,  we see that $\hma = \ma^{\rm h.o.} + O(\gamma^4) = \ma + O(\gamma^4) $.   The parameter $\ma$ is thus essentially equivalent to the north-pole action $\ma$ in the perturbative regime, and actually remains numerically very close  to it in the whole domain of $\gamma $ of interest.  We shall thus use this notation to indicate that, although it is a slightly different quantity,  for all practical purposes, one can think of $\hma$ as the action of the north-pole trajectory.  With this notation, and keeping again only the leading  $\heff$ order 
\begin{equation}
    \begin{aligned}
    \delta m      & =-  2 \sqrt{\hma \delta s} \,  \cos \theta\\
    \varphi_m   & =  +\dfrac{\delta s}{\hma} \, \sin \theta \; .
\end{aligned}   \label{eq:st-to-mp}
\end{equation}
As a check, the Jacobian of the transformation is indeed
\begin{equation}
\left |
\begin{matrix}
    \frac{\partial m }{\partial s} & \frac{\partial  m}{\partial \theta}  \\
    \frac{\partial \varphi_m }{\partial s} & \frac{\partial  \varphi_m}{\partial \theta}
\end{matrix}
\right|
= 
\left |
\begin{matrix}
    - \sqrt{\frac{\hma}{ \delta s}} \cos\theta   &    2 \sqrt{\hma  \delta s} \sin\theta  \\
   \frac{1}{\sqrt{2 \hma \delta s} }  \sin \theta & \sqrt{\frac{\delta s}{\hma}} \cos \theta 
\end{matrix}
\right| = -1 \ .
\end{equation}

Equation~\eqref{eq:st-to-mp} gives, in particular, that 
\begin{equation} \label{eq:delta-s}
    \delta s = {\hma \varphi_m^2} + \frac14 \dfrac{\delta m^2}{\hma} 
\end{equation}
Note that the north pole, in which the initial coherent state is localized, is a singular point for the $(m,\theta)$ representation since the angle $\theta$ is undefined there.  Therefore, it cannot be assumed that the variation of $\theta$ near the north pole is small, which is why the full $(\sin \theta, \cos \theta)$ dependence has to be kept.  Specifically, 
\begin{equation} \label{eq:theta-of-deltamphim}
    \theta(\delta m,\varphi_m) = - \arctan \left(2 \varphi_m\dfrac{\hma}{\delta m} \right) \, 
\end{equation}
so that in the immediate neighborhood of the north pole, $\theta$ varies in the full range $[0,2\pi]$.  This singularity is also the reason why $\delta s = O(\heff)$ instead of the expected $ O(\heff^{1/2})$.

\section{Dynamics on the north-pole trajectory}
\label{app:DynSeparatrix}

The semiclassical calculation of $\mean{a_2}(t)$ in Sec.~\ref{sec:a2dimer} follows essentially the same route as for $\mean{n_2}$, except basically that the classical symbol associated with $\hat n_2$, namely $s_2/\heff$, has to be replaced with the one associated with $\hat a_2$, namely  $\sqrt{s_2/\heff} \exp(-i\theta_2)$.  There is, however, one non-trivial technicality, which is to relate the ``physical'' angle  $\theta_2$ to the angles $(\varphi_S,\varphi_m)$ of the action-angle coordinates as expressed by Eq.~\eqref{eq:theta2-phis}.  The goal of this appendix is to derive this relation.

Therefore, focus on the trajectory $(\ell^*_x(t),\ell^*_y(t))$ starting from $(\ell^*_x(0),\ell^*_y(0)) = (0,0)$, i.e.\ going exactly through the north pole, and which thus corresponds to $\delta m = 0$ (also use the notation $\theta^*(\varphi_m)$, $\Theta^*(\varphi_S,\varphi_m)$, and $\theta_{1,2}^*(\varphi_S,\varphi_m)$, for the variation of these angles on the north-pole trajectory). Consider what is happening near $t=0$, in which case the dynamics corresponds to Eq.~\eqref{eq:ho-lxly} with $\ell_x^0 = 0$, namely
\begin{align*}
    \ell^*_x(t)  &=  \gamma( \cos(St)  -  1)     = \gamma \left( \cos\left(\frac{S \varphi_m}{\omega_m^\alpha}\right)  -  1\right)  = \sqrt{S\delta S} \cos \theta \;  \\
   \ell^*_y(t)  & =  \gamma \sin(St)  =  \gamma \sin \left(\frac{S \varphi_m}{\omega_m^\alpha}\right)  = \sqrt{S\delta S} \sin \theta \;  ,
\end{align*}
so that , keeping in mind that $\omega_m^\alpha < 0$,
\[ 
\tan \theta = \frac{\sin\left(\frac{S \varphi_m}{\omega_m^\alpha}\right)}{\cos\left(\frac{S \varphi_m}{\omega_m^\alpha}\right)  -  1} \simeq \frac{2 \omega_m^\alpha}{S \varphi_m} \xrightarrow{\varphi_m^\alpha \to 0^{\pm}} = \mp \infty \; .
\]
The angle theta therefore switches from $3\pi/2$ to $+\pi/2$ when traversing the north pole, and therefore is subject to a $-\pi$ discontinuity (which is rather obvious when looking at the trajectory).  

Going back to the original $(s_1,s_2,\theta_1,\theta_2)$ variables with
\begin{align*}
    \theta_1 & = \Theta + \frac{\theta}{2} \\
    \theta_2 & = \Theta - \frac{\theta}{2}  \; ,    
\end{align*}
the north pole corresponds to $s_2 = 0$, where of course $\theta_2$ can show $\pm \pi$ discontinuities along a trajectory, but $s_1 \neq 0$, and thus $\theta_1$ cannot be singular.  Therefore, whenever $\theta \rightarrow \theta - \pi$, $\Theta \rightarrow \Theta + \pi/2$, leading to $(\theta_1 \rightarrow \theta_1,\theta_2 \rightarrow \theta_2 +\pi)$.    Thus, each time $\varphi_m$ goes through $0$, the angle $\Theta$ picks up a $+\pi/2$ discontinuity.

Now, for instance, from the form of Eq.~\eqref{eq:Htilde} of the Hamiltonian, 
\[
\dot \Theta^* = \frac{\partial \tH }{\partial S} = \left(\frac{S}{2} - \heff \right) - \frac{S}{2}\frac{\gamma \ell^*_x}{(\ell^*_x)^2 + (\ell^*_y)^2} \; ,
\]
with $(\ell^*_x)^2 + (\ell^*_y)^2 = (S/2)^2 - s^2$.     However, from Eq.\eqref{eq:ham-toupie2}, the energy on the north-pole trajectory is $\tH  (S,\ell^*_x ,\ell^*_y ) = \tH  (S,\ell^*_x \smeq 0,\ell^*_y \smeq 0) = S(S-2\heff)/2$, which implies $((\ell^*_x)^2 + (\ell^*_y)^2) = - 2\gamma \ell^*_x$, and thus 
\begin{equation} \label{eq:magic}
\frac{\gamma \ell^*_x}{(\ell^*_x)^2 + (\ell^*_y)^2} = - \frac12 = \text{const.}    
\end{equation}
Therefore,
\begin{equation} 
    \dot \Theta^*  =  \left(\frac{3}{4}S - \heff \right)  + \frac{\pi }{2} \sum_n {|\omega_n|}\delta(\varphi_m - 2\pi n) \label{eq:dotTheta} \ .
\end{equation}
Equation~\eqref{eq:omegaS-omegam} in Sec.~\ref{eq:input} can be interpreted as understanding $\omega_S$ as the average of $\dot \Theta^*$ over a period $T^*_m$, namely
\begin{equation} \label{eq:omega-S}
\omega_S = \left(\frac{3}{4}S - \heff \right)  - \frac{1}{4} \omega_n \; .
\end{equation}
Furthermore, from Eq.~\eqref{eq:omega-S} it can be deduced that (on the north-pole trajectory, but away from the north pole)
\[
\dot \Theta^* = \dot \varphi_S + \frac14 \dot \varphi_m \; .
\]
Assuming the origin of $\varphi_m$ is on the north pole and the origin of $\varphi_S$ is such that $\theta_1^*(\varphi_S \smeq 0, \varphi_m \smeq 0) = 0$ (which implies $\Theta_1^*(\varphi_S \smeq 0, \varphi_m \smeq 0^-) =  -\pi/4$),  and taking into account the $\pi/2$ discontinuities of $\Theta^*$ at each traversal of the north pole, gives
\begin{align*}
    \Theta^*   & = \varphi_S +  \frac14 \varphi_m + \frac{\pi}{2} \floor{-\varphi_m/2\pi}  - \frac{\pi}{4}\; , \\
    \theta^* &= \ttheta(\varphi_m)  = \tttheta(\varphi_m) - \pi \floor{-\varphi_m/2\pi}  + \frac{\pi}{2}\; ,
\end{align*}
where  $\ttheta(\varphi_m)$ is the (discontinuous) function,  Eq.~\eqref{eq:tilde-theta}, introduced in Sec.~\ref{sec:sth-dyn} and $\tttheta(\varphi_m)$ given by Eq.~\eqref{eq:tilde-tilde-theta} its continuous version shifted by $\pi/2$ so that $\tttheta(\varphi_m \smeq 0) = 0$.
With $\theta^*_2 = \Theta^* - \theta^*/2$, and  $\theta^*_1 = \Theta^* + \theta^*/2$, one has {\em on the north-pole torus}
\begin{align} 
 \theta^*_1(\varphi_S,\varphi_m)& = \varphi_S + \frac14(\varphi_m  + 2 \tttheta(\varphi_m) )\; .
            \label{eq:theta1-phis-app} \\
      \theta^*_2(\varphi_S,\varphi_m) & = \varphi_S + \frac14(\varphi_m  - 2 \tttheta(\varphi_m) ) +  \pi \floor{-\varphi_m/2\pi} - \frac{\pi}{2}\; ,
      \label{eq:theta2-phis-app} 
\end{align}

 When computing the initial Wigner function in action-angle coordinates, what is required is the function $\tttheta(\varphi_m)$ near the north pole, i.e.\ for $\varphi_m = O(\hbar^{1/2})$.  For this, the form of Eq.~\eqref{eq:Htilde} of the Hamiltonian and Eq.~\eqref{eq:magic} can be used to obtain
\[
\dot \theta^* = \frac{\partial \tH }{\partial s} = 2s^* \left( 1 - \frac{\gamma \ell^*_x}{(\ell^*_x)^2 + (\ell^*_y)^2} \right) = s^* \; ,
\]
and therefore $\dot \theta^* (\varphi_m \smeq 0^+) + (S/2)$.  Using the chain rule, near the north pole 
$\dot \theta^* = \omega^*_m (d\tttheta/d \varphi_m) \simeq S/2 $, so that $\tttheta(\varphi_m) = \frac{S}{2 \omega^*_m} \varphi_m + O(\varphi_m^2)$, and within the Wigner function
\begin{equation} \label{eq:theta1-Wigner}
    \theta^*_1(\varphi_S,\varphi_m)  = \varphi_S + \frac14\varphi_m \left(1 + \frac{S}{\omega^*_m} \right)\; .
\end{equation}

\section{Behavior of \texorpdfstring{$\GGSe(\trevo)$}{Delta Phim} in the perturbative regime}
\label{app:DeltaPhiS}

In this appendix, we show that, neglecting $O(\heff \gamma^2)$ corrections,  $\GGSe(\trevo) = 0 + O(\gamma^4)$,  which is why $\trevauto $, the time-location of the maximum of $\autocor(t)$, remains essentially constant  in all the perturbative regime :  $\trevauto \to  \trevo + O(\gamma^4)$.
We first remark that for the unperturbed (scaled) Hamiltonian $\tilde H(\gamma \smeq 0)$, $\partial_S \omega_S^0 = -\partial_m \omega_S^0 = -\partial_S \omega_m^0 = 1$ (see Eq.~\eqref{eq:Omegas0}) can be identified with $\partial_{s_1} \omega_1^0 =  \partial_{s_2} \omega_2^0 = 1 $,  and thus we can understand the unperturbed revival time as 
\[
\ttrevo = \frac{1}{\partial_{s_1} \omega_1^0}\frac{2\pi}{\heff}   = - \frac{1}{\partial_m \omega^0_S}\frac{2\pi}{\heff} \; .
\]

Second, we note that from Eq.~\eqref{eq:toupie}, the (exact) fixed point for a finite value of gamma corresponds in the spherical representation to $(\ell_x,\ell_y) = (-\gamma, 0)$, which, reinserted into the Hamiltonian $\tilde H$ Eq.~\eqref{eq:ham-toupie2} gives for all $S$
\[
\tilde H(S,m \smeq 0) = \frac{ S(S-2\heff)}{2} + \gamma^2 
\]
and thus $\omega_S(S,m \smeq 0) = (S-\heff) \equiv \omega_S(S,\ma;\gamma \smeq 0)$.  In other words, the angular frequency $\omega_S$ at the fixed point is independent of $\gamma$, and therefore equal to its value for the north-pole trajectory at $\gamma=0$ (since then the north-pole trajectory is the fixed point).  As a consequence
\[
\omega_S(S,\ma)  =  (\omega_S^0)  + (\partial_m \omega^0_S) \, \ma + O(\gamma^4)\; .
\]

Now, up to $O(\gamma^4)$ terms we have 
\[\GGSe(\ttrevo)  \equiv (\omega_S^\alpha) \trevo + \Delta_S  \simeq \cmod{((\omega_S^0 ) + (\partial_m \omega^0_S) \,  \ma) \ttrevo  +  2\pi \ma/\heff} \; ,
\]
so that, noting that $\omega_S^0 \ttrevo = (\Sa-\heff) \cdot (2\pi/\heff) = 2 \pi \Nav $ gives
\[
\GGSe(\ttrevo) = O(\gamma^4) \; .
\]
This is negligible in the perturbative regime, and remains rather small in whole $\gamma$-domain of interest.

\bibliographystyle{SciPost_bibstyle}
\bibliography{general_ref,quantumchaos,manybody,molecular,rmtmodify}

\end{document}